\documentclass[%
 reprint,
 superscriptaddress,preprintnumbers,
 nofootinbib,
 amssymb,
 aps,
]{revtex4-1}

\usepackage{amsmath}
\usepackage{graphicx}
\usepackage{dcolumn}
\usepackage{bm}
\usepackage{xcolor}
\usepackage[normalem]{ulem}
\usepackage{xspace}
\usepackage{slashed}
\usepackage{xfrac}
\usepackage{tabularx}
\newcolumntype{Y}{>{\centering\arraybackslash}X}

\usepackage{hyperref}
\definecolor{red}{rgb}{0.6,.0706,.1373}
\definecolor{blue}{rgb}{0,0.396,0.741}
\colorlet{blueRef}{blue!80!black}
\hypersetup{
	colorlinks,
	bookmarksopen,
	bookmarksnumbered,
	citecolor=blueRef, 		
	linkcolor=blue,	
	urlcolor=blue,			
}

\usepackage{siunitx}
\renewcommand{\L}{\mathcal{L}}
\newcommand{\LL}{\mathrm{L}}
\newcommand{\RR}{\mathrm{R}}
\newcommand{\U}{\mathrm{U}}
\newcommand{\SU}{\mathrm{SU}}
\newcommand{\rep}[1]{\mathbf{#1}}
\newcommand{\repbar}[1]{\overline{\mathbf{#1}}}
\newcommand{\eminus}{\vcenter{\hbox{\scalebox{0.6}[1]{$ - $}}}}	
\newcommand{\dd}{\mathop{}\!\mathrm{d}}

\newcommand{\RK}{$ R_{K^{(\ast)}} $\xspace}
\newcommand{\RD}{$ R_{D^{(\ast)}} $\xspace}
\newcommand{\Umt}{$ \U(1)_{L_\mu\eminus L_\tau} $\xspace}
\newcommand{\Ubl}{$ \U(1)_{B\eminus 3L_\mu} $\xspace}

\begin{document}

\title{A Model of Muon Anomalies}

\author{Admir Greljo}
\author{Peter Stangl}
\author{Anders Eller Thomsen}
\affiliation{Albert Einstein Center for Fundamental Physics, Institute for Theoretical Physics, University of Bern, CH-3012 Bern, Switzerland}

\begin{abstract}

The Standard Model (SM) is augmented with a \Ubl gauge symmetry spontaneously broken above the TeV scale when an SM-singlet scalar condenses.
Scalar leptoquarks $S_{1(3)} = (\repbar{3},\, \rep{1} (\rep{3}),\, \sfrac{1}{3})$ charged under \Ubl mediate the intriguing effects observed in muon $(g-2)$, $R_{K^{(*)}}$, and $b \to s \mu^+ \mu^-$, while generically evading all other phenomenological constraints.
The fermionic sector is minimally extended with three right-handed neutrinos, and a successful type-I seesaw mechanism is realized.
Charged lepton flavor violation is effectively suppressed, and proton decay---a common prediction of leptoquarks---is postponed to the dimension-6 effective Lagrangian.
Unavoidable radiative corrections in the Higgs mass and muon Yukawa favor leptoquark masses interesting for collider searches.
The parameters of the model are radiatively stable and can be evolved by the renormalization group to the Planck scale without inconsistencies.
Alternative lepton-flavored gauge extensions of the SM, under which leptoquarks become \textit{muoquarks}, are proposed for comparison.
\end{abstract}

\pacs{Valid PACS appear here}

\maketitle

\newcommand{\fref}[1]{Fig.~\ref{fig:#1}} 
\newcommand{\eref}[1]{Eq.~\eqref{eq:#1}} 
\newcommand{\erefn}[1]{ (\ref{eq:#1})}
\newcommand{\erefs}[2]{Eqs.~(\ref{eq:#1}) - (\ref{eq:#2}) } 
\newcommand{\aref}[1]{Appendix~\ref{app:#1}}
\newcommand{\sref}[1]{Section~\ref{sec:#1}}
\newcommand{\cref}[1]{Chapter~\ref{ch:.#1}}
\newcommand{\tref}[1]{Table~\ref{tab:#1}}

\newcommand{\nn}{\nonumber \\}  
\newcommand{\nnl}{\nonumber \\}  
\newcommand{\nl}{& \nonumber \\ &}
\newcommand{\bnl}{\right .  \nonumber \\  \left .}
\newcommand{\dbnl}{\right .\right . & \nonumber \\ & \left .\left .}

\newcommand{\beq}{\begin{equation}} 
\newcommand{\eeq}{\end{equation}} 
\newcommand{\ba}{\begin{array}}  
\newcommand{\ea}{\end{array}} 
\newcommand{\bea}{\begin{eqnarray}}  
\newcommand{\eea}{\end{eqnarray} }  
\newcommand{\be}{\begin{eqnarray}}  
\newcommand{\ee}{\end{eqnarray} }  
\newcommand{\bal}{\begin{align}}
\newcommand{\eal}{\end{align}}   
\newcommand{\bi}{\begin{itemize}}  
\newcommand{\ei}{\end{itemize}}  
\newcommand{\ben}{\begin{enumerate}}  
\newcommand{\een}{\end{enumerate}}  
\newcommand{\bc}{\begin{center}}
\newcommand{\ec}{\end{center}} 
\newcommand{\bt}{\begin{table}}
\newcommand{\et}{\end{table}}  
\newcommand{\btb}{\begin{tabular}}
\newcommand{\etb}{\end{tabular}}  
\newcommand{\bvec}{\left ( \ba{c}}
\newcommand{\evec}{\ea \right )}

\newcommand{\cO}{{\mathcal O}} 
\newcommand{\co}{{\mathcal O}} 
\newcommand{\cL}{{\mathcal L}} 
\newcommand{\cl}{{\mathcal L}} 
\newcommand{\cM}{{\mathcal M}}

\newcommand{\const}{\mathrm{const}}

\newcommand{\ev}{ \mathrm{eV}}
\newcommand{\kev}{\mathrm{keV}}
\newcommand{\mev}{\mathrm{MeV}}
\newcommand{\gev}{\mathrm{GeV}}
\newcommand{\tev}{\mathrm{TeV}}

\newcommand{\mpl}{M_{\mathrm Pl}}

\def\mgut{\, M_{\rm GUT}}
\def\tgut{\, t_{\rm GUT}}
\def\mpl{\, M_{\rm Pl}}
\def\mkk{\, M_{\rm KK}}
\newcommand{\msusy}{M_{\rm soft}}

\newcommand{\dslash}[1]{#1 \! \! \! {\bf /}}
\newcommand{\ddslash}[1]{#1 \! \! \! \!  {\bf /}}

\def\ads{AdS$_5$\,}
\def\adse{AdS$_5$}
\def\intdk{\int {d^4 k \over (2 \pi)^4}} 

\def\ra{\rangle}
\def\la{\langle}  

\def\sgn{{\rm sgn}}
\def\pa{\partial}  
\newcommand{\dlr}{\overleftrightarrow{\partial}}
\newcommand{\Dlr}{\overleftrightarrow{D}}
\newcommand{\re}{{\mathrm{Re}} \,}
\newcommand{\im}{{\mathrm{Im}} \,}
\newcommand{\tr}{\mathrm T \mathrm r}  

\newcommand{\Ra}{\Rightarrow}
\newcommand{\lra}{\leftrightarrow}
\newcommand{\llra}{\longleftrightarrow}

\newcommand\simlt{\stackrel{<}{{}_\sim}}
\newcommand\simgt{\stackrel{>}{{}_\sim}}   
\newcommand{\zt}{$\mathbb Z_2$ }

\newcommand{\ha}{{\hat a}}
\newcommand{\hab}{{\hat b}}
\newcommand{\hac}{{\hat c}} 

\newcommand{\ti}{\tilde}  
\def\hc{{\rm h.c.}} 
\def\ov{\overline}  
  
\newcommand{\eps}{\epsilon}

\def\cog{\color{OliveGreen}}
\def\cor{\color{Red}}
\def\copu{\color{purple}}
\def\coro{\color{RedOrange}}
\def\coma{\color{Maroon}}
\def\cob{\color{Blue}}
\def\cobr{\color{Brown}}
\def\cobl{\color{Black}}
\def\cost{\color{WildStrawberry}}

\newcommand{\eL}{\epsilon_L}
\newcommand{\eR}{\epsilon_R}
\newcommand{\eSL}{\epsilon_{S_L}}
\newcommand{\eSR}{\epsilon_{S_R}}
\newcommand{\eT}{\epsilon_T}
\def\slashp{p \!\!\! \slash}
\def\slashs{s \!\!\! \slash}
\def\slashE{E \!\!\! \slash_T}

\section{Introduction}

The discovery of the muon in 1936 by Anderson and Neddermeyer marked the birth of modern particle physics.
Today, we are facing a robust theory of elementary particles and their interactions able to describe the vast phenomena observed in laboratories.
The Standard Model (SM) of particle physics is a great success story, with (nearly) all of its relevant and marginal operators measured.
In addition, it provides a rationale for the absence of long-sought, yet unobserved, phenomena such as lepton flavor violation (LFV).
Truncating the SM operators at the canonical dimension-4 level, leads to an emergence of protective accidental symmetries.
These symmetries are broken by irrelevant operators whose effects are suppressed by the next scale in physics, suggesting a long ride in front of the SM.

While we have indeed come a long way in our understanding of particle physics, there are still striking unresolved issues.
Recently, several muon measurements have sparked new hope of making progress with tantalizing hints of physics beyond the SM.
These are $i)$ the anomalous magnetic moment of the muon $(g-2)_\mu$~\cite{Bennett:2006fi,Aoyama:2020ynm} and $ii)$ the rare $B$ meson decay observables \RK and $b \to s \mu^+ \mu^-$ angular distributions~\cite{Aaij:2014ora, Aaij:2017vbb, Aaij:2013qta, Aaij:2015oid, Aaij:2019wad,CMS:2014xfa,Aaij:2017vad}.
The SM prediction for \RK is extremely clean~\cite{Hiller:2003js,Bordone:2016gaq,Isidori:2020acz} and requires much less effort compared to $(g-2)_\mu$, where we use the prediction from the Muon $g-2$ Theory Initiative~\cite{Aoyama:2020ynm,Colangelo:2020lcg,aoyama:2012wk,Aoyama:2019ryr,czarnecki:2002nt,gnendiger:2013pva,davier:2017zfy,keshavarzi:2018mgv,colangelo:2018mtw,hoferichter:2019gzf,davier:2019can,keshavarzi:2019abf,kurz:2014wya,melnikov:2003xd,masjuan:2017tvw,Colangelo:2017fiz,hoferichter:2018kwz,gerardin:2019vio,bijnens:2019ghy,colangelo:2019uex,Blum:2019ugy,colangelo:2014qya}, but see also~\cite{Borsanyi:2020mff}.
Some of the anomalous measurements imply (while others are consistent with) lepton flavor universality violation (LFUV) between muons and their cousins, electrons and taus.
The most recent update of $R_K$ increased the significance of the anomaly and, for the first time, LHCb declared evidence for LFUV~\cite{Aaij:2021vac}.
Additionally, the most recent update on $(g-2)_\mu$ from the Fermilab Muon $g-2$ experiment~\cite{Abi:2021gix} confirmed the Brookhaven measurement~\cite{Bennett:2006fi} and strengthened the tension with the SM prediction~\cite{Aoyama:2020ynm}.

On general grounds, LFUV is expected to be linked to LFV~\cite{Glashow:2014iga,Giudice:2012ms}.
Consider, for instance, the operator
\begin{equation}
\mathcal{L} \supset - \frac{\sqrt{2} e \,v}{(4 \pi \Lambda_{ij})^2}\, \bar \ell^i_{\LL} \sigma^{\mu \nu} \ell^j_{\RR } F_{\mu \nu} + {\rm h.c.}~,
\end{equation}
where $\Lambda_{ij}$ is the cutoff scale and $v  =\SI{174}{GeV} $.
The $(g-2)_\mu$ measurement is explained for $\Lambda_{22} \sim \SI{15}{TeV}$, whereas the absence of $\mu \to e \gamma$ decays~\cite{TheMEG:2016wtm} sets a limit on $\Lambda_{12 (21)}  \gtrsim \SI{3600}{TeV}$.
Even when the couplings to electrons are flavor-suppressed: $\Lambda_{12 (21)} \sqrt{m_e / m_\mu}  \gtrsim \SI{250}{TeV}$.
Such a dramatic difference in scales is most naturally explained by a high-quality muon-number symmetry, $\U(1)_{L_\mu}$, which forbids flavor violation but allows for non-universality.

An important step in this direction is gauging an anomaly-free lepton-flavored symmetry group such as \Umt~\cite{Baek:2001kca,Ma:2001md,Harigaya:2013twa,Altmannshofer:2014pba,Altmannshofer:2019zhy,Crivellin:2016ejn,Crivellin:2015mga,Crivellin:2018qmi,Altmannshofer:2014cfa,Altmannshofer:2015mqa}.
This idea triggered a fruitful line of research into the anomalies: for example, a very light \Umt gauge boson $X$ can at 1-loop order give the correct effect in $(g-2)_\mu$ while remaining unconstrained by the complementary experiments~\cite{Altmannshofer:2014pba,Altmannshofer:2019zhy}.
In parallel, successful \Umt gauge models for \RK were put forward using a contrasting heavy $X$ field.
The latter models also employ a set of vector-like quarks to mediate the $X$ interactions with the SM quarks~\cite{Altmannshofer:2014cfa,Altmannshofer:2015mqa} (see also~\cite{Bonilla:2017lsq,Allanach:2020kss,Alonso:2017uky,Allanach:2018lvl,Altmannshofer:2019xda}).
Unfortunately, the two anomalies cannot be reconciled within the same parameter space.

A popular alternative is a class of leptoquark models, which provide a simple dynamical solution to both anomalies (see e.g.~\cite{Dorsner:2016wpm,Gripaios:2009dq,Hiller:2014yaa,Bauer:2015knc,Barbieri:2015yvd,Buttazzo:2017ixm,Angelescu:2018tyl,Dorsner:2019itg,Angelescu:2021lln,Hiller:2021pul,Marzocca:2018wcf,Dorsner:2017ufx,Becirevic:2018afm}).
Their advantage is that they induce semileptonic transitions at the tree level, while dangerous $\Delta F= 2$ transitions occur only at the 1-loop order.
The contribution to $(g-2)_\mu$ enters at the 1-loop but can have a chiral enhancement.
In fact, there is a simple leptoquark model featuring two scalar leptoquark fields that can simultaneously explain both anomalies~\cite{Crivellin:2017zlb,Gherardi:2020qhc}.
However, the main drawback of leptoquark models is that they grossly break the SM accidental symmetries at the renormalizable level.
Specifically, marginal dimension-4 operators of the quark-quark ($q q S$) and quark-lepton ($q \ell S$) type predict excessive proton decay and LFV in stark contrast with observations.
A resolution is the idea of combining leptoquarks with a gauged lepton symmetry as in Ref.~\cite{Hambye:2017qix,Davighi:2020qqa}.

Lepton-flavored $\U(1)$ gauge symmetries impose nontrivial restrictions on the structure of the neutrino mass matrices.
This has been thoroughly studied in the literature in the context of two-zero-texture (-minor) structures, aiming at predicting the remaining parameters in the neutrino sector~\cite{Lavoura:2004tu,Ma:2005py,Asai:2017ryy,Asai:2018ocx,Asai:2019ciz}.
Building on Refs.~\cite{Gherardi:2020qhc,Davighi:2020qqa,Asai:2019ciz,Altmannshofer:2019zhy}, we show how to naturally reconcile all muon anomalies in a single framework and rationalize why these are the first signs of physics beyond the SM.
We face the challenge of generating a phenomenologically acceptable neutrino sector in leptoquark models with gauged lepton flavor while keeping the proton stable.
It turns out that the \Ubl gauge symmetry has the desired property of also forbidding dangerous baryon-number-violating dimension-5 operators.

In Section~\ref{sec:minimal} we present the model and discuss the parameter space capable of addressing the anomalies.
Section~\ref{sec:muta} outlines alternative models and a connection with $b \to c \tau \nu$ decays.
We show that no tuned cancellations is needed in the observables, the scalar masses respect finite naturalness, and the couplings can be extrapolated to high-energies without inconsistencies.

\section{A Model for $(g-2)_\mu$ and $b \to s \mu^+ \mu^-$ }
\label{sec:minimal}
We extend the SM+$3\nu_\RR$ with a gauged \Ubl symmetry under which the leptons $\ell_\LL^2, \mu_\RR, \nu_{\mu,\RR} $ have charge $ \eminus 3 $, while $q^i_\LL, u^i_\RR, d^i_\RR$ all have charge $ + \sfrac{1}{3} $.
With this fermion content, the symmetry is anomaly-free.
An SM singlet $\Phi $ with \Ubl charge $q_\Phi = +3$ is responsible for the spontaneous breaking of the new symmetry.
In addition, the matter field content is extended with two scalar leptoquarks $ S_1 = (\repbar{3},\, \rep{1},\, \sfrac{1}{3} ) $ and $ S_3 = (\repbar{3},\, \rep{3},\, \sfrac{1}{3} ) $ of charge $ +\sfrac{8}{3}  $ under \Ubl.

The renormalizable Lagrangian for this model is
\begin{align} \label{eq:Lagrangian}
	\L =&\, \L_{\mathrm{SM} -V_H} + |D_\mu \Phi|^2 + |D_\mu S_1|^2 + |D_\mu S_3|^2 - \tfrac{1}{4} X_{\mu \nu}^2 \nonumber\\
	&- \big(\eta^{3\LL}_i \, \overline{q}^{c\,i}_\LL  \ell^2_\LL \, S_3 - \eta^{1\LL}_i \overline{q}^{c\,i}_\LL \ell^2_\LL S_1 - \eta^{1\RR}_i \overline{u}^{c\,i}_\RR \mu_\RR S_1 \nonumber\\
	&- \tilde{\eta}^{1\RR}_i \overline{d}^{c\,i}_\RR \nu_{\mu,\RR} S_1  + \hc \big)
	+\tfrac{1}{2}\varepsilon_{BX} B_{\mu \nu} X^{\mu \nu} \\
	&- V_{H\Phi}(H,\Phi) - V_{13}(H,\Phi,S_1,S_3) + \bar \nu^{i}_\RR i \slashed{D} \nu_\RR^i \nonumber\\
	&- \big( y_\nu^{i j} \bar \ell^i_\LL \tilde H \nu_\RR^j + M^{ij}_\RR \bar \nu^{c i}_\RR \nu_\RR^j + y_\Phi^{ij} \Phi\,  \bar \nu^{c i}_\RR \nu_\RR^j + \hc \big)~,\nonumber
\end{align}
where the $ \SU(2)_\LL$ contraction in the left-handed Yukawa couplings is with $i \sigma^2\sigma^a $ for $S_3$ and $ i \sigma^2$ for $S_1$ with Pauli matrices $ \sigma^a$.

The gauge symmetry ensures that leptoquarks couple exclusively to 2nd generation leptons through the Yukawa couplings $\eta^x$. Leptoquarks coupling exclusively to 2nd generation leptons are properly referred to as \textit{muoquarks}.
We will show how this structure reconciles the muon anomalies with the complementary constraints.
We then separately address the scalar potential and the kinetic mixing, as it has minimal baring on the flavor analysis: in 1-loop matching it only gives corrections on top of the operators already generated at tree-level.

Finally, due to the extra gauge symmetry, the model has accidental baryon and individual lepton number symmetries at the renormalizable level just like the SM.
Furthermore, there is an accidental baryon number conservation at the level of dimension-5 operators.
It is an intricate relation between $i)$ neutrino masses and mixings, $ii)$ matter stability, and $iii)$ the high-quality $ \U(1)_{L_\mu}$ global symmetry, which ultimately leads to the choice of the \Ubl gauge symmetry.
An alternative choice is discussed in Section~\ref{sec:muta}.

\subsection{Muoquark solution of the muon anomalies} \label{sec:LQ_pheno}
We assume that $ \Phi $ develops a large VEV so as to break \Ubl and decouple the $ X $, $\nu_R^i$, and $\Phi$ fields for the moment.
The remnant of the \Ubl symmetry provides an effective $\U(1)_{L_\mu} $ global symmetry under which the muoquarks are charged.
This forbids LFV processes such as $\mu \to e \gamma$ but introduces new lepton non-universal muophilic interactions.
The idea is to use a tree-level $ S_3 $ exchange to explain the \RK anomalies and an $ S_1 $ loop for the $ (g-2)_\mu $.

The gauge symmetry fixes the lepton flavor coupling to $S_{1,3}$ but not the quark flavor structure of $\eta^x_i$.
The SM Yukawa interactions exhibit a good approximate flavor symmetry $\U(2)_q \times \U(2)_u \times \U(2)_d$ under which the first two generations transform as doublets, while the third generation is a singlet~\cite{Barbieri:2011ci} (see also~\cite{Kagan:2009bn}).
When this symmetry is exact, only the top and bottom quarks are massive and the CKM matrix is the identity.
A slight breaking, needed to fit data, is minimally provided by the leading breaking spurion $V = (V_{td}, V_{ts})^\mathrm{T}$, which is a doublet of $\U(2)_q$, together with two bidoublets $\Delta_{u,d}$~\cite{Barbieri:2011ci,Fuentes-Martin:2019mun}.
Thinking about this symmetry as a remnant of deep UV dynamics, it is reasonable to assume the muoquark Yukawa couplings share a similar structure.
In particular, we expect the left-handed couplings to be $\eta^{1(3)\LL} \propto \mathcal{O}(V)\oplus 1 $ and the relevant right-handed ones to be $\eta^{1\RR} \propto \mathcal{O}(\Delta_{u}^\dagger V) \oplus 1 $.
This sets the relative size between different quark flavors.
On general grounds we expect the absolute sizes of the couplings and the muoquark masses $M_{1,3}$ to be similar.
Remarkably, when $\eta^{x}_3 = \mathcal{O}(0.1)$ and $M_{1,3} = \mathcal{O}$(TeV), this setup explains $b \to s \mu^+ \mu^-$ and $(g-2)_\mu$ anomalies with negligible corrections to any other complementary constraints. (Note that $\U(2)^3$ is just one example of a CKM-like flavor structure in the quark sector.)

The most general $S_1 + S_3$ renormalizable model is matched to the SM effective field theory at the 1-loop level in Ref.~\cite{Gherardi:2020det}.
We implement these results in a code that is interfaced with the Python package {\tt smelli} (the SMEFT likelihood tool)~\cite{Aebischer:2018iyb,Stangl:2020lbh}.
After we pass the SMEFT Wilson coefficients, which we compute from the parameters of our Lagrangian~\eqref{eq:Lagrangian} at the matching scale $\mu_M$, to {\tt smelli}, this tool automatically takes care of the renormalization group running down to the meson scale as well as the intermediate matching to the low-energy EFT~\cite{Alonso:2013hga,Jenkins:2013wua,Jenkins:2013zja,Dekens:2019ept,Jenkins:2017dyc} thanks to the {\tt wilson}~\cite{Aebischer:2018bkb} package.
It further uses {\tt flavio}~\cite{Straub:2018kue} to compute a large list of electroweak-scale and low-energy precision observables, including charged LFV and LFU, magnetic moments, neutral meson mixings, semileptonic and rare meson decays, etc.
The full list of observables included in the initial version of {\tt smelli} can be found in the appendix of~\cite{Aebischer:2018iyb}, but this list has been extended~\cite{Falkowski:2019hvp}, and we refer to~\cite{smelli} for the up-to-date version.
We update the measurements included in {\tt smelli} and take into account the most recent results for $R_K$~\cite{Aaij:2021vac} and $(g-2)_\mu$~\cite{Abi:2021gix} as well as the current world average of BR$(B_{s,d}\to\mu\mu)$ from~\cite{Altmannshofer:2021qrr}, which includes the most recent LHCb measurement~\cite{LHCb_Bsmumu}.
With this setup, we are now in position to perform a global fit in the parameter space of our model.

\begin{figure}
	\centering
	\includegraphics[width=\columnwidth]{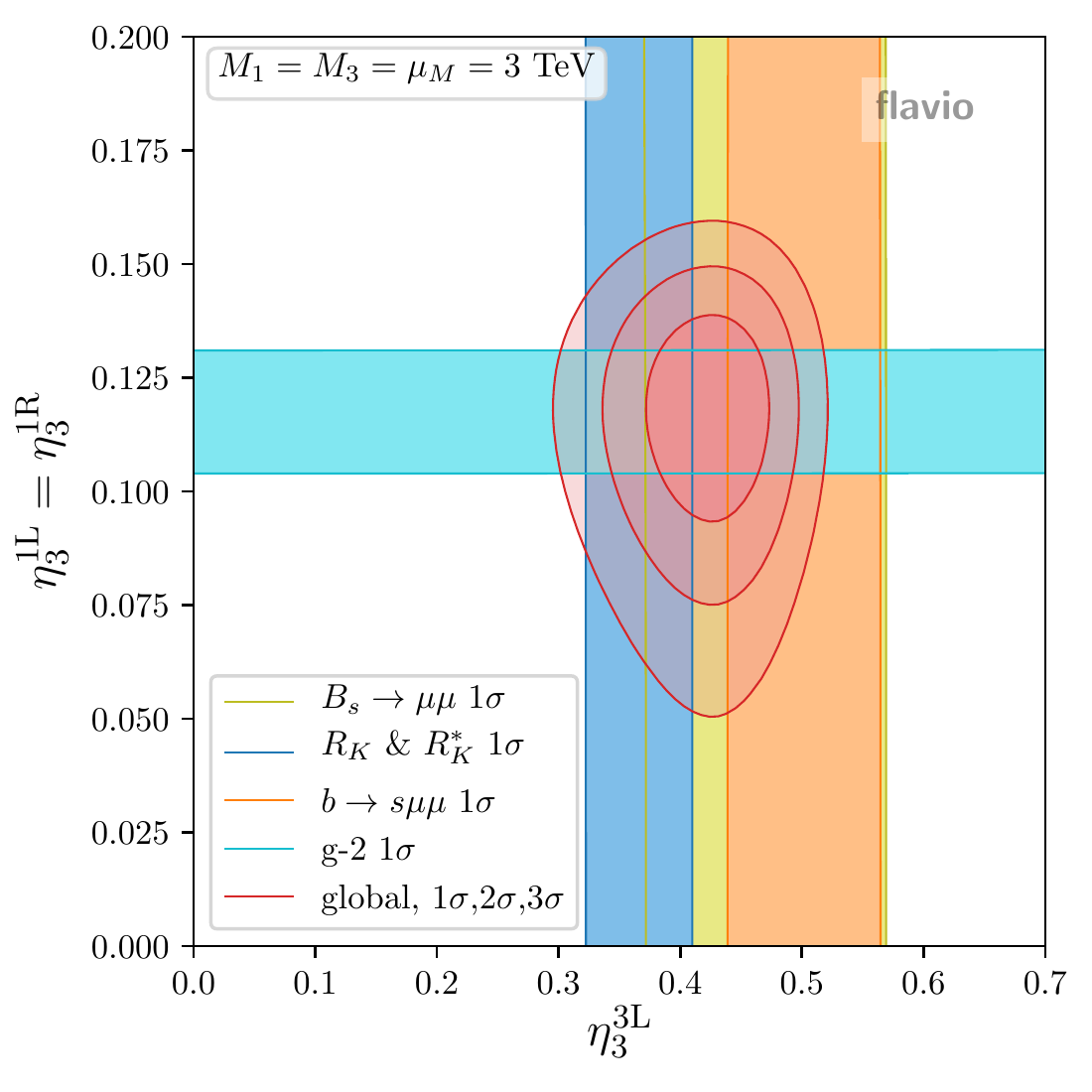}
	\caption{The preferred muoquark Yukawa couplings from the global fit to low-energy data.
		Here we choose $ \eta^{3\LL}_i = (V_{td}, \, V_{ts},\, 1) \, \eta^{3\LL}_{3} $, $ \eta^{1\LL}_i = (V_{td}, \, V_{ts},\, 1)\, \eta^{1\LL}_{3} $, and $  \eta^{1\RR}_i = (0,\, 0,\, 1)\, \eta^{1\RR}_3 $.
		The muoquark masses are set to $ M_1 = M_3 = \SI{3}{TeV} $.
	}
	\label{fig:muon_fit}
\end{figure}

Shown in Fig.~\ref{fig:muon_fit} is the preferred region in the $\eta^{3\LL}_{3}$ versus $\eta^{1\LL}_{3}=\eta^{1\RR}_{3}$ plane for $ M_1 = M_3 = \SI{3}{TeV} $.
We take $ \tilde{\eta}^{1\RR} = 0 $, as loop-induced contributions from the heavy right-handed neutrinos are expected to be negligible in the fit.
Muon anomalies clearly prefer the parameter space far away from the SM limit $\eta^{x}_{3} = 0$.
The best fit point is $(\eta^{3\LL}_{3}, \eta^{1\LL}_{3}=\eta^{1\RR}_{3})\simeq (0.43,0.12)$ with a $\Delta \chi^2\simeq62$ compared to the SM point.
The current limits from direct searches at the LHC are $M_3 \gtrsim \SI{1.7}{TeV}$~\cite{Aad:2020iuy} and $M_1\gtrsim \SI{1.4}{TeV}$~\cite{ATLAS:2020qoc}, while the final reach of HL-LHC is projected in~\cite{Cerri:2018ypt}.
The indirect effects in the high-$p_T$ lepton tails are also beyond the HL-LHC projections for the best fit couplings~\cite{Greljo:2017vvb}.
The change in the mass is accommodated by an approximate linear change in the couplings keeping the same low-energy Wilson coefficients.
However, the finite naturalness of the Higgs mass and muon Yukawa, disfavors heavier muoquarks, as discussed later.

While in principle both muoquarks contribute to all anomalies, there is a clear factorization, namely $S_1$ dominates in the $(g-2)_\mu$ thanks to the chiral enhancement from the top quark, whereas $S_3$ dominates in $b \to s \mu^+ \mu^-$ since it gives a tree-level contribution unlike $S_1$.
The $\U(2)$ flavor structure provides sufficient suppression in all other complementary processes such as $K_L \to \mu^+ \mu^-$~\cite{Gherardi:2019zil}.
When varying the $\mathcal{O}(1)$ coefficients in front of the spurions we find the same goodness of fit:
the best fit region is shifted to accommodate for $b \to s \mu^+ \mu^-$, but none of the complementary observables listed above receive a large pull.

\subsection{Symmetry breaking}
Heavy vector resonances with couplings to both quarks and leptons have been extensively searched for at the LHC.
The most recent ATLAS \SI{13}{TeV} search with $ \SI{139}{fb^{-1}} $ of data~\cite{Aad:2019fac} reports the exclusions on the couplings as a function of the mass in their Fig.~4\,(b).
A viable benchmark example in our case is gauge coupling  $g_X = 0.1$ and mass $m_X = \SI{3}{TeV} $.
The high-$p_T$ dimuon tails~\cite{Greljo:2017vvb} set an upper limit on $g_X /m_X$ for large $m_X$.
In the opposite limit, the bounds are avoided when $g_X$ is small enough (see Fig.~5 of~\cite{Greljo:2017vvb}).
It is, however, always possible to take the decoupling limit, namely large $m_X$ and small $g_X$, without conflicting the muoquark solution of muon anomalies.
We expect $X$ to have negligible effects in flavor physics through suppressed penguins, which decouple in the same limit.\footnote{A \Ubl model with vector-like quarks and $X$ as the main mediator of $b \to s \mu^+ \mu^-$ anomaly cannot reconcile the $B_s$ meson-mixing constraints with the high-$p_T$ dimuon tails~\cite{Greljo:2017vvb}.}

The symmetry breaking scalar $ \Phi $ develops a VEV $ \langle \Phi \rangle = v_\Phi $ related to the $ X $ mass by $ v_\Phi = \sqrt{2} m_X/ 3 g_X $ or \SI{14}{TeV} for the benchmark point.
Taking $ M_{1,3}^2 >0 $ and small cross-quartic couplings, guarantees that $ S_{1,3} $ do not develop a VEV, and the part of the scalar potential relevant for symmetry-breaking is
    \begin{multline}\label{eq:mixing}
	V_{H\Phi}= -\mu_H^2 |H|^2 - \mu_\Phi^2 |\Phi|^2 + \tfrac{1}{2} \lambda_H |H|^4 \\+ \tfrac{1}{4}\lambda_\Phi |\Phi|^4 + \lambda_{\Phi H} |\Phi|^2 |H|^2\, .
	\end{multline}
We can directly relate the potential parameters for the Higgs VEV $ v= \langle H \rangle$; $v_\Phi $; the masses of the radial modes $ m_h, m_\phi $; and the mixing angle, which has to satisfy $ \theta \ll 1 $ by assumption for viable phenomenology:
	\begin{align}
	\lambda_H &= \tfrac{1}{2} v^{\eminus 2} \big[ m_h^2 - \theta^2 \Delta m^2 \big], \nonumber\\
	\lambda_\Phi &= v_\Phi^{\eminus 2} \big[ m_\phi^2 + \theta^2 \Delta m^2\big], \nonumber\\
	\lambda_{\Phi H} &= \tfrac{1}{2} \theta \Delta m^2/ (v v_\Phi), \\
	\mu_H^2 &= \tfrac{1}{2} \big[m_h^2 - \theta^2 \Delta m^2 \big] + \tfrac{1}{2} \theta \Delta m^2 v_\Phi/ v, \nonumber\\
	\mu_\Phi^2 &= \tfrac{1}{2} \big[m_\phi^2 + \theta^2 \Delta m^2 \big] + \tfrac{1}{2} \theta \Delta m^2 v/ v_\Phi, \nonumber
	\end{align}
where $ \Delta m^2 = m_h^2 - m_\phi^2 $.
For $ \lambda_H, \lambda_\Phi \sim 0.1$, the radial mode $\phi$ has a mass around the TeV scale for the benchmark point.
A small $\lambda_{\Phi H} = 10^{\eminus 3} $ consistent with the finite naturalness discussed below leads to unobservable mixing with the Higgs boson~\cite{Chalons:2016jeu,Adhikari:2020vqo}.

\subsection{Naturalness} \label{sec:naturalness}
\begin{figure}
	\centering
	\includegraphics[width=\columnwidth]{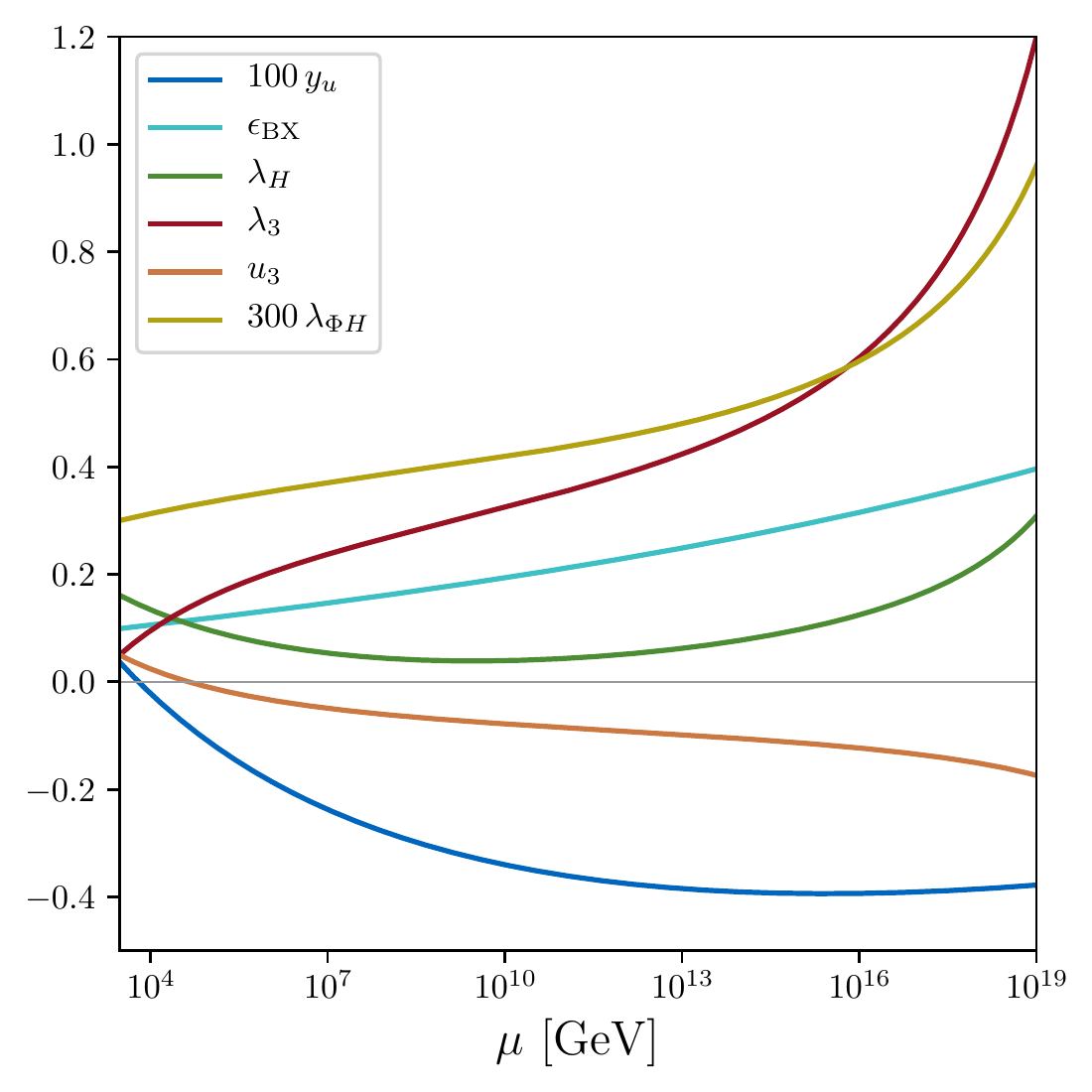}
	\caption{The RG flow of a selection of couplings from the benchmark point to the Planck scale.
All couplings were included in the running, and none of them develops a Landau pole in this range.}
	\label{fig:RG_flow}
\end{figure}

The contribution to the scalar potential involving the muoquarks is
\begin{widetext}
	\begin{equation} \begin{split}
	V_{13} =\, & M_1^2 |S_1|^2 + M_3^2 |S_3|^2
	+\lambda_{\Phi 1} |\Phi|^2 |S_1|^2
	+ \lambda_{\Phi 3} |\Phi|^2 |S_3|^2
	+ \tfrac{1}{2} \lambda_1 (S_1^\dagger S_1)^2
	+\lambda_{H 1} |H|^2 |S_1|^2 + \lambda_{H 3} |H|^2 |S_3|^2  \\
	&+ \kappa_{H3} H^\dagger \sigma^I \sigma^J H (S_3^{\dagger I} S_3^J)
	+ (\kappa_{H13} H^\dagger \sigma^I H (S_1^{\dagger} S_3^I) + \hc )
	+ \tfrac{1}{2} \lambda_3 (S_3^\dagger S_3)^2
	+ \tfrac{1}{2} \kappa_3 (S_3^{\dagger I} S_3^J) (S_3^{\dagger J} S_3^I) \\
	&+ \tfrac{1}{2} \upsilon_3 (S_3^{\dagger I} S_3^J) (S_3^{\dagger I} S_3^J) + \lambda_{13} |S_1|^2 |S_3|^2 + \kappa_{13} (S_3^{\dagger I} S_1) (S_1^\dagger S_3^I) + (\upsilon_{13} (S_1^\dagger S_3^I) (S_1^\dagger S_3^I) + \hc ).
	\end{split} \end{equation}
\end{widetext}
We qualitatively examine the UV consistency of the model through the RG flow of the couplings from the best fit discussed in Sec.~\ref{sec:LQ_pheno}.
In particular, we explore the running of the full model with $ \beta $-functions at 3-loop order for the gauge and 2-loop order for Yukawa and quartic couplings derived with \texttt{RGBeta}~\cite{Thomsen:2021ncy,Pickering:2001aq,Poole:2019kcm}.
The SM couplings have been fixed at $ \mu_M $ as part of the matching process.
The resulting flows for a few selected couplings are shown in Fig.~\ref{fig:RG_flow}.

All terms in the 1-loop $\beta$-functions of the $ \Phi $ cross-quartic couplings $\lambda_{\Phi H}$, $\lambda_{\Phi 1}$, and $\lambda_{\Phi 3} $ involve this set of $ \Phi$ cross quartics or are \Ubl gauge contributions (with/without kinetic mixing).
These three couplings can, consequently, be taken simultaneously small while limiting radiative correction with a floor determined by gauge and 2-loop contribution.
While taking all $ \lambda_{\Phi x} (\mu_M) = 0 $ gives $ \lambda_{\Phi x} (M_\mathrm{Pl}) \sim 10^{\eminus 3} $, $ \lambda_{\Phi x} (\mu_M) = 10^{\eminus 3} $ gives slow running couplings (see Fig.~\ref{fig:RG_flow}).
We, therefore, take the latter as a natural minimal $ \lambda_{\Phi x} $ that prevents a large tuning in the Higgs mass (with a contribution $ \lambda_{\Phi H} v_\Phi^2 $) as $ \Phi $ condenses.

As in most models with many scalar degrees of freedom, the quartic couplings tend to reinforce each others running in such a way that they drive each other to Landau poles quickly.
Avoiding any such before the Planck scale in our case tends to favor small quartic couplings $ \lesssim 0.05 $ at the matching scale.
Individual couplings can be larger, but in particular the muoquark self-couplings have fast running due to their large multiplicity, leading to poles (typically driven by $ \lambda_3 $).
The constraints from absence of Landau poles are much stronger than those obtained from electroweak precision data and Higgs signal strengths~\cite{Gherardi:2020qhc,Crivellin:2020ukd}, which constrain the couplings at $ \mathcal{O}(1) $.
For the benchmark point, we take all remaining quartic couplings of $ V_{13} $ to be $ 0.05 $ at $ \mu_M $.

It is worth pointing out that the potential is stable all the way up to the Planck scale in the benchmark scenario; although $ \upsilon_3 $ runs negative, the condition $ \lambda_3+ \kappa_3 + \upsilon_3 \geq 0 $ is satisfied, ensuring stability.
The minimum of the potential discussed in the previous section is thus the true vacuum.

The large charges of the muoquarks under the \Ubl symmetry has profound impact on the RG flow.
The $ g_X^4 $ contribution to the muoquark quartic couplings scale with their charge to the fourth power, and so is extremely sensitive to the value of $ g_X $.\footnote{E.g.~the 1-loop $ \lambda_3 $ $ \beta $-function has a contribution $ \dd \lambda_3/\dd \ln \mu \supset 1024 g_X^4/ (27 \pi^2) $.}
For instance, $ g_X(\mu_M)= 0.15 $ leads to a Landau pole at $ M_\mathrm{Pl} $, whereas $ g_X(\mu_M)= 0.25 $ pulls the pole to $ \sim \SI{e11}{GeV} $.
The same large charges (with that of the muon) also cause sizable running in the kinetic-mixing parameter $ \varepsilon_{BX} $.
For our benchmark point $ g_X(\mu_M)= 0.1 $ we observe $ \varepsilon_{BX}(M_\mathrm{Pl}) - \varepsilon_{BX}(\mu_M) \sim 0.2 $ regardless of the exact value $ \varepsilon_{BX}(\mu_M) $ (cf. Fig.~\ref{fig:RG_flow}).
As a natural value for this parameter we take $ \varepsilon_{BX}= 0.1 $, which is perfectly consistent with phenomenology~\cite{Hook:2010tw}.

As in any model with multiple mass scales, there is a risk that the heavier scale will destabilize the lighter through radiative corrections.
In our case the 1-loop correction to the Higgs mass parameter due to the heavy muoquark is~\cite{Gherardi:2020det}
	\begin{equation}
	\begin{split}
	\delta \mu_H^2 = & -\dfrac{9 (\lambda_{H 3} + \kappa_{H 3}) }{(4\pi)^2} M_3^2 \left(1 + \ln \dfrac{\mu_M^2}{M_3^2} \right) \\
	&+ \dfrac{3 \lambda_{H 1} }{(4\pi)^2} M_1^2 \left(1 + \ln \dfrac{\mu_M^2}{M_1^2} \right) + \mathcal{O}(\mu^4/M_{1,3}^2).
	\end{split}
	\end{equation}
With the small quartic couplings of $ \mathcal{O}(0.05) $, as preferred by the RG, the theory is finitely natural for $ M_{1,3} \lesssim \mathcal{O}(\SI{1}{TeV}) $.
Tuning arguments therefore favor light muoquarks which is a great news for collider searches.

Additionally, the $ S_1 $ muoquark generates a non-multiplicative radiative corrections to the muon Yukawa coupling~\cite{Gherardi:2020det,Capdevilla:2020qel,Capdevilla:2021rwo}:
	\begin{equation}
	\delta y_\mu = -\dfrac{3}{(4\pi)^2} \left(1 + \ln \dfrac{\mu_M^2}{M_1^2} \right) \eta_i^{1\LL \ast } y_u^{ij} \eta^{1\RR}_j .
	\end{equation}
For the part of parameter space with large enough couplings to explain the $ (g-2)_\mu $, a tuning argument again favors models with smaller masses.
In our best fit point the change in $ y_\mu $ is roughly $ 50\% $.
The same muoquark loop that gives the threshold correction to $ y_\mu $ also gives rise to a significant running of this Yukawa as shown in Fig.~\ref{fig:RG_flow}.
This is yet another independent argument in favor of lighter muoquarks potentially accessible at high-$p_T$.

\subsection{Neutrino masses and proton decay}

Coming back to the neutrino sector outlined in the last line of the model Lagrangian~\eqref{eq:Lagrangian}, the \Ubl gauge symmetry imposes a flavor structure for $y_\nu$, $M_\RR$ and $y_\Phi$.
Notably, $ y_\nu $ splits into a $2\times2$ electron--tau block and a diagonal muon entry.
When $\Phi$ receives a VEV, the Majorana mass matrix is entirely populated except for the (2,2) entry.
This structure has enough parametric freedom to explain the observed neutrino oscillation data~\cite{Esteban:2018azc}, the limit on the sum of neutrino masses from Planck~\cite{Aghanim:2018eyx}, and the absence of neutrinoless double beta decay~\cite{KamLAND-Zen:2016pfg}.
Ref.~\cite{Asai:2019ciz} performed a careful analysis of a specific limit when the $y_{\nu}^{13,31}$ and $y_{\Phi}^{23}$ are set to zero, arriving at the two-zero minor structure of type $\mathrm{D}_1^\RR$.
This limit perfectly accommodates neutrino oscillations data, predicting $\sum_i m_{\nu_i}$ comfortably below the present limit and no neutrinoless beta decay.
The firm predictions of the $\mathrm{D}_1^\RR$ can be alerted in our case by nonzero  $y_{\nu}^{13, 31}$ and $y_{\Phi}^{23}$ parameters.

The type-I seesaw formula for the masses of the active neutrinos,
	\begin{equation}
	m_{\nu} \simeq - v^2 y_\nu M^{\eminus 1}_{\RR} y^{\mathrm{T}}_\nu\,,
	\end{equation}
suggests that in our chosen benchmark the Dirac Yukawa is in the same ballpark as the electron Yukawa, $\mathcal{O}(10^{\eminus 6})$.
The $S_1$ muoquark, contributing to $(g-2)_\mu$ would radiatively correct the $y_\nu^{22}$ with the bottom quark in the loop~\cite{Brdar:2020quo}.
The $\tilde{\eta}^{1\RR}_3$ coupling is an input parameter, however, if it is of the same order as the $\eta^{1\RR}_3$ coupling, it would contribute comparably to the tree-level.
Hence, no tuning is introduced here.

Finally, the Lagrangian in Eq.~\eqref{eq:Lagrangian} respects baryon number and keeps the proton stable.
However, the absence of $B$ violation for a TeV-scale leptoquark model has to be required also for the leading irrelevant operators arising at dimension-5~\cite{Arnold:2013cva,Assad:2017iib}.
Quantum gravity is expected to break global charges~\cite{Banks:2010zn}, and even if the dimension-5 operator under consideration is suppressed by the Planck scale, it is not enough to evade the stringent bounds on the proton lifetime.
This seems to be a quite generic issue often neglected in the literature, with the notable exception of the Pati--Salam gauge leptoquark, see e.g.~\cite{DiLuzio:2017vat,Greljo:2018tuh,Bordone:2017bld,Bordone:2018nbg,Cornella:2019hct,Fornal:2018dqn,Blanke:2018sro,Fuentes-Martin:2019ign,Guadagnoli:2020tlx,Heeck:2018ntp,Fuentes-Martin:2020bnh,Fuentes-Martin:2019bue,Fuentes-Martin:2020luw,Fuentes-Martin:2020hvc}.

The \Ubl gauge symmetry, however, with the available field content ensure that $B$ number is conserved also at the dimension-5 effective Lagrangian.\footnote{The only way to build color singlets with non-vanishing baryon number at this order is with fields $ SSS $, $ qSS $, or $ qqS $.
These combinations have \Ubl charge $ \pm 8 $, $ \pm 5 $, and $ \pm 2 $, respectively.
It is easy to verify that they cannot be completed to a gauge invariant dimension-5 operator with the available matter fields.}  The leading breaking is expected at dimension 6 similarly to the SM.
It is a nontrivial fact that this is compatible with the minimal realization of neutrino masses.
This is, for instance, not the case for \Umt symmetry where the minimal neutrino sector~\cite{Davighi:2020qqa} allows for a coupling $ 1/M_\mathrm{Pl} \, \overline{q}^c_\LL S_3 \Phi q_\LL $, which, together with the $\overline{q}^c_\LL \ell_\LL S_3$ needed for the anomaly, leads to proton decay in gross violation of the experiment.
We estimate that such leptoquark has to be several orders of magnitude heavier to respect the proton lifetime bound, or, equivalently, the couplings should be smaller.
In either case, the explanation of the anomaly is gone.
Going beyond the minimal neutrino mass realizations in \Umt, even more involved constructions proposed in the literature share this problem, see e.g.~\cite{Heeck:2011wj,Crivellin:2015lwa,Nomura:2018cle,Araki:2019rmw}.

\section{Alternative models}
\label{sec:muta}

We now turn our focus to alternative models for the muon and $ B $-decay anomalies, in some of which the \Ubl symmetry is exchanged for other $ \U(1)_X $ symmetries. These models offer different scenarios of phenomenological interest. 

\subsection{The scenarios for muon anomalies} \label{sec:mu-tau_muon_anomalies}

\Ubl is only one example of many possible lepton-flavored gauge extensions of the SM, under which leptoquarks become \textit{muoquarks}. Variations of the model can use different choices of $ \U(1)_X $ symmetry to ensure the leptoquarks coupling exclusively to second generation leptons and fall into three classes shown in Table~\ref{tab:model_types} based on what mediators are responsible for the \RK and $ (g-2)_\mu $ anomalies. Below we give some specific examples of these variations:

\begin{table}[t] \renewcommand*{\arraystretch}{1.4}
	\centering
	\begin{tabularx}{.9\columnwidth}{|l|Y Y Y|}
		\hline \hline
		& Type A & Type B & Type C$ $ \\ \hline
		\RK, $ b\to s \mu \mu$ & $ S_3 $ & $ S_3 $ & heavy $ X $ \\
		$(g-2)_\mu$ & $ S_1/R_2 $ & light $ X $ & $ S_1/R_2 $ \\ \hline \hline
	\end{tabularx}
	\caption{Three types of \textit{muoquark} models, which can address the muon anomalies for a variety of lepton-flavored $ \U(1)_X $ gauge groups. For each model class, a field responsible for addressing a corresponding anomaly, is listed. The an $ R_2 $ muoquark with SM charges $  (\rep{3},\, \rep{2},\, \sfrac{7}{6})$ can be used as an alternative to $ S_1 $ for addressing the $ (g-2)_\mu $.}
	\label{tab:model_types}
\end{table}

\textbf{Type A} --- As a showcase example, we trade the \Ubl for a \Umt gauge symmetry to obtain an extension of the leptoquark model of Ref.~\cite{Davighi:2020qqa}.
The leptoquarks are assigned charge $ \eminus 1 $ under the symmetry, such that they still couple exclusively to 2nd generation leptons.
The minimal type-I seesaw realization of the neutrino mass with the \Umt-breaking scalar of charge $+1$ predicts the two-zero minor structure $\mathrm{C}^\RR$, which shows some tension in fitting $\theta_{23}$ and $\sum_i m_{\nu_i}$~\cite{Asai:2019ciz}, thus more elaborate model building may be needed~\cite{Araki:2019rmw}.\footnote{We will not explore these constructions in any detail here but merely reiterate the point that a charge-$ 1 $ scalar is potentially problematic since it enables a baryon-number-violating dimension-5 operator.}

The muoquark solution of the muon anomalies discussed above applies equally to this model.
The main phenomenological difference is that the gauge vector $X $ does not couple to quarks and is less constrained at colliders.
Thus, the $X $ field can more easily elude current experimental bound (see Fig.~2 of~\cite{Altmannshofer:2014pba}).
For example, constraints from neutrino trident production requires $m_{X} \gtrsim 60\, (200) \, \si{GeV}$ for $g_X \sim 0.1 \,(0.3)$.
Again, $X $ and $\Phi $ can simply be decoupled in the limit of the large $v_\Phi$  and small gauge coupling. This scenario belongs to Type A class of models as explained in Table~\ref{tab:model_types}.

\textbf{Type B} --- A second avenue to address $ (g-2)_\mu $ arises in this model, invoking a light \Umt gauge boson $X$ as a mediator running in the loop~\cite{Altmannshofer:2014pba,Baek:2001kca,Gninenko:2001hx}.
The discrepancy between the experiment and the SM prediction can be resolved with $ m_X \sim \SI{20}{MeV} $ and gauge coupling $ g_X\sim 5\cdot 10^{\eminus 4} $, nestling snugly in the window allowed by current experimental constraints, such as CCFR and Borexino~\cite{Altmannshofer:2014pba,Altmannshofer:2019zhy}.
In fact, even the future DUNE experiment is not expected to cover the entire window~\cite{Altmannshofer:2019zhy}.
In this scenario, $ S_1 $ is entirely superfluous to the anomalies and can be removed from the model altogether.
Additionally, the small allowed region for $X$ mass and couplings gives a sharp prediction for the \Umt sector.
We have checked that the small gauge coupling and associated small kinetic mixing are stable under radiative corrections.
The \RK anomaly in this scenario is still explained by a tree-level mediation of $ S_3 $, and with a similar allowed parameter space as before.

\begin{figure}[t]
	\centering
	\includegraphics[width=\columnwidth]{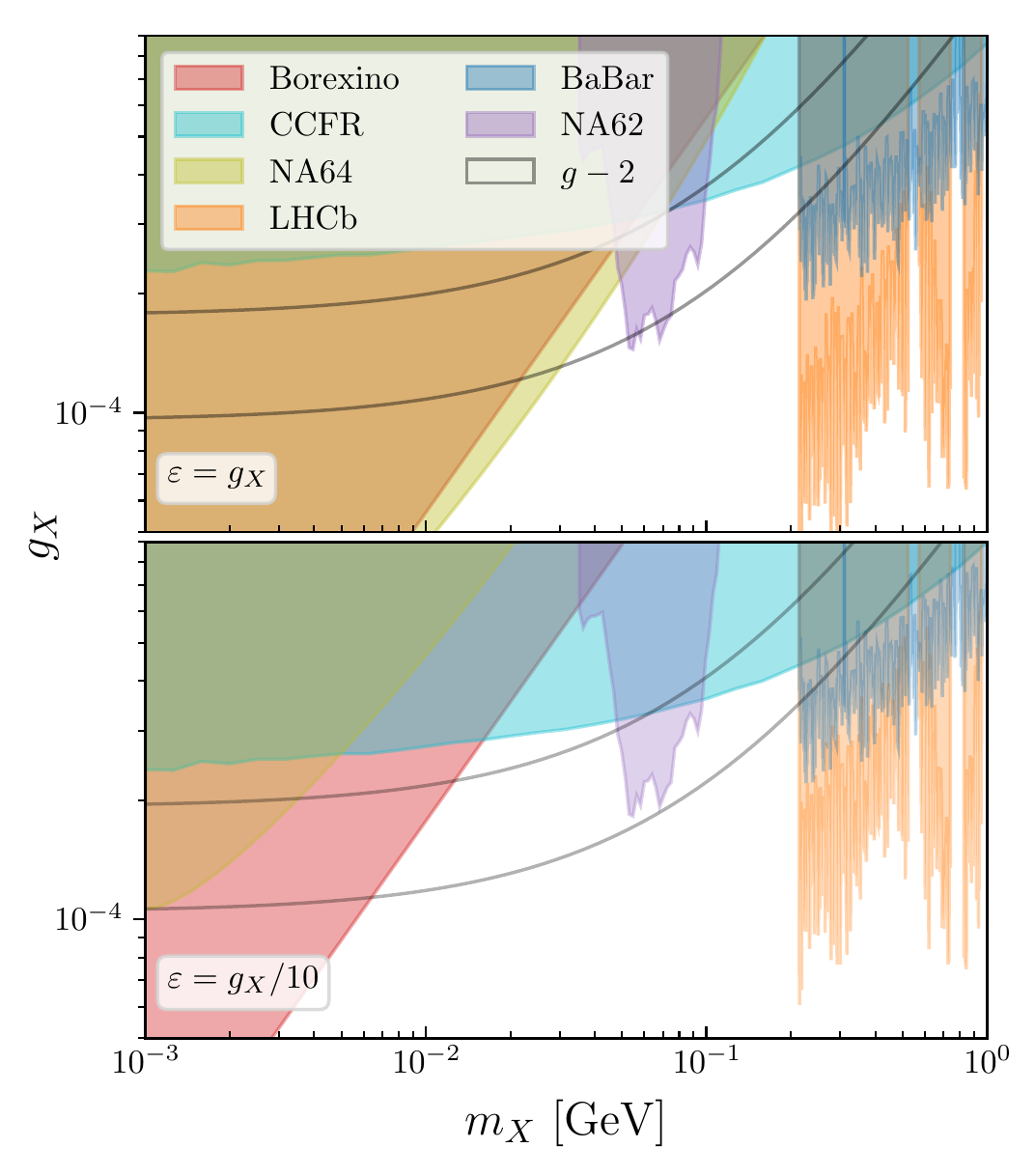}
	\caption{Allowed parameter space for the light $ X $ solution to the $ (g-2)_\mu $ anomaly in the $ \U(1)_{B - 3 L_\mu}$ model. The shaded regions are excluded by various experiments, while the region between the black lines is preferred by $ (g-2)_\mu $. The upper (lower) plot uses kinetic mixing $ \varepsilon_{BX} = g_X $ ($ \varepsilon_{BX} = g_X / 10 $). } \label{fig:light_X} 
\end{figure}

The \Umt-breaking VEV, $ v_\Phi $, is given as $ v_\Phi = \sqrt{2} m_X/ |q_\Phi| g_X \sim \SI{60}{GeV}/ |q_\Phi| $ while the cross-quartic coupling induces mixing between real scalars $h$ and $\phi$.
This scenario has a chance to leave observable imprints in the overall Higgs couplings or in the invisible Higgs decays ($h \to X X$) if the cross quartic in Eq.~\eqref{eq:mixing} is large enough.
On the contrary, as discussed in Sec.~\ref{sec:naturalness}, we verified that a small mixing is stable under radiative corrections.
For sufficiently large $\lambda_\Phi$, the mass of $\phi$ can easily be above $m_h /2$ such that $h \to \phi \phi$ decays are absent.
Even if the mixing is tuned away, $\phi$ decays promptly through $\phi \to X X$, way before BBN takes place, since the decay rate is dominated by the Goldstone mode $\propto |\lambda_\Phi |^2$.
On that note, $X \to \nu \bar \nu$ of muon or tau flavor is safe from the BBN bounds~\cite{Altmannshofer:2019zhy}.\footnote{See also~\cite{Araki:2021xdk} for potential connection with the $H_0$ tension.}

It is interesting that the light $ X $ explanation of the $ (g-2)_\mu $ can also be adapted to the \Ubl model, i.e. the $ S_1 $ field can be removed completely.
In this model, the $ X $ coupling to quarks are not induced by kinetic mixing with the photon but follows directly from the gauge symmetry, increasing the sensitivity of some experiments.
Differently from the \Umt model sans muoquarks, the running of the kinetic mixing is not multiplicative, and we observe that for the small $ g_X, \varepsilon_{BX} $ needed in this scenario, we can expect $ [\varepsilon_{BX}(M_\mathrm{Pl}) - \varepsilon_{BX}(\mu_M)] / g_X(\mu_M) = \mathcal{O}(1) $.\footnote{The running of $ \varepsilon_{BX} $ is at one loop given by 
	\[
	\dfrac{\dd \varepsilon_{BX}}{ \dd \ln \mu}  = \dfrac{192 g_Y g_X + (86 g_Y^2  + 577 g_X^2 ) \varepsilon_{BX}}{192 \pi^2}   
	\] 
and is completely dominated by the first term for $ g_X > \varepsilon_{BX} $ most relevant in this scenario.}
The $ X $ coupling to electrons is completely determined by the mixing, and an enhancement leads to stronger bounds from the Borexino experiment from scattering of solar neutrinos on electrons~\cite{Altmannshofer:2019zhy,Agostini:2017ixy}.
Ultimately, $ \varepsilon_{BX} $ is a free parameter of the model, in the absence of any unification of the gauge groups in the UV, and can be chosen small to weaken this bound.
The neutrino trident production, $ N \nu_\mu \to N \nu_\mu \mu^+ \mu^-$, constraint from the CCFR experiment~\cite{Altmannshofer:2014pba,Mishra:1991bv} scales directly with the effective coupling of $ X $ to 2nd generation leptons, $ -3 g_X$ and $(-3 g_X - e \varepsilon_{BX}) $ for $ \nu_\mu $ and $ \mu $, respectively,  and does not constrain the solution space for $ (g-2)_\mu $ over the \Umt models.
We combine this constraints with the \texttt{Darkcast}~\cite{Ilten:2018crw} bounds on light vectors from a variety of different experiments.
The combined bounds on the \Ubl light $ X $ explanation of the $ (g-2)_\mu $ is shown in Fig.~\ref{fig:light_X} along with the best fit for the anomaly.
We see that for small values of $ \varepsilon_{BX} $ a region of parameter space allow for a light $ X $ explanation of the $ (g-2)_\mu $ anomaly.

\textbf{Type C} --- Finally, when the $X$ field is heavy enough, it can successfully explain \RK at tree-level. To name a few examples of explicit models, Ref.~\cite{Altmannshofer:2014cfa} is based on the \Umt gauge group whereas Ref.~\cite{Allanach:2020kss} works with  $\U(1)_{B_3 - L_2}$. The first model uses vector-like quarks to mediate the $Z' b s$ interactions. The second model directly couples muons with the third generation quarks, but requires additional structure to accommodate for the CKM matrix and to mediate the interaction with the strange quark. These models are not able to simultaneously accommodate for $ (g-2)_\mu $ due to the constraints from the neutrino trident production. A simple solution is to add an $S_1$ leptoquark charged under $\U(1)_X$ such that it only couples to muons as in Eq.~\eqref{eq:Lagrangian}. Alternatively, an $R_2 = (\rep{3},\, \rep{2},\, \sfrac{7}{6} ) $ muoquark representation can be used~\cite{Dorsner:2019itg}. 
While explaining $ (g-2)_\mu $, $ R_2 $ can also give an additional contribution to the $ b s \mu \mu $ effective coupling, potentially improving the fit to $ B_s \to \mu^+ \mu^- $ beyond what can be done with $ X $ alone. 
The muoquark is a compelling alternative to the colorless scalar representation, such as a $H' =(\rep{1},\, \rep{2},\, \sfrac{1}{2})$, due to the top-Yukawa enhancement in the loop.

\subsection{Combined explanation of \RD and muon anomalies}

\begin{figure}
	\centering
	\includegraphics[width=\columnwidth]{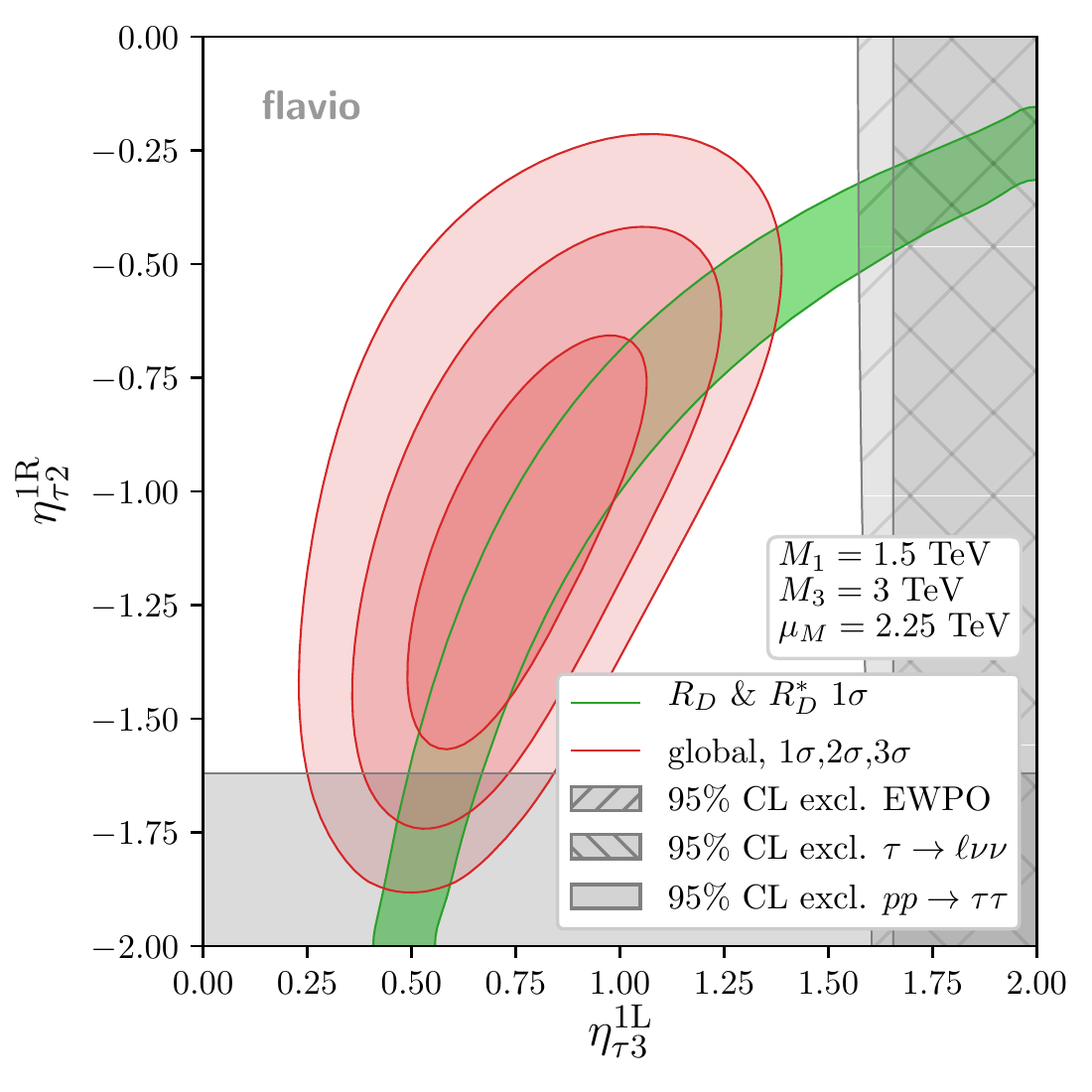}
	\caption{Preferred regions for the Yukawa couplings $ \eta^{1\LL}_{\tau3} $ versus $ \eta^{1\RR}_{\tau2}$ when $S_1$ has $L_\mu- L_\tau$ charge $ +1 $.
	The fit is driven by the $b \to c \tau \nu$ anomaly.}\label{fig:RD}
\end{figure}

In passing, we note that a variation of the model can simultaneously address the anomaly reported in charged-current transitions $b \to c \tau \nu$ together with the muon anomalies.
The anomalous observables \RD~\cite{Lees:2012xj,Lees:2013uzd,Huschle:2015rga,Sato:2016svk,Aaij:2015yra,Hirose:2016wfn,Hirose:2017dxl,Aaij:2017uff,Aaij:2017deq,Aaij:2017tyk,Aoki:2016frl} show a combined tension from the SM at the level of $3.1\,\sigma$.
The $S_1$ leptoquark can resolve the tension when it couples to left-handed bottom quark and tau neutrino as well as right-handed charm quark and tau.
The minimal-coupling scenario is compatible with other complementary observables, as shown in Ref.~\cite{Gherardi:2020qhc}.

The same reference attempts a combined explanation of  \RD,  \RK, and $(g-2)_\mu$ introducing $S_1$ and $S_3$ leptoquarks and allowing for ten independent Yukawa parameters in the fit.
The successful solution in the parameter space requires the $S_1$ contribution to dominate new physics effect in \RD.
The main drawback of this solution is the need for a tuned cancellation in the LFV $\tau \to \mu \gamma$ observable~\cite{Gherardi:2020qhc}.
The contribution to this decay is indeed unavoidable when one requires $S_1$ to resolve both \RD and $(g-2)_\mu$.
The cancellation is achieved by an independent coupling not needed to fit the anomaly.
This is another example suggesting a symmetry protecting against LFV.

In our setup, there is as an elegant resolution of all anomalies, which does not require any tuning.
The idea is to add an additional (tauphilic) $S_1$ with charge $+1$ under \Umt: it couples exclusively to tau and tau neutrino, so it has a minimal set of couplings needed to accommodate the \RD.
The couplings to the left-handed quarks are consistent with the $\U(2)$ symmetry, where the dominant couplings are with the bottom quark.
The couplings to $d$ and $s$, suppressed by $V_{ts}$ and $V_{td}$, are consistent with the constraints from complementary observables.
The weakest point of this setup is the unusually large right-handed charm coupling, not predicted by the minimal breaking of $\U(2)^3$ flavor symmetry.
For $ M_{1} = \SI{1.5}{TeV} $, a benchmark point within $1\sigma$ is for $|\eta^{1\LL}_{\tau 3} \times \eta^{1\RR}_{\tau 2}| \sim 0.7$.
Large couplings to top quark are also allowed by the data but not needed for the anomaly.
Introducing a new spurion $V_u = \rep{2}$ of $\U(2)_u$, such that $V_{u 2} \sim 0.5$, predicts $|\eta^{1\LL ,1\RR}_{\tau 3}| \sim 1.5$ and $|\eta^{1\RR}_{\tau 2}| \sim 0.7$.
We show likelihood contours in the $\eta^{1\LL}_{\tau 3}$ -- $\eta^{1\RR}_{\tau 2}$ plane in Fig.~\ref{fig:RD}, which demonstrate that the above benchmark point is indeed within the $1\sigma$ range around the best-fit point and that it is compatible with all constraints.\footnote{
We consider all constraints implemented in {\tt smelli} and the $pp\to\tau\tau$ constraint is taken from Ref.~\cite{Angelescu:2021lln}.
}

On top of this, muon anomalies can be solved by both scenarios discussed in Sec.~\ref{sec:mu-tau_muon_anomalies}:  with a $S_3$ solution of $b \to s \mu^+ \mu^-$ and $i)$ the light $X$ solution of $(g-2)_\mu$ or $ii)$ a (mounphilic) $S_1$ for $(g-2)_\mu$.
In other words, \Umt provides an effective decoupling of different scenarios for solving $b \to c \tau \nu$,  $b \to s \mu^+ \mu^-$ and $(g-2)_\mu$.
The solutions are standalone, which is important if some anomalies go away, but are easily combined together without any side effects.

\section{Conclusions}
\label{sec:conclusions}
The Standard Model of particle physics is successful in explaining both seen and unseen phenomena.
When viewed as an effective theory, unseen phenomena such as proton decay or charged lepton flavor violation are attributed to the higher-dimensional operators and become statements about the scale of new physics.

Muon anomalies in $(g-2)_\mu$, \RK, and $b \to s \mu^+ \mu^-$ could be the first imprints of physics beyond the SM.
The philosophy followed here is to minimally extend the SM to account for these effects while keeping the same accidental symmetries of the lower-dimensional operators in the EFT expansion of the new model.
In particular, a $\U(1)_X$ lepton-flavored gauge symmetry is used to ensure a high-quality lepton flavor symmetry such that LFUV is allowed when suitable leptoquarks mediators are present, while LFV is not.
Besides, neutrino phenomenology and matter stability, when taken together, provide a nontrivial guide for model building.

Lepton-flavored $ \U(1)_X $ gauge extensions of the SM, under which leptoquarks become \textit{muoquarks}, provide versatile framework for simultaneously addressing muon anomalies as explained in Table~\ref{tab:model_types}. All three types of models (Type A, Type B and Type C) are exemplified in Sec.~\ref{sec:mu-tau_muon_anomalies}, while a detailed investigation is left for the future work.

\section*{Acknowledgements}

We thank Wolfgang Altmannshofer, Joe Davighi, Javier Fuentes-Martin and Julian Heeck for useful comments on v1. We particularly thank Jure Zupan for encouraging us to carefully examine the light $X$ phenomenology in \Ubl model. The work of AG and AET has received funding from the Swiss National Science Foundation (SNF) through the Eccellenza Professorial Fellowship ``Flavor Physics at the High Energy Frontier'' project number 186866.
The work of AG is also partially supported by the European Research Council (ERC) under the European Union’s Horizon 2020 research and innovation programme, grant agreement 833280 (FLAY).

\bibliographystyle{apsrev4-1}

\bibliography{GST.bib}

\begin{thebibliography}{149}%
\makeatletter
\providecommand \@ifxundefined [1]{%
 \@ifx{#1\undefined}
}%
\providecommand \@ifnum [1]{%
 \ifnum #1\expandafter \@firstoftwo
 \else \expandafter \@secondoftwo
 \fi
}%
\providecommand \@ifx [1]{%
 \ifx #1\expandafter \@firstoftwo
 \else \expandafter \@secondoftwo
 \fi
}%
\providecommand \natexlab [1]{#1}%
\providecommand \enquote  [1]{``#1''}%
\providecommand \bibnamefont  [1]{#1}%
\providecommand \bibfnamefont [1]{#1}%
\providecommand \citenamefont [1]{#1}%
\providecommand \href@noop [0]{\@secondoftwo}%
\providecommand \href [0]{\begingroup \@sanitize@url \@href}%
\providecommand \@href[1]{\@@startlink{#1}\@@href}%
\providecommand \@@href[1]{\endgroup#1\@@endlink}%
\providecommand \@sanitize@url [0]{\catcode `\\12\catcode `\$12\catcode
  `\&12\catcode `\#12\catcode `\^12\catcode `\_12\catcode `\%12\relax}%
\providecommand \@@startlink[1]{}%
\providecommand \@@endlink[0]{}%
\providecommand \url  [0]{\begingroup\@sanitize@url \@url }%
\providecommand \@url [1]{\endgroup\@href {#1}{\urlprefix }}%
\providecommand \urlprefix  [0]{URL }%
\providecommand \Eprint [0]{\href }%
\providecommand \doibase [0]{http://dx.doi.org/}%
\providecommand \selectlanguage [0]{\@gobble}%
\providecommand \bibinfo  [0]{\@secondoftwo}%
\providecommand \bibfield  [0]{\@secondoftwo}%
\providecommand \translation [1]{[#1]}%
\providecommand \BibitemOpen [0]{}%
\providecommand \bibitemStop [0]{}%
\providecommand \bibitemNoStop [0]{.\EOS\space}%
\providecommand \EOS [0]{\spacefactor3000\relax}%
\providecommand \BibitemShut  [1]{\csname bibitem#1\endcsname}%
\let\auto@bib@innerbib\@empty
\bibitem [{\citenamefont {Bennett}\ \emph {et~al.}(2006)\citenamefont {Bennett}
  \emph {et~al.}}]{Bennett:2006fi}%
  \BibitemOpen
  \bibfield  {author} {\bibinfo {author} {\bibfnamefont {G.~W.}\ \bibnamefont
  {Bennett}} \emph {et~al.} (\bibinfo {collaboration} {Muon g-2}),\ }\href
  {\doibase 10.1103/PhysRevD.73.072003} {\bibfield  {journal} {\bibinfo
  {journal} {Phys. Rev. D}\ }\textbf {\bibinfo {volume} {73}},\ \bibinfo
  {pages} {072003} (\bibinfo {year} {2006})},\ \Eprint
  {http://arxiv.org/abs/hep-ex/0602035} {arXiv:hep-ex/0602035} \BibitemShut
  {NoStop}%
\bibitem [{\citenamefont {Aoyama}\ \emph {et~al.}(2020)\citenamefont {Aoyama}
  \emph {et~al.}}]{Aoyama:2020ynm}%
  \BibitemOpen
  \bibfield  {author} {\bibinfo {author} {\bibfnamefont {T.}~\bibnamefont
  {Aoyama}} \emph {et~al.},\ }\href {\doibase 10.1016/j.physrep.2020.07.006}
  {\bibfield  {journal} {\bibinfo  {journal} {Phys. Rept.}\ }\textbf {\bibinfo
  {volume} {887}},\ \bibinfo {pages} {1} (\bibinfo {year} {2020})},\ \Eprint
  {http://arxiv.org/abs/2006.04822} {arXiv:2006.04822 [hep-ph]} \BibitemShut
  {NoStop}%
\bibitem [{\citenamefont {Aaij}\ \emph {et~al.}(2014)\citenamefont {Aaij} \emph
  {et~al.}}]{Aaij:2014ora}%
  \BibitemOpen
  \bibfield  {author} {\bibinfo {author} {\bibfnamefont {R.}~\bibnamefont
  {Aaij}} \emph {et~al.} (\bibinfo {collaboration} {LHCb}),\ }\href {\doibase
  10.1103/PhysRevLett.113.151601} {\bibfield  {journal} {\bibinfo  {journal}
  {Phys. Rev. Lett.}\ }\textbf {\bibinfo {volume} {113}},\ \bibinfo {pages}
  {151601} (\bibinfo {year} {2014})},\ \Eprint {http://arxiv.org/abs/1406.6482}
  {arXiv:1406.6482 [hep-ex]} \BibitemShut {NoStop}%
\bibitem [{\citenamefont {Aaij}\ \emph
  {et~al.}(2017{\natexlab{a}})\citenamefont {Aaij} \emph
  {et~al.}}]{Aaij:2017vbb}%
  \BibitemOpen
  \bibfield  {author} {\bibinfo {author} {\bibfnamefont {R.}~\bibnamefont
  {Aaij}} \emph {et~al.} (\bibinfo {collaboration} {LHCb}),\ }\href {\doibase
  10.1007/JHEP08(2017)055} {\bibfield  {journal} {\bibinfo  {journal} {JHEP}\
  }\textbf {\bibinfo {volume} {08}},\ \bibinfo {pages} {055} (\bibinfo {year}
  {2017}{\natexlab{a}})},\ \Eprint {http://arxiv.org/abs/1705.05802}
  {arXiv:1705.05802 [hep-ex]} \BibitemShut {NoStop}%
\bibitem [{\citenamefont {Aaij}\ \emph {et~al.}(2013)\citenamefont {Aaij} \emph
  {et~al.}}]{Aaij:2013qta}%
  \BibitemOpen
  \bibfield  {author} {\bibinfo {author} {\bibfnamefont {R.}~\bibnamefont
  {Aaij}} \emph {et~al.} (\bibinfo {collaboration} {LHCb}),\ }\href {\doibase
  10.1103/PhysRevLett.111.191801} {\bibfield  {journal} {\bibinfo  {journal}
  {Phys. Rev. Lett.}\ }\textbf {\bibinfo {volume} {111}},\ \bibinfo {pages}
  {191801} (\bibinfo {year} {2013})},\ \Eprint {http://arxiv.org/abs/1308.1707}
  {arXiv:1308.1707 [hep-ex]} \BibitemShut {NoStop}%
\bibitem [{\citenamefont {Aaij}\ \emph {et~al.}(2016)\citenamefont {Aaij} \emph
  {et~al.}}]{Aaij:2015oid}%
  \BibitemOpen
  \bibfield  {author} {\bibinfo {author} {\bibfnamefont {R.}~\bibnamefont
  {Aaij}} \emph {et~al.} (\bibinfo {collaboration} {LHCb}),\ }\href {\doibase
  10.1007/JHEP02(2016)104} {\bibfield  {journal} {\bibinfo  {journal} {JHEP}\
  }\textbf {\bibinfo {volume} {02}},\ \bibinfo {pages} {104} (\bibinfo {year}
  {2016})},\ \Eprint {http://arxiv.org/abs/1512.04442} {arXiv:1512.04442
  [hep-ex]} \BibitemShut {NoStop}%
\bibitem [{\citenamefont {Aaij}\ \emph {et~al.}(2019)\citenamefont {Aaij} \emph
  {et~al.}}]{Aaij:2019wad}%
  \BibitemOpen
  \bibfield  {author} {\bibinfo {author} {\bibfnamefont {R.}~\bibnamefont
  {Aaij}} \emph {et~al.} (\bibinfo {collaboration} {LHCb}),\ }\href {\doibase
  10.1103/PhysRevLett.122.191801} {\bibfield  {journal} {\bibinfo  {journal}
  {Phys. Rev. Lett.}\ }\textbf {\bibinfo {volume} {122}},\ \bibinfo {pages}
  {191801} (\bibinfo {year} {2019})},\ \Eprint
  {http://arxiv.org/abs/1903.09252} {arXiv:1903.09252 [hep-ex]} \BibitemShut
  {NoStop}%
\bibitem [{\citenamefont {Khachatryan}\ \emph {et~al.}(2015)\citenamefont
  {Khachatryan} \emph {et~al.}}]{CMS:2014xfa}%
  \BibitemOpen
  \bibfield  {author} {\bibinfo {author} {\bibfnamefont {V.}~\bibnamefont
  {Khachatryan}} \emph {et~al.} (\bibinfo {collaboration} {CMS, LHCb}),\ }\href
  {\doibase 10.1038/nature14474} {\bibfield  {journal} {\bibinfo  {journal}
  {Nature}\ }\textbf {\bibinfo {volume} {522}},\ \bibinfo {pages} {68}
  (\bibinfo {year} {2015})},\ \Eprint {http://arxiv.org/abs/1411.4413}
  {arXiv:1411.4413 [hep-ex]} \BibitemShut {NoStop}%
\bibitem [{\citenamefont {Aaij}\ \emph
  {et~al.}(2017{\natexlab{b}})\citenamefont {Aaij} \emph
  {et~al.}}]{Aaij:2017vad}%
  \BibitemOpen
  \bibfield  {author} {\bibinfo {author} {\bibfnamefont {R.}~\bibnamefont
  {Aaij}} \emph {et~al.} (\bibinfo {collaboration} {LHCb}),\ }\href {\doibase
  10.1103/PhysRevLett.118.191801} {\bibfield  {journal} {\bibinfo  {journal}
  {Phys. Rev. Lett.}\ }\textbf {\bibinfo {volume} {118}},\ \bibinfo {pages}
  {191801} (\bibinfo {year} {2017}{\natexlab{b}})},\ \Eprint
  {http://arxiv.org/abs/1703.05747} {arXiv:1703.05747 [hep-ex]} \BibitemShut
  {NoStop}%
\bibitem [{\citenamefont {Hiller}\ and\ \citenamefont
  {Kruger}(2004)}]{Hiller:2003js}%
  \BibitemOpen
  \bibfield  {author} {\bibinfo {author} {\bibfnamefont {G.}~\bibnamefont
  {Hiller}}\ and\ \bibinfo {author} {\bibfnamefont {F.}~\bibnamefont
  {Kruger}},\ }\href {\doibase 10.1103/PhysRevD.69.074020} {\bibfield
  {journal} {\bibinfo  {journal} {Phys. Rev. D}\ }\textbf {\bibinfo {volume}
  {69}},\ \bibinfo {pages} {074020} (\bibinfo {year} {2004})},\ \Eprint
  {http://arxiv.org/abs/hep-ph/0310219} {arXiv:hep-ph/0310219} \BibitemShut
  {NoStop}%
\bibitem [{\citenamefont {Bordone}\ \emph {et~al.}(2016)\citenamefont
  {Bordone}, \citenamefont {Isidori},\ and\ \citenamefont
  {Pattori}}]{Bordone:2016gaq}%
  \BibitemOpen
  \bibfield  {author} {\bibinfo {author} {\bibfnamefont {M.}~\bibnamefont
  {Bordone}}, \bibinfo {author} {\bibfnamefont {G.}~\bibnamefont {Isidori}}, \
  and\ \bibinfo {author} {\bibfnamefont {A.}~\bibnamefont {Pattori}},\ }\href
  {\doibase 10.1140/epjc/s10052-016-4274-7} {\bibfield  {journal} {\bibinfo
  {journal} {Eur. Phys. J. C}\ }\textbf {\bibinfo {volume} {76}},\ \bibinfo
  {pages} {440} (\bibinfo {year} {2016})},\ \Eprint
  {http://arxiv.org/abs/1605.07633} {arXiv:1605.07633 [hep-ph]} \BibitemShut
  {NoStop}%
\bibitem [{\citenamefont {Isidori}\ \emph {et~al.}(2020)\citenamefont
  {Isidori}, \citenamefont {Nabeebaccus},\ and\ \citenamefont
  {Zwicky}}]{Isidori:2020acz}%
  \BibitemOpen
  \bibfield  {author} {\bibinfo {author} {\bibfnamefont {G.}~\bibnamefont
  {Isidori}}, \bibinfo {author} {\bibfnamefont {S.}~\bibnamefont
  {Nabeebaccus}}, \ and\ \bibinfo {author} {\bibfnamefont {R.}~\bibnamefont
  {Zwicky}},\ }\href {\doibase 10.1007/JHEP12(2020)104} {\bibfield  {journal}
  {\bibinfo  {journal} {JHEP}\ }\textbf {\bibinfo {volume} {12}},\ \bibinfo
  {pages} {104} (\bibinfo {year} {2020})},\ \Eprint
  {http://arxiv.org/abs/2009.00929} {arXiv:2009.00929 [hep-ph]} \BibitemShut
  {NoStop}%
\bibitem [{\citenamefont {Colangelo}\ \emph {et~al.}(2021)\citenamefont
  {Colangelo}, \citenamefont {Hoferichter},\ and\ \citenamefont
  {Stoffer}}]{Colangelo:2020lcg}%
  \BibitemOpen
  \bibfield  {author} {\bibinfo {author} {\bibfnamefont {G.}~\bibnamefont
  {Colangelo}}, \bibinfo {author} {\bibfnamefont {M.}~\bibnamefont
  {Hoferichter}}, \ and\ \bibinfo {author} {\bibfnamefont {P.}~\bibnamefont
  {Stoffer}},\ }\href {\doibase 10.1016/j.physletb.2021.136073} {\bibfield
  {journal} {\bibinfo  {journal} {Phys. Lett. B}\ }\textbf {\bibinfo {volume}
  {814}},\ \bibinfo {pages} {136073} (\bibinfo {year} {2021})},\ \Eprint
  {http://arxiv.org/abs/2010.07943} {arXiv:2010.07943 [hep-ph]} \BibitemShut
  {NoStop}%
\bibitem [{\citenamefont {Aoyama}\ \emph {et~al.}(2012)\citenamefont {Aoyama},
  \citenamefont {Hayakawa}, \citenamefont {Kinoshita},\ and\ \citenamefont
  {Nio}}]{aoyama:2012wk}%
  \BibitemOpen
  \bibfield  {author} {\bibinfo {author} {\bibfnamefont {T.}~\bibnamefont
  {Aoyama}}, \bibinfo {author} {\bibfnamefont {M.}~\bibnamefont {Hayakawa}},
  \bibinfo {author} {\bibfnamefont {T.}~\bibnamefont {Kinoshita}}, \ and\
  \bibinfo {author} {\bibfnamefont {M.}~\bibnamefont {Nio}},\ }\href {\doibase
  10.1103/PhysRevLett.109.111808} {\bibfield  {journal} {\bibinfo  {journal}
  {Phys. Rev. Lett.}\ }\textbf {\bibinfo {volume} {109}},\ \bibinfo {pages}
  {111808} (\bibinfo {year} {2012})},\ \Eprint {http://arxiv.org/abs/1205.5370}
  {arXiv:1205.5370 [hep-ph]} \BibitemShut {NoStop}%
\bibitem [{\citenamefont {Aoyama}\ \emph {et~al.}(2019)\citenamefont {Aoyama},
  \citenamefont {Kinoshita},\ and\ \citenamefont {Nio}}]{Aoyama:2019ryr}%
  \BibitemOpen
  \bibfield  {author} {\bibinfo {author} {\bibfnamefont {T.}~\bibnamefont
  {Aoyama}}, \bibinfo {author} {\bibfnamefont {T.}~\bibnamefont {Kinoshita}}, \
  and\ \bibinfo {author} {\bibfnamefont {M.}~\bibnamefont {Nio}},\ }\href
  {\doibase 10.3390/atoms7010028} {\bibfield  {journal} {\bibinfo  {journal}
  {Atoms}\ }\textbf {\bibinfo {volume} {7}},\ \bibinfo {pages} {28} (\bibinfo
  {year} {2019})}\BibitemShut {NoStop}%
\bibitem [{\citenamefont {Czarnecki}\ \emph {et~al.}(2003)\citenamefont
  {Czarnecki}, \citenamefont {Marciano},\ and\ \citenamefont
  {Vainshtein}}]{czarnecki:2002nt}%
  \BibitemOpen
  \bibfield  {author} {\bibinfo {author} {\bibfnamefont {A.}~\bibnamefont
  {Czarnecki}}, \bibinfo {author} {\bibfnamefont {W.~J.}\ \bibnamefont
  {Marciano}}, \ and\ \bibinfo {author} {\bibfnamefont {A.}~\bibnamefont
  {Vainshtein}},\ }\href {\doibase 10.1103/PhysRevD.67.073006} {\bibfield
  {journal} {\bibinfo  {journal} {Phys. Rev.}\ }\textbf {\bibinfo {volume}
  {D67}},\ \bibinfo {pages} {073006} (\bibinfo {year} {2003})},\ \bibinfo
  {note} {[Erratum: Phys. Rev. {\bf D73}, 119901 (2006)]},\ \Eprint
  {http://arxiv.org/abs/hep-ph/0212229} {arXiv:hep-ph/0212229 [hep-ph]}
  \BibitemShut {NoStop}%
\bibitem [{\citenamefont {Gnendiger}\ \emph {et~al.}(2013)\citenamefont
  {Gnendiger}, \citenamefont {St{\"o}ckinger},\ and\ \citenamefont
  {St{\"o}ckinger-Kim}}]{gnendiger:2013pva}%
  \BibitemOpen
  \bibfield  {author} {\bibinfo {author} {\bibfnamefont {C.}~\bibnamefont
  {Gnendiger}}, \bibinfo {author} {\bibfnamefont {D.}~\bibnamefont
  {St{\"o}ckinger}}, \ and\ \bibinfo {author} {\bibfnamefont {H.}~\bibnamefont
  {St{\"o}ckinger-Kim}},\ }\href {\doibase 10.1103/PhysRevD.88.053005}
  {\bibfield  {journal} {\bibinfo  {journal} {Phys. Rev.}\ }\textbf {\bibinfo
  {volume} {D88}},\ \bibinfo {pages} {053005} (\bibinfo {year} {2013})},\
  \Eprint {http://arxiv.org/abs/1306.5546} {arXiv:1306.5546 [hep-ph]}
  \BibitemShut {NoStop}%
\bibitem [{\citenamefont {Davier}\ \emph {et~al.}(2017)\citenamefont {Davier},
  \citenamefont {Hoecker}, \citenamefont {Malaescu},\ and\ \citenamefont
  {Zhang}}]{davier:2017zfy}%
  \BibitemOpen
  \bibfield  {author} {\bibinfo {author} {\bibfnamefont {M.}~\bibnamefont
  {Davier}}, \bibinfo {author} {\bibfnamefont {A.}~\bibnamefont {Hoecker}},
  \bibinfo {author} {\bibfnamefont {B.}~\bibnamefont {Malaescu}}, \ and\
  \bibinfo {author} {\bibfnamefont {Z.}~\bibnamefont {Zhang}},\ }\href
  {\doibase 10.1140/epjc/s10052-017-5161-6} {\bibfield  {journal} {\bibinfo
  {journal} {Eur. Phys. J.}\ }\textbf {\bibinfo {volume} {C77}},\ \bibinfo
  {pages} {827} (\bibinfo {year} {2017})},\ \Eprint
  {http://arxiv.org/abs/1706.09436} {arXiv:1706.09436 [hep-ph]} \BibitemShut
  {NoStop}%
\bibitem [{\citenamefont {Keshavarzi}\ \emph {et~al.}(2018)\citenamefont
  {Keshavarzi}, \citenamefont {Nomura},\ and\ \citenamefont
  {Teubner}}]{keshavarzi:2018mgv}%
  \BibitemOpen
  \bibfield  {author} {\bibinfo {author} {\bibfnamefont {A.}~\bibnamefont
  {Keshavarzi}}, \bibinfo {author} {\bibfnamefont {D.}~\bibnamefont {Nomura}},
  \ and\ \bibinfo {author} {\bibfnamefont {T.}~\bibnamefont {Teubner}},\ }\href
  {\doibase 10.1103/PhysRevD.97.114025} {\bibfield  {journal} {\bibinfo
  {journal} {Phys. Rev.}\ }\textbf {\bibinfo {volume} {D97}},\ \bibinfo {pages}
  {114025} (\bibinfo {year} {2018})},\ \Eprint
  {http://arxiv.org/abs/1802.02995} {arXiv:1802.02995 [hep-ph]} \BibitemShut
  {NoStop}%
\bibitem [{\citenamefont {Colangelo}\ \emph {et~al.}(2019)\citenamefont
  {Colangelo}, \citenamefont {Hoferichter},\ and\ \citenamefont
  {Stoffer}}]{colangelo:2018mtw}%
  \BibitemOpen
  \bibfield  {author} {\bibinfo {author} {\bibfnamefont {G.}~\bibnamefont
  {Colangelo}}, \bibinfo {author} {\bibfnamefont {M.}~\bibnamefont
  {Hoferichter}}, \ and\ \bibinfo {author} {\bibfnamefont {P.}~\bibnamefont
  {Stoffer}},\ }\href {\doibase 10.1007/JHEP02(2019)006} {\bibfield  {journal}
  {\bibinfo  {journal} {JHEP}\ }\textbf {\bibinfo {volume} {02}},\ \bibinfo
  {pages} {006} (\bibinfo {year} {2019})},\ \Eprint
  {http://arxiv.org/abs/1810.00007} {arXiv:1810.00007 [hep-ph]} \BibitemShut
  {NoStop}%
\bibitem [{\citenamefont {Hoferichter}\ \emph {et~al.}(2019)\citenamefont
  {Hoferichter}, \citenamefont {Hoid},\ and\ \citenamefont
  {Kubis}}]{hoferichter:2019gzf}%
  \BibitemOpen
  \bibfield  {author} {\bibinfo {author} {\bibfnamefont {M.}~\bibnamefont
  {Hoferichter}}, \bibinfo {author} {\bibfnamefont {B.-L.}\ \bibnamefont
  {Hoid}}, \ and\ \bibinfo {author} {\bibfnamefont {B.}~\bibnamefont {Kubis}},\
  }\href {\doibase 10.1007/JHEP08(2019)137} {\bibfield  {journal} {\bibinfo
  {journal} {JHEP}\ }\textbf {\bibinfo {volume} {08}},\ \bibinfo {pages} {137}
  (\bibinfo {year} {2019})},\ \Eprint {http://arxiv.org/abs/1907.01556}
  {arXiv:1907.01556 [hep-ph]} \BibitemShut {NoStop}%
\bibitem [{\citenamefont {Davier}\ \emph {et~al.}(2020)\citenamefont {Davier},
  \citenamefont {Hoecker}, \citenamefont {Malaescu},\ and\ \citenamefont
  {Zhang}}]{davier:2019can}%
  \BibitemOpen
  \bibfield  {author} {\bibinfo {author} {\bibfnamefont {M.}~\bibnamefont
  {Davier}}, \bibinfo {author} {\bibfnamefont {A.}~\bibnamefont {Hoecker}},
  \bibinfo {author} {\bibfnamefont {B.}~\bibnamefont {Malaescu}}, \ and\
  \bibinfo {author} {\bibfnamefont {Z.}~\bibnamefont {Zhang}},\ }\href
  {\doibase 10.1140/epjc/s10052-020-7792-2} {\bibfield  {journal} {\bibinfo
  {journal} {Eur. Phys. J.}\ }\textbf {\bibinfo {volume} {C80}},\ \bibinfo
  {pages} {241} (\bibinfo {year} {2020})},\ \bibinfo {note} {[Erratum: Eur.
  Phys. J. {\bf C80}, 410 (2020)]},\ \Eprint {http://arxiv.org/abs/1908.00921}
  {arXiv:1908.00921 [hep-ph]} \BibitemShut {NoStop}%
\bibitem [{\citenamefont {Keshavarzi}\ \emph {et~al.}(2020)\citenamefont
  {Keshavarzi}, \citenamefont {Nomura},\ and\ \citenamefont
  {Teubner}}]{keshavarzi:2019abf}%
  \BibitemOpen
  \bibfield  {author} {\bibinfo {author} {\bibfnamefont {A.}~\bibnamefont
  {Keshavarzi}}, \bibinfo {author} {\bibfnamefont {D.}~\bibnamefont {Nomura}},
  \ and\ \bibinfo {author} {\bibfnamefont {T.}~\bibnamefont {Teubner}},\ }\href
  {\doibase 10.1103/PhysRevD.101.014029} {\bibfield  {journal} {\bibinfo
  {journal} {Phys. Rev.}\ }\textbf {\bibinfo {volume} {D101}},\ \bibinfo
  {pages} {014029} (\bibinfo {year} {2020})},\ \Eprint
  {http://arxiv.org/abs/1911.00367} {arXiv:1911.00367 [hep-ph]} \BibitemShut
  {NoStop}%
\bibitem [{\citenamefont {Kurz}\ \emph {et~al.}(2014)\citenamefont {Kurz},
  \citenamefont {Liu}, \citenamefont {Marquard},\ and\ \citenamefont
  {Steinhauser}}]{kurz:2014wya}%
  \BibitemOpen
  \bibfield  {author} {\bibinfo {author} {\bibfnamefont {A.}~\bibnamefont
  {Kurz}}, \bibinfo {author} {\bibfnamefont {T.}~\bibnamefont {Liu}}, \bibinfo
  {author} {\bibfnamefont {P.}~\bibnamefont {Marquard}}, \ and\ \bibinfo
  {author} {\bibfnamefont {M.}~\bibnamefont {Steinhauser}},\ }\href {\doibase
  10.1016/j.physletb.2014.05.043} {\bibfield  {journal} {\bibinfo  {journal}
  {Phys. Lett.}\ }\textbf {\bibinfo {volume} {B734}},\ \bibinfo {pages} {144}
  (\bibinfo {year} {2014})},\ \Eprint {http://arxiv.org/abs/1403.6400}
  {arXiv:1403.6400 [hep-ph]} \BibitemShut {NoStop}%
\bibitem [{\citenamefont {Melnikov}\ and\ \citenamefont
  {Vainshtein}(2004)}]{melnikov:2003xd}%
  \BibitemOpen
  \bibfield  {author} {\bibinfo {author} {\bibfnamefont {K.}~\bibnamefont
  {Melnikov}}\ and\ \bibinfo {author} {\bibfnamefont {A.}~\bibnamefont
  {Vainshtein}},\ }\href {\doibase 10.1103/PhysRevD.70.113006} {\bibfield
  {journal} {\bibinfo  {journal} {Phys. Rev.}\ }\textbf {\bibinfo {volume}
  {D70}},\ \bibinfo {pages} {113006} (\bibinfo {year} {2004})},\ \Eprint
  {http://arxiv.org/abs/hep-ph/0312226} {arXiv:hep-ph/0312226 [hep-ph]}
  \BibitemShut {NoStop}%
\bibitem [{\citenamefont {Masjuan}\ and\ \citenamefont
  {S{\'a}nchez-Puertas}(2017)}]{masjuan:2017tvw}%
  \BibitemOpen
  \bibfield  {author} {\bibinfo {author} {\bibfnamefont {P.}~\bibnamefont
  {Masjuan}}\ and\ \bibinfo {author} {\bibfnamefont {P.}~\bibnamefont
  {S{\'a}nchez-Puertas}},\ }\href {\doibase 10.1103/PhysRevD.95.054026}
  {\bibfield  {journal} {\bibinfo  {journal} {Phys. Rev.}\ }\textbf {\bibinfo
  {volume} {D95}},\ \bibinfo {pages} {054026} (\bibinfo {year} {2017})},\
  \Eprint {http://arxiv.org/abs/1701.05829} {arXiv:1701.05829 [hep-ph]}
  \BibitemShut {NoStop}%
\bibitem [{\citenamefont {Colangelo}\ \emph {et~al.}(2017)\citenamefont
  {Colangelo}, \citenamefont {Hoferichter}, \citenamefont {Procura},\ and\
  \citenamefont {Stoffer}}]{Colangelo:2017fiz}%
  \BibitemOpen
  \bibfield  {author} {\bibinfo {author} {\bibfnamefont {G.}~\bibnamefont
  {Colangelo}}, \bibinfo {author} {\bibfnamefont {M.}~\bibnamefont
  {Hoferichter}}, \bibinfo {author} {\bibfnamefont {M.}~\bibnamefont
  {Procura}}, \ and\ \bibinfo {author} {\bibfnamefont {P.}~\bibnamefont
  {Stoffer}},\ }\href {\doibase 10.1007/JHEP04(2017)161} {\bibfield  {journal}
  {\bibinfo  {journal} {JHEP}\ }\textbf {\bibinfo {volume} {04}},\ \bibinfo
  {pages} {161} (\bibinfo {year} {2017})},\ \Eprint
  {http://arxiv.org/abs/1702.07347} {arXiv:1702.07347 [hep-ph]} \BibitemShut
  {NoStop}%
\bibitem [{\citenamefont {Hoferichter}\ \emph {et~al.}(2018)\citenamefont
  {Hoferichter}, \citenamefont {Hoid}, \citenamefont {Kubis}, \citenamefont
  {Leupold},\ and\ \citenamefont {Schneider}}]{hoferichter:2018kwz}%
  \BibitemOpen
  \bibfield  {author} {\bibinfo {author} {\bibfnamefont {M.}~\bibnamefont
  {Hoferichter}}, \bibinfo {author} {\bibfnamefont {B.-L.}\ \bibnamefont
  {Hoid}}, \bibinfo {author} {\bibfnamefont {B.}~\bibnamefont {Kubis}},
  \bibinfo {author} {\bibfnamefont {S.}~\bibnamefont {Leupold}}, \ and\
  \bibinfo {author} {\bibfnamefont {S.~P.}\ \bibnamefont {Schneider}},\ }\href
  {\doibase 10.1007/JHEP10(2018)141} {\bibfield  {journal} {\bibinfo  {journal}
  {JHEP}\ }\textbf {\bibinfo {volume} {10}},\ \bibinfo {pages} {141} (\bibinfo
  {year} {2018})},\ \Eprint {http://arxiv.org/abs/1808.04823} {arXiv:1808.04823
  [hep-ph]} \BibitemShut {NoStop}%
\bibitem [{\citenamefont {G{\'e}rardin}\ \emph {et~al.}(2019)\citenamefont
  {G{\'e}rardin}, \citenamefont {Meyer},\ and\ \citenamefont
  {Nyffeler}}]{gerardin:2019vio}%
  \BibitemOpen
  \bibfield  {author} {\bibinfo {author} {\bibfnamefont {A.}~\bibnamefont
  {G{\'e}rardin}}, \bibinfo {author} {\bibfnamefont {H.~B.}\ \bibnamefont
  {Meyer}}, \ and\ \bibinfo {author} {\bibfnamefont {A.}~\bibnamefont
  {Nyffeler}},\ }\href {\doibase 10.1103/PhysRevD.100.034520} {\bibfield
  {journal} {\bibinfo  {journal} {Phys. Rev.}\ }\textbf {\bibinfo {volume}
  {D100}},\ \bibinfo {pages} {034520} (\bibinfo {year} {2019})},\ \Eprint
  {http://arxiv.org/abs/1903.09471} {arXiv:1903.09471 [hep-lat]} \BibitemShut
  {NoStop}%
\bibitem [{\citenamefont {Bijnens}\ \emph {et~al.}(2019)\citenamefont
  {Bijnens}, \citenamefont {Hermansson-Truedsson},\ and\ \citenamefont
  {Rodr{\'i}guez-S{\'a}nchez}}]{bijnens:2019ghy}%
  \BibitemOpen
  \bibfield  {author} {\bibinfo {author} {\bibfnamefont {J.}~\bibnamefont
  {Bijnens}}, \bibinfo {author} {\bibfnamefont {N.}~\bibnamefont
  {Hermansson-Truedsson}}, \ and\ \bibinfo {author} {\bibfnamefont
  {A.}~\bibnamefont {Rodr{\'i}guez-S{\'a}nchez}},\ }\href {\doibase
  10.1016/j.physletb.2019.134994} {\bibfield  {journal} {\bibinfo  {journal}
  {Phys. Lett.}\ }\textbf {\bibinfo {volume} {B798}},\ \bibinfo {pages}
  {134994} (\bibinfo {year} {2019})},\ \Eprint
  {http://arxiv.org/abs/1908.03331} {arXiv:1908.03331 [hep-ph]} \BibitemShut
  {NoStop}%
\bibitem [{\citenamefont {Colangelo}\ \emph {et~al.}(2020)\citenamefont
  {Colangelo}, \citenamefont {Hagelstein}, \citenamefont {Hoferichter},
  \citenamefont {Laub},\ and\ \citenamefont {Stoffer}}]{colangelo:2019uex}%
  \BibitemOpen
  \bibfield  {author} {\bibinfo {author} {\bibfnamefont {G.}~\bibnamefont
  {Colangelo}}, \bibinfo {author} {\bibfnamefont {F.}~\bibnamefont
  {Hagelstein}}, \bibinfo {author} {\bibfnamefont {M.}~\bibnamefont
  {Hoferichter}}, \bibinfo {author} {\bibfnamefont {L.}~\bibnamefont {Laub}}, \
  and\ \bibinfo {author} {\bibfnamefont {P.}~\bibnamefont {Stoffer}},\ }\href
  {\doibase 10.1007/JHEP03(2020)101} {\bibfield  {journal} {\bibinfo  {journal}
  {JHEP}\ }\textbf {\bibinfo {volume} {03}},\ \bibinfo {pages} {101} (\bibinfo
  {year} {2020})},\ \Eprint {http://arxiv.org/abs/1910.13432} {arXiv:1910.13432
  [hep-ph]} \BibitemShut {NoStop}%
\bibitem [{\citenamefont {Blum}\ \emph {et~al.}(2020)\citenamefont {Blum},
  \citenamefont {Christ}, \citenamefont {Hayakawa}, \citenamefont {Izubuchi},
  \citenamefont {Jin}, \citenamefont {Jung},\ and\ \citenamefont
  {Lehner}}]{Blum:2019ugy}%
  \BibitemOpen
  \bibfield  {author} {\bibinfo {author} {\bibfnamefont {T.}~\bibnamefont
  {Blum}}, \bibinfo {author} {\bibfnamefont {N.}~\bibnamefont {Christ}},
  \bibinfo {author} {\bibfnamefont {M.}~\bibnamefont {Hayakawa}}, \bibinfo
  {author} {\bibfnamefont {T.}~\bibnamefont {Izubuchi}}, \bibinfo {author}
  {\bibfnamefont {L.}~\bibnamefont {Jin}}, \bibinfo {author} {\bibfnamefont
  {C.}~\bibnamefont {Jung}}, \ and\ \bibinfo {author} {\bibfnamefont
  {C.}~\bibnamefont {Lehner}},\ }\href {\doibase
  10.1103/PhysRevLett.124.132002} {\bibfield  {journal} {\bibinfo  {journal}
  {Phys. Rev. Lett.}\ }\textbf {\bibinfo {volume} {124}},\ \bibinfo {pages}
  {132002} (\bibinfo {year} {2020})},\ \Eprint
  {http://arxiv.org/abs/1911.08123} {arXiv:1911.08123 [hep-lat]} \BibitemShut
  {NoStop}%
\bibitem [{\citenamefont {Colangelo}\ \emph {et~al.}(2014)\citenamefont
  {Colangelo}, \citenamefont {Hoferichter}, \citenamefont {Nyffeler},
  \citenamefont {Passera},\ and\ \citenamefont {Stoffer}}]{colangelo:2014qya}%
  \BibitemOpen
  \bibfield  {author} {\bibinfo {author} {\bibfnamefont {G.}~\bibnamefont
  {Colangelo}}, \bibinfo {author} {\bibfnamefont {M.}~\bibnamefont
  {Hoferichter}}, \bibinfo {author} {\bibfnamefont {A.}~\bibnamefont
  {Nyffeler}}, \bibinfo {author} {\bibfnamefont {M.}~\bibnamefont {Passera}}, \
  and\ \bibinfo {author} {\bibfnamefont {P.}~\bibnamefont {Stoffer}},\ }\href
  {\doibase 10.1016/j.physletb.2014.06.012} {\bibfield  {journal} {\bibinfo
  {journal} {Phys. Lett.}\ }\textbf {\bibinfo {volume} {B735}},\ \bibinfo
  {pages} {90} (\bibinfo {year} {2014})},\ \Eprint
  {http://arxiv.org/abs/1403.7512} {arXiv:1403.7512 [hep-ph]} \BibitemShut
  {NoStop}%
\bibitem [{\citenamefont {Borsanyi}\ \emph {et~al.}(2020)\citenamefont
  {Borsanyi} \emph {et~al.}}]{Borsanyi:2020mff}%
  \BibitemOpen
  \bibfield  {author} {\bibinfo {author} {\bibfnamefont {S.}~\bibnamefont
  {Borsanyi}} \emph {et~al.},\ }\href@noop {} {\  (\bibinfo {year} {2020})},\
  \Eprint {http://arxiv.org/abs/2002.12347} {arXiv:2002.12347 [hep-lat]}
  \BibitemShut {NoStop}%
\bibitem [{\citenamefont {Aaij}\ \emph {et~al.}(2021)\citenamefont {Aaij} \emph
  {et~al.}}]{Aaij:2021vac}%
  \BibitemOpen
  \bibfield  {author} {\bibinfo {author} {\bibfnamefont {R.}~\bibnamefont
  {Aaij}} \emph {et~al.} (\bibinfo {collaboration} {LHCb}),\ }\href@noop {} {\
  (\bibinfo {year} {2021})},\ \Eprint {http://arxiv.org/abs/2103.11769}
  {arXiv:2103.11769 [hep-ex]} \BibitemShut {NoStop}%
\bibitem [{\citenamefont {Abi}\ \emph {et~al.}(2021)\citenamefont {Abi} \emph
  {et~al.}}]{Abi:2021gix}%
  \BibitemOpen
  \bibfield  {author} {\bibinfo {author} {\bibfnamefont {B.}~\bibnamefont
  {Abi}} \emph {et~al.},\ }\href {\doibase 10.1103/PhysRevLett.126.141801} {\
  (\bibinfo {year} {2021}),\ 10.1103/PhysRevLett.126.141801},\ \Eprint
  {http://arxiv.org/abs/2104.03281} {arXiv:2104.03281 [hep-ex]} \BibitemShut
  {NoStop}%
\bibitem [{\citenamefont {Glashow}\ \emph {et~al.}(2015)\citenamefont
  {Glashow}, \citenamefont {Guadagnoli},\ and\ \citenamefont
  {Lane}}]{Glashow:2014iga}%
  \BibitemOpen
  \bibfield  {author} {\bibinfo {author} {\bibfnamefont {S.~L.}\ \bibnamefont
  {Glashow}}, \bibinfo {author} {\bibfnamefont {D.}~\bibnamefont {Guadagnoli}},
  \ and\ \bibinfo {author} {\bibfnamefont {K.}~\bibnamefont {Lane}},\ }\href
  {\doibase 10.1103/PhysRevLett.114.091801} {\bibfield  {journal} {\bibinfo
  {journal} {Phys. Rev. Lett.}\ }\textbf {\bibinfo {volume} {114}},\ \bibinfo
  {pages} {091801} (\bibinfo {year} {2015})},\ \Eprint
  {http://arxiv.org/abs/1411.0565} {arXiv:1411.0565 [hep-ph]} \BibitemShut
  {NoStop}%
\bibitem [{\citenamefont {Giudice}\ \emph {et~al.}(2012)\citenamefont
  {Giudice}, \citenamefont {Paradisi},\ and\ \citenamefont
  {Passera}}]{Giudice:2012ms}%
  \BibitemOpen
  \bibfield  {author} {\bibinfo {author} {\bibfnamefont {G.~F.}\ \bibnamefont
  {Giudice}}, \bibinfo {author} {\bibfnamefont {P.}~\bibnamefont {Paradisi}}, \
  and\ \bibinfo {author} {\bibfnamefont {M.}~\bibnamefont {Passera}},\ }\href
  {\doibase 10.1007/JHEP11(2012)113} {\bibfield  {journal} {\bibinfo  {journal}
  {JHEP}\ }\textbf {\bibinfo {volume} {11}},\ \bibinfo {pages} {113} (\bibinfo
  {year} {2012})},\ \Eprint {http://arxiv.org/abs/1208.6583} {arXiv:1208.6583
  [hep-ph]} \BibitemShut {NoStop}%
\bibitem [{\citenamefont {Baldini}\ \emph {et~al.}(2016)\citenamefont {Baldini}
  \emph {et~al.}}]{TheMEG:2016wtm}%
  \BibitemOpen
  \bibfield  {author} {\bibinfo {author} {\bibfnamefont {A.~M.}\ \bibnamefont
  {Baldini}} \emph {et~al.} (\bibinfo {collaboration} {MEG}),\ }\href {\doibase
  10.1140/epjc/s10052-016-4271-x} {\bibfield  {journal} {\bibinfo  {journal}
  {Eur. Phys. J. C}\ }\textbf {\bibinfo {volume} {76}},\ \bibinfo {pages} {434}
  (\bibinfo {year} {2016})},\ \Eprint {http://arxiv.org/abs/1605.05081}
  {arXiv:1605.05081 [hep-ex]} \BibitemShut {NoStop}%
\bibitem [{\citenamefont {Baek}\ \emph {et~al.}(2001)\citenamefont {Baek},
  \citenamefont {Deshpande}, \citenamefont {He},\ and\ \citenamefont
  {Ko}}]{Baek:2001kca}%
  \BibitemOpen
  \bibfield  {author} {\bibinfo {author} {\bibfnamefont {S.}~\bibnamefont
  {Baek}}, \bibinfo {author} {\bibfnamefont {N.~G.}\ \bibnamefont {Deshpande}},
  \bibinfo {author} {\bibfnamefont {X.~G.}\ \bibnamefont {He}}, \ and\ \bibinfo
  {author} {\bibfnamefont {P.}~\bibnamefont {Ko}},\ }\href {\doibase
  10.1103/PhysRevD.64.055006} {\bibfield  {journal} {\bibinfo  {journal} {Phys.
  Rev. D}\ }\textbf {\bibinfo {volume} {64}},\ \bibinfo {pages} {055006}
  (\bibinfo {year} {2001})},\ \Eprint {http://arxiv.org/abs/hep-ph/0104141}
  {arXiv:hep-ph/0104141} \BibitemShut {NoStop}%
\bibitem [{\citenamefont {Ma}\ \emph {et~al.}(2002)\citenamefont {Ma},
  \citenamefont {Roy},\ and\ \citenamefont {Roy}}]{Ma:2001md}%
  \BibitemOpen
  \bibfield  {author} {\bibinfo {author} {\bibfnamefont {E.}~\bibnamefont
  {Ma}}, \bibinfo {author} {\bibfnamefont {D.~P.}\ \bibnamefont {Roy}}, \ and\
  \bibinfo {author} {\bibfnamefont {S.}~\bibnamefont {Roy}},\ }\href {\doibase
  10.1016/S0370-2693(01)01428-9} {\bibfield  {journal} {\bibinfo  {journal}
  {Phys. Lett. B}\ }\textbf {\bibinfo {volume} {525}},\ \bibinfo {pages} {101}
  (\bibinfo {year} {2002})},\ \Eprint {http://arxiv.org/abs/hep-ph/0110146}
  {arXiv:hep-ph/0110146} \BibitemShut {NoStop}%
\bibitem [{\citenamefont {Harigaya}\ \emph {et~al.}(2014)\citenamefont
  {Harigaya}, \citenamefont {Igari}, \citenamefont {Nojiri}, \citenamefont
  {Takeuchi},\ and\ \citenamefont {Tobe}}]{Harigaya:2013twa}%
  \BibitemOpen
  \bibfield  {author} {\bibinfo {author} {\bibfnamefont {K.}~\bibnamefont
  {Harigaya}}, \bibinfo {author} {\bibfnamefont {T.}~\bibnamefont {Igari}},
  \bibinfo {author} {\bibfnamefont {M.~M.}\ \bibnamefont {Nojiri}}, \bibinfo
  {author} {\bibfnamefont {M.}~\bibnamefont {Takeuchi}}, \ and\ \bibinfo
  {author} {\bibfnamefont {K.}~\bibnamefont {Tobe}},\ }\href {\doibase
  10.1007/JHEP03(2014)105} {\bibfield  {journal} {\bibinfo  {journal} {JHEP}\
  }\textbf {\bibinfo {volume} {03}},\ \bibinfo {pages} {105} (\bibinfo {year}
  {2014})},\ \Eprint {http://arxiv.org/abs/1311.0870} {arXiv:1311.0870
  [hep-ph]} \BibitemShut {NoStop}%
\bibitem [{\citenamefont {Altmannshofer}\ \emph
  {et~al.}(2014{\natexlab{a}})\citenamefont {Altmannshofer}, \citenamefont
  {Gori}, \citenamefont {Pospelov},\ and\ \citenamefont
  {Yavin}}]{Altmannshofer:2014pba}%
  \BibitemOpen
  \bibfield  {author} {\bibinfo {author} {\bibfnamefont {W.}~\bibnamefont
  {Altmannshofer}}, \bibinfo {author} {\bibfnamefont {S.}~\bibnamefont {Gori}},
  \bibinfo {author} {\bibfnamefont {M.}~\bibnamefont {Pospelov}}, \ and\
  \bibinfo {author} {\bibfnamefont {I.}~\bibnamefont {Yavin}},\ }\href
  {\doibase 10.1103/PhysRevLett.113.091801} {\bibfield  {journal} {\bibinfo
  {journal} {Phys. Rev. Lett.}\ }\textbf {\bibinfo {volume} {113}},\ \bibinfo
  {pages} {091801} (\bibinfo {year} {2014}{\natexlab{a}})},\ \Eprint
  {http://arxiv.org/abs/1406.2332} {arXiv:1406.2332 [hep-ph]} \BibitemShut
  {NoStop}%
\bibitem [{\citenamefont {Altmannshofer}\ \emph {et~al.}(2019)\citenamefont
  {Altmannshofer}, \citenamefont {Gori}, \citenamefont {Mart\'\i{}n-Albo},
  \citenamefont {Sousa},\ and\ \citenamefont
  {Wallbank}}]{Altmannshofer:2019zhy}%
  \BibitemOpen
  \bibfield  {author} {\bibinfo {author} {\bibfnamefont {W.}~\bibnamefont
  {Altmannshofer}}, \bibinfo {author} {\bibfnamefont {S.}~\bibnamefont {Gori}},
  \bibinfo {author} {\bibfnamefont {J.}~\bibnamefont {Mart\'\i{}n-Albo}},
  \bibinfo {author} {\bibfnamefont {A.}~\bibnamefont {Sousa}}, \ and\ \bibinfo
  {author} {\bibfnamefont {M.}~\bibnamefont {Wallbank}},\ }\href {\doibase
  10.1103/PhysRevD.100.115029} {\bibfield  {journal} {\bibinfo  {journal}
  {Phys. Rev. D}\ }\textbf {\bibinfo {volume} {100}},\ \bibinfo {pages}
  {115029} (\bibinfo {year} {2019})},\ \Eprint
  {http://arxiv.org/abs/1902.06765} {arXiv:1902.06765 [hep-ph]} \BibitemShut
  {NoStop}%
\bibitem [{\citenamefont {Crivellin}\ \emph
  {et~al.}(2017{\natexlab{a}})\citenamefont {Crivellin}, \citenamefont
  {Fuentes-Martin}, \citenamefont {Greljo},\ and\ \citenamefont
  {Isidori}}]{Crivellin:2016ejn}%
  \BibitemOpen
  \bibfield  {author} {\bibinfo {author} {\bibfnamefont {A.}~\bibnamefont
  {Crivellin}}, \bibinfo {author} {\bibfnamefont {J.}~\bibnamefont
  {Fuentes-Martin}}, \bibinfo {author} {\bibfnamefont {A.}~\bibnamefont
  {Greljo}}, \ and\ \bibinfo {author} {\bibfnamefont {G.}~\bibnamefont
  {Isidori}},\ }\href {\doibase 10.1016/j.physletb.2016.12.057} {\bibfield
  {journal} {\bibinfo  {journal} {Phys. Lett. B}\ }\textbf {\bibinfo {volume}
  {766}},\ \bibinfo {pages} {77} (\bibinfo {year} {2017}{\natexlab{a}})},\
  \Eprint {http://arxiv.org/abs/1611.02703} {arXiv:1611.02703 [hep-ph]}
  \BibitemShut {NoStop}%
\bibitem [{\citenamefont {Crivellin}\ \emph
  {et~al.}(2015{\natexlab{a}})\citenamefont {Crivellin}, \citenamefont
  {D'Ambrosio},\ and\ \citenamefont {Heeck}}]{Crivellin:2015mga}%
  \BibitemOpen
  \bibfield  {author} {\bibinfo {author} {\bibfnamefont {A.}~\bibnamefont
  {Crivellin}}, \bibinfo {author} {\bibfnamefont {G.}~\bibnamefont
  {D'Ambrosio}}, \ and\ \bibinfo {author} {\bibfnamefont {J.}~\bibnamefont
  {Heeck}},\ }\href {\doibase 10.1103/PhysRevLett.114.151801} {\bibfield
  {journal} {\bibinfo  {journal} {Phys. Rev. Lett.}\ }\textbf {\bibinfo
  {volume} {114}},\ \bibinfo {pages} {151801} (\bibinfo {year}
  {2015}{\natexlab{a}})},\ \Eprint {http://arxiv.org/abs/1501.00993}
  {arXiv:1501.00993 [hep-ph]} \BibitemShut {NoStop}%
\bibitem [{\citenamefont {Crivellin}\ \emph {et~al.}(2018)\citenamefont
  {Crivellin}, \citenamefont {Hoferichter},\ and\ \citenamefont
  {Schmidt-Wellenburg}}]{Crivellin:2018qmi}%
  \BibitemOpen
  \bibfield  {author} {\bibinfo {author} {\bibfnamefont {A.}~\bibnamefont
  {Crivellin}}, \bibinfo {author} {\bibfnamefont {M.}~\bibnamefont
  {Hoferichter}}, \ and\ \bibinfo {author} {\bibfnamefont {P.}~\bibnamefont
  {Schmidt-Wellenburg}},\ }\href {\doibase 10.1103/PhysRevD.98.113002}
  {\bibfield  {journal} {\bibinfo  {journal} {Phys. Rev. D}\ }\textbf {\bibinfo
  {volume} {98}},\ \bibinfo {pages} {113002} (\bibinfo {year} {2018})},\
  \Eprint {http://arxiv.org/abs/1807.11484} {arXiv:1807.11484 [hep-ph]}
  \BibitemShut {NoStop}%
\bibitem [{\citenamefont {Altmannshofer}\ \emph
  {et~al.}(2014{\natexlab{b}})\citenamefont {Altmannshofer}, \citenamefont
  {Gori}, \citenamefont {Pospelov},\ and\ \citenamefont
  {Yavin}}]{Altmannshofer:2014cfa}%
  \BibitemOpen
  \bibfield  {author} {\bibinfo {author} {\bibfnamefont {W.}~\bibnamefont
  {Altmannshofer}}, \bibinfo {author} {\bibfnamefont {S.}~\bibnamefont {Gori}},
  \bibinfo {author} {\bibfnamefont {M.}~\bibnamefont {Pospelov}}, \ and\
  \bibinfo {author} {\bibfnamefont {I.}~\bibnamefont {Yavin}},\ }\href
  {\doibase 10.1103/PhysRevD.89.095033} {\bibfield  {journal} {\bibinfo
  {journal} {Phys. Rev. D}\ }\textbf {\bibinfo {volume} {89}},\ \bibinfo
  {pages} {095033} (\bibinfo {year} {2014}{\natexlab{b}})},\ \Eprint
  {http://arxiv.org/abs/1403.1269} {arXiv:1403.1269 [hep-ph]} \BibitemShut
  {NoStop}%
\bibitem [{\citenamefont {Altmannshofer}\ and\ \citenamefont
  {Yavin}(2015)}]{Altmannshofer:2015mqa}%
  \BibitemOpen
  \bibfield  {author} {\bibinfo {author} {\bibfnamefont {W.}~\bibnamefont
  {Altmannshofer}}\ and\ \bibinfo {author} {\bibfnamefont {I.}~\bibnamefont
  {Yavin}},\ }\href {\doibase 10.1103/PhysRevD.92.075022} {\bibfield  {journal}
  {\bibinfo  {journal} {Phys. Rev. D}\ }\textbf {\bibinfo {volume} {92}},\
  \bibinfo {pages} {075022} (\bibinfo {year} {2015})},\ \Eprint
  {http://arxiv.org/abs/1508.07009} {arXiv:1508.07009 [hep-ph]} \BibitemShut
  {NoStop}%
\bibitem [{\citenamefont {Bonilla}\ \emph {et~al.}(2018)\citenamefont
  {Bonilla}, \citenamefont {Modak}, \citenamefont {Srivastava},\ and\
  \citenamefont {Valle}}]{Bonilla:2017lsq}%
  \BibitemOpen
  \bibfield  {author} {\bibinfo {author} {\bibfnamefont {C.}~\bibnamefont
  {Bonilla}}, \bibinfo {author} {\bibfnamefont {T.}~\bibnamefont {Modak}},
  \bibinfo {author} {\bibfnamefont {R.}~\bibnamefont {Srivastava}}, \ and\
  \bibinfo {author} {\bibfnamefont {J.~W.~F.}\ \bibnamefont {Valle}},\ }\href
  {\doibase 10.1103/PhysRevD.98.095002} {\bibfield  {journal} {\bibinfo
  {journal} {Phys. Rev. D}\ }\textbf {\bibinfo {volume} {98}},\ \bibinfo
  {pages} {095002} (\bibinfo {year} {2018})},\ \Eprint
  {http://arxiv.org/abs/1705.00915} {arXiv:1705.00915 [hep-ph]} \BibitemShut
  {NoStop}%
\bibitem [{\citenamefont {Allanach}(2021)}]{Allanach:2020kss}%
  \BibitemOpen
  \bibfield  {author} {\bibinfo {author} {\bibfnamefont {B.~C.}\ \bibnamefont
  {Allanach}},\ }\href {\doibase 10.1140/epjc/s10052-021-08855-w} {\bibfield
  {journal} {\bibinfo  {journal} {Eur. Phys. J. C}\ }\textbf {\bibinfo {volume}
  {81}},\ \bibinfo {pages} {56} (\bibinfo {year} {2021})},\ \Eprint
  {http://arxiv.org/abs/2009.02197} {arXiv:2009.02197 [hep-ph]} \BibitemShut
  {NoStop}%
\bibitem [{\citenamefont {Alonso}\ \emph {et~al.}(2017)\citenamefont {Alonso},
  \citenamefont {Cox}, \citenamefont {Han},\ and\ \citenamefont
  {Yanagida}}]{Alonso:2017uky}%
  \BibitemOpen
  \bibfield  {author} {\bibinfo {author} {\bibfnamefont {R.}~\bibnamefont
  {Alonso}}, \bibinfo {author} {\bibfnamefont {P.}~\bibnamefont {Cox}},
  \bibinfo {author} {\bibfnamefont {C.}~\bibnamefont {Han}}, \ and\ \bibinfo
  {author} {\bibfnamefont {T.~T.}\ \bibnamefont {Yanagida}},\ }\href {\doibase
  10.1016/j.physletb.2017.10.027} {\bibfield  {journal} {\bibinfo  {journal}
  {Phys. Lett. B}\ }\textbf {\bibinfo {volume} {774}},\ \bibinfo {pages} {643}
  (\bibinfo {year} {2017})},\ \Eprint {http://arxiv.org/abs/1705.03858}
  {arXiv:1705.03858 [hep-ph]} \BibitemShut {NoStop}%
\bibitem [{\citenamefont {Allanach}\ and\ \citenamefont
  {Davighi}(2018)}]{Allanach:2018lvl}%
  \BibitemOpen
  \bibfield  {author} {\bibinfo {author} {\bibfnamefont {B.~C.}\ \bibnamefont
  {Allanach}}\ and\ \bibinfo {author} {\bibfnamefont {J.}~\bibnamefont
  {Davighi}},\ }\href {\doibase 10.1007/JHEP12(2018)075} {\bibfield  {journal}
  {\bibinfo  {journal} {JHEP}\ }\textbf {\bibinfo {volume} {12}},\ \bibinfo
  {pages} {075} (\bibinfo {year} {2018})},\ \Eprint
  {http://arxiv.org/abs/1809.01158} {arXiv:1809.01158 [hep-ph]} \BibitemShut
  {NoStop}%
\bibitem [{\citenamefont {Altmannshofer}\ \emph {et~al.}(2020)\citenamefont
  {Altmannshofer}, \citenamefont {Davighi},\ and\ \citenamefont
  {Nardecchia}}]{Altmannshofer:2019xda}%
  \BibitemOpen
  \bibfield  {author} {\bibinfo {author} {\bibfnamefont {W.}~\bibnamefont
  {Altmannshofer}}, \bibinfo {author} {\bibfnamefont {J.}~\bibnamefont
  {Davighi}}, \ and\ \bibinfo {author} {\bibfnamefont {M.}~\bibnamefont
  {Nardecchia}},\ }\href {\doibase 10.1103/PhysRevD.101.015004} {\bibfield
  {journal} {\bibinfo  {journal} {Phys. Rev. D}\ }\textbf {\bibinfo {volume}
  {101}},\ \bibinfo {pages} {015004} (\bibinfo {year} {2020})},\ \Eprint
  {http://arxiv.org/abs/1909.02021} {arXiv:1909.02021 [hep-ph]} \BibitemShut
  {NoStop}%
\bibitem [{\citenamefont {Dor\v{s}ner}\ \emph {et~al.}(2016)\citenamefont
  {Dor\v{s}ner}, \citenamefont {Fajfer}, \citenamefont {Greljo}, \citenamefont
  {Kamenik},\ and\ \citenamefont {Ko\v{s}nik}}]{Dorsner:2016wpm}%
  \BibitemOpen
  \bibfield  {author} {\bibinfo {author} {\bibfnamefont {I.}~\bibnamefont
  {Dor\v{s}ner}}, \bibinfo {author} {\bibfnamefont {S.}~\bibnamefont {Fajfer}},
  \bibinfo {author} {\bibfnamefont {A.}~\bibnamefont {Greljo}}, \bibinfo
  {author} {\bibfnamefont {J.~F.}\ \bibnamefont {Kamenik}}, \ and\ \bibinfo
  {author} {\bibfnamefont {N.}~\bibnamefont {Ko\v{s}nik}},\ }\href {\doibase
  10.1016/j.physrep.2016.06.001} {\bibfield  {journal} {\bibinfo  {journal}
  {Phys. Rept.}\ }\textbf {\bibinfo {volume} {641}},\ \bibinfo {pages} {1}
  (\bibinfo {year} {2016})},\ \Eprint {http://arxiv.org/abs/1603.04993}
  {arXiv:1603.04993 [hep-ph]} \BibitemShut {NoStop}%
\bibitem [{\citenamefont {Gripaios}(2010)}]{Gripaios:2009dq}%
  \BibitemOpen
  \bibfield  {author} {\bibinfo {author} {\bibfnamefont {B.}~\bibnamefont
  {Gripaios}},\ }\href {\doibase 10.1007/JHEP02(2010)045} {\bibfield  {journal}
  {\bibinfo  {journal} {JHEP}\ }\textbf {\bibinfo {volume} {02}},\ \bibinfo
  {pages} {045} (\bibinfo {year} {2010})},\ \Eprint
  {http://arxiv.org/abs/0910.1789} {arXiv:0910.1789 [hep-ph]} \BibitemShut
  {NoStop}%
\bibitem [{\citenamefont {Hiller}\ and\ \citenamefont
  {Schmaltz}(2014)}]{Hiller:2014yaa}%
  \BibitemOpen
  \bibfield  {author} {\bibinfo {author} {\bibfnamefont {G.}~\bibnamefont
  {Hiller}}\ and\ \bibinfo {author} {\bibfnamefont {M.}~\bibnamefont
  {Schmaltz}},\ }\href {\doibase 10.1103/PhysRevD.90.054014} {\bibfield
  {journal} {\bibinfo  {journal} {Phys. Rev. D}\ }\textbf {\bibinfo {volume}
  {90}},\ \bibinfo {pages} {054014} (\bibinfo {year} {2014})},\ \Eprint
  {http://arxiv.org/abs/1408.1627} {arXiv:1408.1627 [hep-ph]} \BibitemShut
  {NoStop}%
\bibitem [{\citenamefont {Bauer}\ and\ \citenamefont
  {Neubert}(2016)}]{Bauer:2015knc}%
  \BibitemOpen
  \bibfield  {author} {\bibinfo {author} {\bibfnamefont {M.}~\bibnamefont
  {Bauer}}\ and\ \bibinfo {author} {\bibfnamefont {M.}~\bibnamefont
  {Neubert}},\ }\href {\doibase 10.1103/PhysRevLett.116.141802} {\bibfield
  {journal} {\bibinfo  {journal} {Phys. Rev. Lett.}\ }\textbf {\bibinfo
  {volume} {116}},\ \bibinfo {pages} {141802} (\bibinfo {year} {2016})},\
  \Eprint {http://arxiv.org/abs/1511.01900} {arXiv:1511.01900 [hep-ph]}
  \BibitemShut {NoStop}%
\bibitem [{\citenamefont {Barbieri}\ \emph {et~al.}(2016)\citenamefont
  {Barbieri}, \citenamefont {Isidori}, \citenamefont {Pattori},\ and\
  \citenamefont {Senia}}]{Barbieri:2015yvd}%
  \BibitemOpen
  \bibfield  {author} {\bibinfo {author} {\bibfnamefont {R.}~\bibnamefont
  {Barbieri}}, \bibinfo {author} {\bibfnamefont {G.}~\bibnamefont {Isidori}},
  \bibinfo {author} {\bibfnamefont {A.}~\bibnamefont {Pattori}}, \ and\
  \bibinfo {author} {\bibfnamefont {F.}~\bibnamefont {Senia}},\ }\href
  {\doibase 10.1140/epjc/s10052-016-3905-3} {\bibfield  {journal} {\bibinfo
  {journal} {Eur. Phys. J. C}\ }\textbf {\bibinfo {volume} {76}},\ \bibinfo
  {pages} {67} (\bibinfo {year} {2016})},\ \Eprint
  {http://arxiv.org/abs/1512.01560} {arXiv:1512.01560 [hep-ph]} \BibitemShut
  {NoStop}%
\bibitem [{\citenamefont {Buttazzo}\ \emph {et~al.}(2017)\citenamefont
  {Buttazzo}, \citenamefont {Greljo}, \citenamefont {Isidori},\ and\
  \citenamefont {Marzocca}}]{Buttazzo:2017ixm}%
  \BibitemOpen
  \bibfield  {author} {\bibinfo {author} {\bibfnamefont {D.}~\bibnamefont
  {Buttazzo}}, \bibinfo {author} {\bibfnamefont {A.}~\bibnamefont {Greljo}},
  \bibinfo {author} {\bibfnamefont {G.}~\bibnamefont {Isidori}}, \ and\
  \bibinfo {author} {\bibfnamefont {D.}~\bibnamefont {Marzocca}},\ }\href
  {\doibase 10.1007/JHEP11(2017)044} {\bibfield  {journal} {\bibinfo  {journal}
  {JHEP}\ }\textbf {\bibinfo {volume} {11}},\ \bibinfo {pages} {044} (\bibinfo
  {year} {2017})},\ \Eprint {http://arxiv.org/abs/1706.07808} {arXiv:1706.07808
  [hep-ph]} \BibitemShut {NoStop}%
\bibitem [{\citenamefont {Angelescu}\ \emph {et~al.}(2018)\citenamefont
  {Angelescu}, \citenamefont {Be\v{c}irevi\'c}, \citenamefont {Faroughy},\ and\
  \citenamefont {Sumensari}}]{Angelescu:2018tyl}%
  \BibitemOpen
  \bibfield  {author} {\bibinfo {author} {\bibfnamefont {A.}~\bibnamefont
  {Angelescu}}, \bibinfo {author} {\bibfnamefont {D.}~\bibnamefont
  {Be\v{c}irevi\'c}}, \bibinfo {author} {\bibfnamefont {D.~A.}\ \bibnamefont
  {Faroughy}}, \ and\ \bibinfo {author} {\bibfnamefont {O.}~\bibnamefont
  {Sumensari}},\ }\href {\doibase 10.1007/JHEP10(2018)183} {\bibfield
  {journal} {\bibinfo  {journal} {JHEP}\ }\textbf {\bibinfo {volume} {10}},\
  \bibinfo {pages} {183} (\bibinfo {year} {2018})},\ \Eprint
  {http://arxiv.org/abs/1808.08179} {arXiv:1808.08179 [hep-ph]} \BibitemShut
  {NoStop}%
\bibitem [{\citenamefont {Dor\v{s}ner}\ \emph {et~al.}(2020)\citenamefont
  {Dor\v{s}ner}, \citenamefont {Fajfer},\ and\ \citenamefont
  {Sumensari}}]{Dorsner:2019itg}%
  \BibitemOpen
  \bibfield  {author} {\bibinfo {author} {\bibfnamefont {I.}~\bibnamefont
  {Dor\v{s}ner}}, \bibinfo {author} {\bibfnamefont {S.}~\bibnamefont {Fajfer}},
  \ and\ \bibinfo {author} {\bibfnamefont {O.}~\bibnamefont {Sumensari}},\
  }\href {\doibase 10.1007/JHEP06(2020)089} {\bibfield  {journal} {\bibinfo
  {journal} {JHEP}\ }\textbf {\bibinfo {volume} {06}},\ \bibinfo {pages} {089}
  (\bibinfo {year} {2020})},\ \Eprint {http://arxiv.org/abs/1910.03877}
  {arXiv:1910.03877 [hep-ph]} \BibitemShut {NoStop}%
\bibitem [{\citenamefont {Angelescu}\ \emph {et~al.}(2021)\citenamefont
  {Angelescu}, \citenamefont {Be\v{c}irevi\'c}, \citenamefont {Faroughy},
  \citenamefont {Jaffredo},\ and\ \citenamefont
  {Sumensari}}]{Angelescu:2021lln}%
  \BibitemOpen
  \bibfield  {author} {\bibinfo {author} {\bibfnamefont {A.}~\bibnamefont
  {Angelescu}}, \bibinfo {author} {\bibfnamefont {D.}~\bibnamefont
  {Be\v{c}irevi\'c}}, \bibinfo {author} {\bibfnamefont {D.~A.}\ \bibnamefont
  {Faroughy}}, \bibinfo {author} {\bibfnamefont {F.}~\bibnamefont {Jaffredo}},
  \ and\ \bibinfo {author} {\bibfnamefont {O.}~\bibnamefont {Sumensari}},\
  }\href@noop {} {\  (\bibinfo {year} {2021})},\ \Eprint
  {http://arxiv.org/abs/2103.12504} {arXiv:2103.12504 [hep-ph]} \BibitemShut
  {NoStop}%
\bibitem [{\citenamefont {Hiller}\ \emph {et~al.}(2021)\citenamefont {Hiller},
  \citenamefont {Loose},\ and\ \citenamefont
  {Ni\v{s}and\v{z}i\'c}}]{Hiller:2021pul}%
  \BibitemOpen
  \bibfield  {author} {\bibinfo {author} {\bibfnamefont {G.}~\bibnamefont
  {Hiller}}, \bibinfo {author} {\bibfnamefont {D.}~\bibnamefont {Loose}}, \
  and\ \bibinfo {author} {\bibfnamefont {I.}~\bibnamefont
  {Ni\v{s}and\v{z}i\'c}},\ }\href@noop {} {\  (\bibinfo {year} {2021})},\
  \Eprint {http://arxiv.org/abs/2103.12724} {arXiv:2103.12724 [hep-ph]}
  \BibitemShut {NoStop}%
\bibitem [{\citenamefont {Marzocca}(2018)}]{Marzocca:2018wcf}%
  \BibitemOpen
  \bibfield  {author} {\bibinfo {author} {\bibfnamefont {D.}~\bibnamefont
  {Marzocca}},\ }\href {\doibase 10.1007/JHEP07(2018)121} {\bibfield  {journal}
  {\bibinfo  {journal} {JHEP}\ }\textbf {\bibinfo {volume} {07}},\ \bibinfo
  {pages} {121} (\bibinfo {year} {2018})},\ \Eprint
  {http://arxiv.org/abs/1803.10972} {arXiv:1803.10972 [hep-ph]} \BibitemShut
  {NoStop}%
\bibitem [{\citenamefont {Dor\v{s}ner}\ \emph {et~al.}(2017)\citenamefont
  {Dor\v{s}ner}, \citenamefont {Fajfer}, \citenamefont {Faroughy},\ and\
  \citenamefont {Ko\v{s}nik}}]{Dorsner:2017ufx}%
  \BibitemOpen
  \bibfield  {author} {\bibinfo {author} {\bibfnamefont {I.}~\bibnamefont
  {Dor\v{s}ner}}, \bibinfo {author} {\bibfnamefont {S.}~\bibnamefont {Fajfer}},
  \bibinfo {author} {\bibfnamefont {D.~A.}\ \bibnamefont {Faroughy}}, \ and\
  \bibinfo {author} {\bibfnamefont {N.}~\bibnamefont {Ko\v{s}nik}},\ }\href
  {\doibase 10.1007/JHEP10(2017)188} {\bibfield  {journal} {\bibinfo  {journal}
  {JHEP}\ }\textbf {\bibinfo {volume} {10}},\ \bibinfo {pages} {188} (\bibinfo
  {year} {2017})},\ \Eprint {http://arxiv.org/abs/1706.07779} {arXiv:1706.07779
  [hep-ph]} \BibitemShut {NoStop}%
\bibitem [{\citenamefont {Be\v{c}irevi\'c}\ \emph {et~al.}(2018)\citenamefont
  {Be\v{c}irevi\'c}, \citenamefont {Dor\v{s}ner}, \citenamefont {Fajfer},
  \citenamefont {Ko\v{s}nik}, \citenamefont {Faroughy},\ and\ \citenamefont
  {Sumensari}}]{Becirevic:2018afm}%
  \BibitemOpen
  \bibfield  {author} {\bibinfo {author} {\bibfnamefont {D.}~\bibnamefont
  {Be\v{c}irevi\'c}}, \bibinfo {author} {\bibfnamefont {I.}~\bibnamefont
  {Dor\v{s}ner}}, \bibinfo {author} {\bibfnamefont {S.}~\bibnamefont {Fajfer}},
  \bibinfo {author} {\bibfnamefont {N.}~\bibnamefont {Ko\v{s}nik}}, \bibinfo
  {author} {\bibfnamefont {D.~A.}\ \bibnamefont {Faroughy}}, \ and\ \bibinfo
  {author} {\bibfnamefont {O.}~\bibnamefont {Sumensari}},\ }\href {\doibase
  10.1103/PhysRevD.98.055003} {\bibfield  {journal} {\bibinfo  {journal} {Phys.
  Rev. D}\ }\textbf {\bibinfo {volume} {98}},\ \bibinfo {pages} {055003}
  (\bibinfo {year} {2018})},\ \Eprint {http://arxiv.org/abs/1806.05689}
  {arXiv:1806.05689 [hep-ph]} \BibitemShut {NoStop}%
\bibitem [{\citenamefont {Crivellin}\ \emph
  {et~al.}(2017{\natexlab{b}})\citenamefont {Crivellin}, \citenamefont
  {M\"uller},\ and\ \citenamefont {Ota}}]{Crivellin:2017zlb}%
  \BibitemOpen
  \bibfield  {author} {\bibinfo {author} {\bibfnamefont {A.}~\bibnamefont
  {Crivellin}}, \bibinfo {author} {\bibfnamefont {D.}~\bibnamefont {M\"uller}},
  \ and\ \bibinfo {author} {\bibfnamefont {T.}~\bibnamefont {Ota}},\ }\href
  {\doibase 10.1007/JHEP09(2017)040} {\bibfield  {journal} {\bibinfo  {journal}
  {JHEP}\ }\textbf {\bibinfo {volume} {09}},\ \bibinfo {pages} {040} (\bibinfo
  {year} {2017}{\natexlab{b}})},\ \Eprint {http://arxiv.org/abs/1703.09226}
  {arXiv:1703.09226 [hep-ph]} \BibitemShut {NoStop}%
\bibitem [{\citenamefont {Gherardi}\ \emph {et~al.}(2021)\citenamefont
  {Gherardi}, \citenamefont {Marzocca},\ and\ \citenamefont
  {Venturini}}]{Gherardi:2020qhc}%
  \BibitemOpen
  \bibfield  {author} {\bibinfo {author} {\bibfnamefont {V.}~\bibnamefont
  {Gherardi}}, \bibinfo {author} {\bibfnamefont {D.}~\bibnamefont {Marzocca}},
  \ and\ \bibinfo {author} {\bibfnamefont {E.}~\bibnamefont {Venturini}},\
  }\href {\doibase 10.1007/JHEP01(2021)138} {\bibfield  {journal} {\bibinfo
  {journal} {JHEP}\ }\textbf {\bibinfo {volume} {01}},\ \bibinfo {pages} {138}
  (\bibinfo {year} {2021})},\ \Eprint {http://arxiv.org/abs/2008.09548}
  {arXiv:2008.09548 [hep-ph]} \BibitemShut {NoStop}%
\bibitem [{\citenamefont {Hambye}\ and\ \citenamefont
  {Heeck}(2018)}]{Hambye:2017qix}%
  \BibitemOpen
  \bibfield  {author} {\bibinfo {author} {\bibfnamefont {T.}~\bibnamefont
  {Hambye}}\ and\ \bibinfo {author} {\bibfnamefont {J.}~\bibnamefont {Heeck}},\
  }\href {\doibase 10.1103/PhysRevLett.120.171801} {\bibfield  {journal}
  {\bibinfo  {journal} {Phys. Rev. Lett.}\ }\textbf {\bibinfo {volume} {120}},\
  \bibinfo {pages} {171801} (\bibinfo {year} {2018})},\ \Eprint
  {http://arxiv.org/abs/1712.04871} {arXiv:1712.04871 [hep-ph]} \BibitemShut
  {NoStop}%
\bibitem [{\citenamefont {Davighi}\ \emph {et~al.}(2020)\citenamefont
  {Davighi}, \citenamefont {Kirk},\ and\ \citenamefont
  {Nardecchia}}]{Davighi:2020qqa}%
  \BibitemOpen
  \bibfield  {author} {\bibinfo {author} {\bibfnamefont {J.}~\bibnamefont
  {Davighi}}, \bibinfo {author} {\bibfnamefont {M.}~\bibnamefont {Kirk}}, \
  and\ \bibinfo {author} {\bibfnamefont {M.}~\bibnamefont {Nardecchia}},\
  }\href {\doibase 10.1007/JHEP12(2020)111} {\bibfield  {journal} {\bibinfo
  {journal} {JHEP}\ }\textbf {\bibinfo {volume} {12}},\ \bibinfo {pages} {111}
  (\bibinfo {year} {2020})},\ \Eprint {http://arxiv.org/abs/2007.15016}
  {arXiv:2007.15016 [hep-ph]} \BibitemShut {NoStop}%
\bibitem [{\citenamefont {Lavoura}(2005)}]{Lavoura:2004tu}%
  \BibitemOpen
  \bibfield  {author} {\bibinfo {author} {\bibfnamefont {L.}~\bibnamefont
  {Lavoura}},\ }\href {\doibase 10.1016/j.physletb.2005.01.047} {\bibfield
  {journal} {\bibinfo  {journal} {Phys. Lett. B}\ }\textbf {\bibinfo {volume}
  {609}},\ \bibinfo {pages} {317} (\bibinfo {year} {2005})},\ \Eprint
  {http://arxiv.org/abs/hep-ph/0411232} {arXiv:hep-ph/0411232} \BibitemShut
  {NoStop}%
\bibitem [{\citenamefont {Ma}(2005)}]{Ma:2005py}%
  \BibitemOpen
  \bibfield  {author} {\bibinfo {author} {\bibfnamefont {E.}~\bibnamefont
  {Ma}},\ }\href {\doibase 10.1103/PhysRevD.71.111301} {\bibfield  {journal}
  {\bibinfo  {journal} {Phys. Rev. D}\ }\textbf {\bibinfo {volume} {71}},\
  \bibinfo {pages} {111301} (\bibinfo {year} {2005})},\ \Eprint
  {http://arxiv.org/abs/hep-ph/0501056} {arXiv:hep-ph/0501056} \BibitemShut
  {NoStop}%
\bibitem [{\citenamefont {Asai}\ \emph {et~al.}(2017)\citenamefont {Asai},
  \citenamefont {Hamaguchi},\ and\ \citenamefont {Nagata}}]{Asai:2017ryy}%
  \BibitemOpen
  \bibfield  {author} {\bibinfo {author} {\bibfnamefont {K.}~\bibnamefont
  {Asai}}, \bibinfo {author} {\bibfnamefont {K.}~\bibnamefont {Hamaguchi}}, \
  and\ \bibinfo {author} {\bibfnamefont {N.}~\bibnamefont {Nagata}},\ }\href
  {\doibase 10.1140/epjc/s10052-017-5348-x} {\bibfield  {journal} {\bibinfo
  {journal} {Eur. Phys. J. C}\ }\textbf {\bibinfo {volume} {77}},\ \bibinfo
  {pages} {763} (\bibinfo {year} {2017})},\ \Eprint
  {http://arxiv.org/abs/1705.00419} {arXiv:1705.00419 [hep-ph]} \BibitemShut
  {NoStop}%
\bibitem [{\citenamefont {Asai}\ \emph {et~al.}(2019)\citenamefont {Asai},
  \citenamefont {Hamaguchi}, \citenamefont {Nagata}, \citenamefont {Tseng},\
  and\ \citenamefont {Tsumura}}]{Asai:2018ocx}%
  \BibitemOpen
  \bibfield  {author} {\bibinfo {author} {\bibfnamefont {K.}~\bibnamefont
  {Asai}}, \bibinfo {author} {\bibfnamefont {K.}~\bibnamefont {Hamaguchi}},
  \bibinfo {author} {\bibfnamefont {N.}~\bibnamefont {Nagata}}, \bibinfo
  {author} {\bibfnamefont {S.-Y.}\ \bibnamefont {Tseng}}, \ and\ \bibinfo
  {author} {\bibfnamefont {K.}~\bibnamefont {Tsumura}},\ }\href {\doibase
  10.1103/PhysRevD.99.055029} {\bibfield  {journal} {\bibinfo  {journal} {Phys.
  Rev. D}\ }\textbf {\bibinfo {volume} {99}},\ \bibinfo {pages} {055029}
  (\bibinfo {year} {2019})},\ \Eprint {http://arxiv.org/abs/1811.07571}
  {arXiv:1811.07571 [hep-ph]} \BibitemShut {NoStop}%
\bibitem [{\citenamefont {Asai}(2020)}]{Asai:2019ciz}%
  \BibitemOpen
  \bibfield  {author} {\bibinfo {author} {\bibfnamefont {K.}~\bibnamefont
  {Asai}},\ }\href {\doibase 10.1140/epjc/s10052-020-7622-6} {\bibfield
  {journal} {\bibinfo  {journal} {Eur. Phys. J. C}\ }\textbf {\bibinfo {volume}
  {80}},\ \bibinfo {pages} {76} (\bibinfo {year} {2020})},\ \Eprint
  {http://arxiv.org/abs/1907.04042} {arXiv:1907.04042 [hep-ph]} \BibitemShut
  {NoStop}%
\bibitem [{\citenamefont {Barbieri}\ \emph {et~al.}(2011)\citenamefont
  {Barbieri}, \citenamefont {Isidori}, \citenamefont {Jones-Perez},
  \citenamefont {Lodone},\ and\ \citenamefont {Straub}}]{Barbieri:2011ci}%
  \BibitemOpen
  \bibfield  {author} {\bibinfo {author} {\bibfnamefont {R.}~\bibnamefont
  {Barbieri}}, \bibinfo {author} {\bibfnamefont {G.}~\bibnamefont {Isidori}},
  \bibinfo {author} {\bibfnamefont {J.}~\bibnamefont {Jones-Perez}}, \bibinfo
  {author} {\bibfnamefont {P.}~\bibnamefont {Lodone}}, \ and\ \bibinfo {author}
  {\bibfnamefont {D.~M.}\ \bibnamefont {Straub}},\ }\href {\doibase
  10.1140/epjc/s10052-011-1725-z} {\bibfield  {journal} {\bibinfo  {journal}
  {Eur. Phys. J. C}\ }\textbf {\bibinfo {volume} {71}},\ \bibinfo {pages}
  {1725} (\bibinfo {year} {2011})},\ \Eprint {http://arxiv.org/abs/1105.2296}
  {arXiv:1105.2296 [hep-ph]} \BibitemShut {NoStop}%
\bibitem [{\citenamefont {Kagan}\ \emph {et~al.}(2009)\citenamefont {Kagan},
  \citenamefont {Perez}, \citenamefont {Volansky},\ and\ \citenamefont
  {Zupan}}]{Kagan:2009bn}%
  \BibitemOpen
  \bibfield  {author} {\bibinfo {author} {\bibfnamefont {A.~L.}\ \bibnamefont
  {Kagan}}, \bibinfo {author} {\bibfnamefont {G.}~\bibnamefont {Perez}},
  \bibinfo {author} {\bibfnamefont {T.}~\bibnamefont {Volansky}}, \ and\
  \bibinfo {author} {\bibfnamefont {J.}~\bibnamefont {Zupan}},\ }\href
  {\doibase 10.1103/PhysRevD.80.076002} {\bibfield  {journal} {\bibinfo
  {journal} {Phys. Rev. D}\ }\textbf {\bibinfo {volume} {80}},\ \bibinfo
  {pages} {076002} (\bibinfo {year} {2009})},\ \Eprint
  {http://arxiv.org/abs/0903.1794} {arXiv:0903.1794 [hep-ph]} \BibitemShut
  {NoStop}%
\bibitem [{\citenamefont {Fuentes-Mart\'\i{}n}\ \emph
  {et~al.}(2020{\natexlab{a}})\citenamefont {Fuentes-Mart\'\i{}n},
  \citenamefont {Isidori}, \citenamefont {Pag\`es},\ and\ \citenamefont
  {Yamamoto}}]{Fuentes-Martin:2019mun}%
  \BibitemOpen
  \bibfield  {author} {\bibinfo {author} {\bibfnamefont {J.}~\bibnamefont
  {Fuentes-Mart\'\i{}n}}, \bibinfo {author} {\bibfnamefont {G.}~\bibnamefont
  {Isidori}}, \bibinfo {author} {\bibfnamefont {J.}~\bibnamefont {Pag\`es}}, \
  and\ \bibinfo {author} {\bibfnamefont {K.}~\bibnamefont {Yamamoto}},\ }\href
  {\doibase 10.1016/j.physletb.2019.135080} {\bibfield  {journal} {\bibinfo
  {journal} {Phys. Lett. B}\ }\textbf {\bibinfo {volume} {800}},\ \bibinfo
  {pages} {135080} (\bibinfo {year} {2020}{\natexlab{a}})},\ \Eprint
  {http://arxiv.org/abs/1909.02519} {arXiv:1909.02519 [hep-ph]} \BibitemShut
  {NoStop}%
\bibitem [{\citenamefont {Gherardi}\ \emph {et~al.}(2020)\citenamefont
  {Gherardi}, \citenamefont {Marzocca},\ and\ \citenamefont
  {Venturini}}]{Gherardi:2020det}%
  \BibitemOpen
  \bibfield  {author} {\bibinfo {author} {\bibfnamefont {V.}~\bibnamefont
  {Gherardi}}, \bibinfo {author} {\bibfnamefont {D.}~\bibnamefont {Marzocca}},
  \ and\ \bibinfo {author} {\bibfnamefont {E.}~\bibnamefont {Venturini}},\
  }\href {\doibase 10.1007/JHEP07(2020)225} {\bibfield  {journal} {\bibinfo
  {journal} {JHEP}\ }\textbf {\bibinfo {volume} {07}},\ \bibinfo {pages} {225}
  (\bibinfo {year} {2020})},\ \bibinfo {note} {[Erratum: JHEP 01, 006
  (2021)]},\ \Eprint {http://arxiv.org/abs/2003.12525} {arXiv:2003.12525
  [hep-ph]} \BibitemShut {NoStop}%
\bibitem [{\citenamefont {Aebischer}\ \emph {et~al.}(2019)\citenamefont
  {Aebischer}, \citenamefont {Kumar}, \citenamefont {Stangl},\ and\
  \citenamefont {Straub}}]{Aebischer:2018iyb}%
  \BibitemOpen
  \bibfield  {author} {\bibinfo {author} {\bibfnamefont {J.}~\bibnamefont
  {Aebischer}}, \bibinfo {author} {\bibfnamefont {J.}~\bibnamefont {Kumar}},
  \bibinfo {author} {\bibfnamefont {P.}~\bibnamefont {Stangl}}, \ and\ \bibinfo
  {author} {\bibfnamefont {D.~M.}\ \bibnamefont {Straub}},\ }\href {\doibase
  10.1140/epjc/s10052-019-6977-z} {\bibfield  {journal} {\bibinfo  {journal}
  {Eur. Phys. J. C}\ }\textbf {\bibinfo {volume} {79}},\ \bibinfo {pages} {509}
  (\bibinfo {year} {2019})},\ \Eprint {http://arxiv.org/abs/1810.07698}
  {arXiv:1810.07698 [hep-ph]} \BibitemShut {NoStop}%
\bibitem [{\citenamefont {Stangl}(2020)}]{Stangl:2020lbh}%
  \BibitemOpen
  \bibfield  {author} {\bibinfo {author} {\bibfnamefont {P.}~\bibnamefont
  {Stangl}},\ }in\ \href@noop {} {\emph {\bibinfo {booktitle} {{Tools for High
  Energy Physics and Cosmology}}}}\ (\bibinfo {year} {2020})\ \Eprint
  {http://arxiv.org/abs/2012.12211} {arXiv:2012.12211 [hep-ph]} \BibitemShut
  {NoStop}%
\bibitem [{\citenamefont {Alonso}\ \emph {et~al.}(2014)\citenamefont {Alonso},
  \citenamefont {Jenkins}, \citenamefont {Manohar},\ and\ \citenamefont
  {Trott}}]{Alonso:2013hga}%
  \BibitemOpen
  \bibfield  {author} {\bibinfo {author} {\bibfnamefont {R.}~\bibnamefont
  {Alonso}}, \bibinfo {author} {\bibfnamefont {E.~E.}\ \bibnamefont {Jenkins}},
  \bibinfo {author} {\bibfnamefont {A.~V.}\ \bibnamefont {Manohar}}, \ and\
  \bibinfo {author} {\bibfnamefont {M.}~\bibnamefont {Trott}},\ }\href
  {\doibase 10.1007/JHEP04(2014)159} {\bibfield  {journal} {\bibinfo  {journal}
  {JHEP}\ }\textbf {\bibinfo {volume} {04}},\ \bibinfo {pages} {159} (\bibinfo
  {year} {2014})},\ \Eprint {http://arxiv.org/abs/1312.2014} {arXiv:1312.2014
  [hep-ph]} \BibitemShut {NoStop}%
\bibitem [{\citenamefont {Jenkins}\ \emph {et~al.}(2014)\citenamefont
  {Jenkins}, \citenamefont {Manohar},\ and\ \citenamefont
  {Trott}}]{Jenkins:2013wua}%
  \BibitemOpen
  \bibfield  {author} {\bibinfo {author} {\bibfnamefont {E.~E.}\ \bibnamefont
  {Jenkins}}, \bibinfo {author} {\bibfnamefont {A.~V.}\ \bibnamefont
  {Manohar}}, \ and\ \bibinfo {author} {\bibfnamefont {M.}~\bibnamefont
  {Trott}},\ }\href {\doibase 10.1007/JHEP01(2014)035} {\bibfield  {journal}
  {\bibinfo  {journal} {JHEP}\ }\textbf {\bibinfo {volume} {01}},\ \bibinfo
  {pages} {035} (\bibinfo {year} {2014})},\ \Eprint
  {http://arxiv.org/abs/1310.4838} {arXiv:1310.4838 [hep-ph]} \BibitemShut
  {NoStop}%
\bibitem [{\citenamefont {Jenkins}\ \emph {et~al.}(2013)\citenamefont
  {Jenkins}, \citenamefont {Manohar},\ and\ \citenamefont
  {Trott}}]{Jenkins:2013zja}%
  \BibitemOpen
  \bibfield  {author} {\bibinfo {author} {\bibfnamefont {E.~E.}\ \bibnamefont
  {Jenkins}}, \bibinfo {author} {\bibfnamefont {A.~V.}\ \bibnamefont
  {Manohar}}, \ and\ \bibinfo {author} {\bibfnamefont {M.}~\bibnamefont
  {Trott}},\ }\href {\doibase 10.1007/JHEP10(2013)087} {\bibfield  {journal}
  {\bibinfo  {journal} {JHEP}\ }\textbf {\bibinfo {volume} {10}},\ \bibinfo
  {pages} {087} (\bibinfo {year} {2013})},\ \Eprint
  {http://arxiv.org/abs/1308.2627} {arXiv:1308.2627 [hep-ph]} \BibitemShut
  {NoStop}%
\bibitem [{\citenamefont {Dekens}\ and\ \citenamefont
  {Stoffer}(2019)}]{Dekens:2019ept}%
  \BibitemOpen
  \bibfield  {author} {\bibinfo {author} {\bibfnamefont {W.}~\bibnamefont
  {Dekens}}\ and\ \bibinfo {author} {\bibfnamefont {P.}~\bibnamefont
  {Stoffer}},\ }\href {\doibase 10.1007/JHEP10(2019)197} {\bibfield  {journal}
  {\bibinfo  {journal} {JHEP}\ }\textbf {\bibinfo {volume} {10}},\ \bibinfo
  {pages} {197} (\bibinfo {year} {2019})},\ \Eprint
  {http://arxiv.org/abs/1908.05295} {arXiv:1908.05295 [hep-ph]} \BibitemShut
  {NoStop}%
\bibitem [{\citenamefont {Jenkins}\ \emph {et~al.}(2018)\citenamefont
  {Jenkins}, \citenamefont {Manohar},\ and\ \citenamefont
  {Stoffer}}]{Jenkins:2017dyc}%
  \BibitemOpen
  \bibfield  {author} {\bibinfo {author} {\bibfnamefont {E.~E.}\ \bibnamefont
  {Jenkins}}, \bibinfo {author} {\bibfnamefont {A.~V.}\ \bibnamefont
  {Manohar}}, \ and\ \bibinfo {author} {\bibfnamefont {P.}~\bibnamefont
  {Stoffer}},\ }\href {\doibase 10.1007/JHEP01(2018)084} {\bibfield  {journal}
  {\bibinfo  {journal} {JHEP}\ }\textbf {\bibinfo {volume} {01}},\ \bibinfo
  {pages} {084} (\bibinfo {year} {2018})},\ \Eprint
  {http://arxiv.org/abs/1711.05270} {arXiv:1711.05270 [hep-ph]} \BibitemShut
  {NoStop}%
\bibitem [{\citenamefont {Aebischer}\ \emph {et~al.}(2018)\citenamefont
  {Aebischer}, \citenamefont {Kumar},\ and\ \citenamefont
  {Straub}}]{Aebischer:2018bkb}%
  \BibitemOpen
  \bibfield  {author} {\bibinfo {author} {\bibfnamefont {J.}~\bibnamefont
  {Aebischer}}, \bibinfo {author} {\bibfnamefont {J.}~\bibnamefont {Kumar}}, \
  and\ \bibinfo {author} {\bibfnamefont {D.~M.}\ \bibnamefont {Straub}},\
  }\href {\doibase 10.1140/epjc/s10052-018-6492-7} {\bibfield  {journal}
  {\bibinfo  {journal} {Eur. Phys. J. C}\ }\textbf {\bibinfo {volume} {78}},\
  \bibinfo {pages} {1026} (\bibinfo {year} {2018})},\ \Eprint
  {http://arxiv.org/abs/1804.05033} {arXiv:1804.05033 [hep-ph]} \BibitemShut
  {NoStop}%
\bibitem [{\citenamefont {Straub}(2018)}]{Straub:2018kue}%
  \BibitemOpen
  \bibfield  {author} {\bibinfo {author} {\bibfnamefont {D.~M.}\ \bibnamefont
  {Straub}},\ }\href@noop {} {\  (\bibinfo {year} {2018})},\ \Eprint
  {http://arxiv.org/abs/1810.08132} {arXiv:1810.08132 [hep-ph]} \BibitemShut
  {NoStop}%
\bibitem [{\citenamefont {Falkowski}\ and\ \citenamefont
  {Straub}(2020)}]{Falkowski:2019hvp}%
  \BibitemOpen
  \bibfield  {author} {\bibinfo {author} {\bibfnamefont {A.}~\bibnamefont
  {Falkowski}}\ and\ \bibinfo {author} {\bibfnamefont {D.}~\bibnamefont
  {Straub}},\ }\href {\doibase 10.1007/JHEP04(2020)066} {\bibfield  {journal}
  {\bibinfo  {journal} {JHEP}\ }\textbf {\bibinfo {volume} {04}},\ \bibinfo
  {pages} {066} (\bibinfo {year} {2020})},\ \Eprint
  {http://arxiv.org/abs/1911.07866} {arXiv:1911.07866 [hep-ph]} \BibitemShut
  {NoStop}%
\bibitem [{sme()}]{smelli}%
  \BibitemOpen
  \href@noop {} {}\bibinfo {howpublished}
  {\url{https://github.com/smelli/smelli}}\BibitemShut {NoStop}%
\bibitem [{\citenamefont {Altmannshofer}\ and\ \citenamefont
  {Stangl}(2021)}]{Altmannshofer:2021qrr}%
  \BibitemOpen
  \bibfield  {author} {\bibinfo {author} {\bibfnamefont {W.}~\bibnamefont
  {Altmannshofer}}\ and\ \bibinfo {author} {\bibfnamefont {P.}~\bibnamefont
  {Stangl}},\ }\href@noop {} {\  (\bibinfo {year} {2021})},\ \Eprint
  {http://arxiv.org/abs/2103.13370} {arXiv:2103.13370 [hep-ph]} \BibitemShut
  {NoStop}%
\bibitem [{\citenamefont {Santimaria}()}]{LHCb_Bsmumu}%
  \BibitemOpen
  \bibfield  {author} {\bibinfo {author} {\bibfnamefont {M.}~\bibnamefont
  {Santimaria}} (\bibinfo {collaboration} {LHCb}),\ }\href@noop {} {}\bibinfo
  {howpublished}
  {\url{https://indico.cern.ch/event/976688/attachments/2213706/3747159/santimaria_LHC_seminar_2021.pdf}}\BibitemShut
  {NoStop}%
\bibitem [{\citenamefont {Aad}\ \emph {et~al.}(2020)\citenamefont {Aad} \emph
  {et~al.}}]{Aad:2020iuy}%
  \BibitemOpen
  \bibfield  {author} {\bibinfo {author} {\bibfnamefont {G.}~\bibnamefont
  {Aad}} \emph {et~al.} (\bibinfo {collaboration} {ATLAS}),\ }\href {\doibase
  10.1007/JHEP10(2020)112} {\bibfield  {journal} {\bibinfo  {journal} {JHEP}\
  }\textbf {\bibinfo {volume} {10}},\ \bibinfo {pages} {112} (\bibinfo {year}
  {2020})},\ \Eprint {http://arxiv.org/abs/2006.05872} {arXiv:2006.05872
  [hep-ex]} \BibitemShut {NoStop}%
\bibitem [{ATL(2020)}]{ATLAS:2020qoc}%
  \BibitemOpen
  \href@noop {} {\  (\bibinfo {year} {2020})}\BibitemShut {NoStop}%
\bibitem [{\citenamefont {Cerri}\ \emph {et~al.}(2019)\citenamefont {Cerri}
  \emph {et~al.}}]{Cerri:2018ypt}%
  \BibitemOpen
  \bibfield  {author} {\bibinfo {author} {\bibfnamefont {A.}~\bibnamefont
  {Cerri}} \emph {et~al.},\ }\href {\doibase 10.23731/CYRM-2019-007.867}
  {\bibfield  {journal} {\bibinfo  {journal} {CERN Yellow Rep. Monogr.}\
  }\textbf {\bibinfo {volume} {7}},\ \bibinfo {pages} {867} (\bibinfo {year}
  {2019})},\ \Eprint {http://arxiv.org/abs/1812.07638} {arXiv:1812.07638
  [hep-ph]} \BibitemShut {NoStop}%
\bibitem [{\citenamefont {Greljo}\ and\ \citenamefont
  {Marzocca}(2017)}]{Greljo:2017vvb}%
  \BibitemOpen
  \bibfield  {author} {\bibinfo {author} {\bibfnamefont {A.}~\bibnamefont
  {Greljo}}\ and\ \bibinfo {author} {\bibfnamefont {D.}~\bibnamefont
  {Marzocca}},\ }\href {\doibase 10.1140/epjc/s10052-017-5119-8} {\bibfield
  {journal} {\bibinfo  {journal} {Eur. Phys. J. C}\ }\textbf {\bibinfo {volume}
  {77}},\ \bibinfo {pages} {548} (\bibinfo {year} {2017})},\ \Eprint
  {http://arxiv.org/abs/1704.09015} {arXiv:1704.09015 [hep-ph]} \BibitemShut
  {NoStop}%
\bibitem [{\citenamefont {Gherardi}\ \emph {et~al.}(2019)\citenamefont
  {Gherardi}, \citenamefont {Marzocca}, \citenamefont {Nardecchia},\ and\
  \citenamefont {Romanino}}]{Gherardi:2019zil}%
  \BibitemOpen
  \bibfield  {author} {\bibinfo {author} {\bibfnamefont {V.}~\bibnamefont
  {Gherardi}}, \bibinfo {author} {\bibfnamefont {D.}~\bibnamefont {Marzocca}},
  \bibinfo {author} {\bibfnamefont {M.}~\bibnamefont {Nardecchia}}, \ and\
  \bibinfo {author} {\bibfnamefont {A.}~\bibnamefont {Romanino}},\ }\href
  {\doibase 10.1007/JHEP10(2019)112} {\bibfield  {journal} {\bibinfo  {journal}
  {JHEP}\ }\textbf {\bibinfo {volume} {10}},\ \bibinfo {pages} {112} (\bibinfo
  {year} {2019})},\ \Eprint {http://arxiv.org/abs/1903.10954} {arXiv:1903.10954
  [hep-ph]} \BibitemShut {NoStop}%
\bibitem [{\citenamefont {Aad}\ \emph {et~al.}(2019)\citenamefont {Aad} \emph
  {et~al.}}]{Aad:2019fac}%
  \BibitemOpen
  \bibfield  {author} {\bibinfo {author} {\bibfnamefont {G.}~\bibnamefont
  {Aad}} \emph {et~al.} (\bibinfo {collaboration} {ATLAS}),\ }\href {\doibase
  10.1016/j.physletb.2019.07.016} {\bibfield  {journal} {\bibinfo  {journal}
  {Phys. Lett. B}\ }\textbf {\bibinfo {volume} {796}},\ \bibinfo {pages} {68}
  (\bibinfo {year} {2019})},\ \Eprint {http://arxiv.org/abs/1903.06248}
  {arXiv:1903.06248 [hep-ex]} \BibitemShut {NoStop}%
\bibitem [{\citenamefont {Chalons}\ \emph {et~al.}(2016)\citenamefont
  {Chalons}, \citenamefont {Lopez-Val}, \citenamefont {Robens},\ and\
  \citenamefont {Stefaniak}}]{Chalons:2016jeu}%
  \BibitemOpen
  \bibfield  {author} {\bibinfo {author} {\bibfnamefont {G.}~\bibnamefont
  {Chalons}}, \bibinfo {author} {\bibfnamefont {D.}~\bibnamefont {Lopez-Val}},
  \bibinfo {author} {\bibfnamefont {T.}~\bibnamefont {Robens}}, \ and\ \bibinfo
  {author} {\bibfnamefont {T.}~\bibnamefont {Stefaniak}},\ }\href {\doibase
  10.22323/1.282.1180} {\bibfield  {journal} {\bibinfo  {journal} {PoS}\
  }\textbf {\bibinfo {volume} {ICHEP2016}},\ \bibinfo {pages} {1180} (\bibinfo
  {year} {2016})},\ \Eprint {http://arxiv.org/abs/1611.03007} {arXiv:1611.03007
  [hep-ph]} \BibitemShut {NoStop}%
\bibitem [{\citenamefont {Adhikari}\ \emph {et~al.}(2020)\citenamefont
  {Adhikari}, \citenamefont {Lewis},\ and\ \citenamefont
  {Sullivan}}]{Adhikari:2020vqo}%
  \BibitemOpen
  \bibfield  {author} {\bibinfo {author} {\bibfnamefont {S.}~\bibnamefont
  {Adhikari}}, \bibinfo {author} {\bibfnamefont {I.~M.}\ \bibnamefont {Lewis}},
  \ and\ \bibinfo {author} {\bibfnamefont {M.}~\bibnamefont {Sullivan}},\
  }\href@noop {} {\  (\bibinfo {year} {2020})},\ \Eprint
  {http://arxiv.org/abs/2003.10449} {arXiv:2003.10449 [hep-ph]} \BibitemShut
  {NoStop}%
\bibitem [{\citenamefont {Thomsen}(2021)}]{Thomsen:2021ncy}%
  \BibitemOpen
  \bibfield  {author} {\bibinfo {author} {\bibfnamefont {A.~E.}\ \bibnamefont
  {Thomsen}},\ }\href@noop {} {\  (\bibinfo {year} {2021})},\ \Eprint
  {http://arxiv.org/abs/2101.08265} {arXiv:2101.08265 [hep-ph]} \BibitemShut
  {NoStop}%
\bibitem [{\citenamefont {Pickering}\ \emph {et~al.}(2001)\citenamefont
  {Pickering}, \citenamefont {Gracey},\ and\ \citenamefont
  {Jones}}]{Pickering:2001aq}%
  \BibitemOpen
  \bibfield  {author} {\bibinfo {author} {\bibfnamefont {A.~G.~M.}\
  \bibnamefont {Pickering}}, \bibinfo {author} {\bibfnamefont {J.~A.}\
  \bibnamefont {Gracey}}, \ and\ \bibinfo {author} {\bibfnamefont {D.~R.~T.}\
  \bibnamefont {Jones}},\ }\href {\doibase 10.1016/S0370-2693(01)00624-4}
  {\bibfield  {journal} {\bibinfo  {journal} {Phys. Lett. B}\ }\textbf
  {\bibinfo {volume} {510}},\ \bibinfo {pages} {347} (\bibinfo {year}
  {2001})},\ \bibinfo {note} {[Erratum: Phys.Lett.B 535, 377 (2002)]},\ \Eprint
  {http://arxiv.org/abs/hep-ph/0104247} {arXiv:hep-ph/0104247} \BibitemShut
  {NoStop}%
\bibitem [{\citenamefont {Poole}\ and\ \citenamefont
  {Thomsen}(2019)}]{Poole:2019kcm}%
  \BibitemOpen
  \bibfield  {author} {\bibinfo {author} {\bibfnamefont {C.}~\bibnamefont
  {Poole}}\ and\ \bibinfo {author} {\bibfnamefont {A.~E.}\ \bibnamefont
  {Thomsen}},\ }\href {\doibase 10.1007/JHEP09(2019)055} {\bibfield  {journal}
  {\bibinfo  {journal} {JHEP}\ }\textbf {\bibinfo {volume} {09}},\ \bibinfo
  {pages} {055} (\bibinfo {year} {2019})},\ \Eprint
  {http://arxiv.org/abs/1906.04625} {arXiv:1906.04625 [hep-th]} \BibitemShut
  {NoStop}%
\bibitem [{\citenamefont {Crivellin}\ \emph {et~al.}(2020)\citenamefont
  {Crivellin}, \citenamefont {M\"uller},\ and\ \citenamefont
  {Saturnino}}]{Crivellin:2020ukd}%
  \BibitemOpen
  \bibfield  {author} {\bibinfo {author} {\bibfnamefont {A.}~\bibnamefont
  {Crivellin}}, \bibinfo {author} {\bibfnamefont {D.}~\bibnamefont {M\"uller}},
  \ and\ \bibinfo {author} {\bibfnamefont {F.}~\bibnamefont {Saturnino}},\
  }\href {\doibase 10.1007/JHEP11(2020)094} {\bibfield  {journal} {\bibinfo
  {journal} {JHEP}\ }\textbf {\bibinfo {volume} {11}},\ \bibinfo {pages} {094}
  (\bibinfo {year} {2020})},\ \Eprint {http://arxiv.org/abs/2006.10758}
  {arXiv:2006.10758 [hep-ph]} \BibitemShut {NoStop}%
\bibitem [{\citenamefont {Hook}\ \emph {et~al.}(2011)\citenamefont {Hook},
  \citenamefont {Izaguirre},\ and\ \citenamefont {Wacker}}]{Hook:2010tw}%
  \BibitemOpen
  \bibfield  {author} {\bibinfo {author} {\bibfnamefont {A.}~\bibnamefont
  {Hook}}, \bibinfo {author} {\bibfnamefont {E.}~\bibnamefont {Izaguirre}}, \
  and\ \bibinfo {author} {\bibfnamefont {J.~G.}\ \bibnamefont {Wacker}},\
  }\href {\doibase 10.1155/2011/859762} {\bibfield  {journal} {\bibinfo
  {journal} {Adv. High Energy Phys.}\ }\textbf {\bibinfo {volume} {2011}},\
  \bibinfo {pages} {859762} (\bibinfo {year} {2011})},\ \Eprint
  {http://arxiv.org/abs/1006.0973} {arXiv:1006.0973 [hep-ph]} \BibitemShut
  {NoStop}%
\bibitem [{\citenamefont {Capdevilla}\ \emph {et~al.}(2020)\citenamefont
  {Capdevilla}, \citenamefont {Curtin}, \citenamefont {Kahn},\ and\
  \citenamefont {Krnjaic}}]{Capdevilla:2020qel}%
  \BibitemOpen
  \bibfield  {author} {\bibinfo {author} {\bibfnamefont {R.}~\bibnamefont
  {Capdevilla}}, \bibinfo {author} {\bibfnamefont {D.}~\bibnamefont {Curtin}},
  \bibinfo {author} {\bibfnamefont {Y.}~\bibnamefont {Kahn}}, \ and\ \bibinfo
  {author} {\bibfnamefont {G.}~\bibnamefont {Krnjaic}},\ }\href@noop {} {\
  (\bibinfo {year} {2020})},\ \Eprint {http://arxiv.org/abs/2006.16277}
  {arXiv:2006.16277 [hep-ph]} \BibitemShut {NoStop}%
\bibitem [{\citenamefont {Capdevilla}\ \emph {et~al.}(2021)\citenamefont
  {Capdevilla}, \citenamefont {Curtin}, \citenamefont {Kahn},\ and\
  \citenamefont {Krnjaic}}]{Capdevilla:2021rwo}%
  \BibitemOpen
  \bibfield  {author} {\bibinfo {author} {\bibfnamefont {R.}~\bibnamefont
  {Capdevilla}}, \bibinfo {author} {\bibfnamefont {D.}~\bibnamefont {Curtin}},
  \bibinfo {author} {\bibfnamefont {Y.}~\bibnamefont {Kahn}}, \ and\ \bibinfo
  {author} {\bibfnamefont {G.}~\bibnamefont {Krnjaic}},\ }\href@noop {} {\
  (\bibinfo {year} {2021})},\ \Eprint {http://arxiv.org/abs/2101.10334}
  {arXiv:2101.10334 [hep-ph]} \BibitemShut {NoStop}%
\bibitem [{\citenamefont {Esteban}\ \emph {et~al.}(2019)\citenamefont
  {Esteban}, \citenamefont {Gonzalez-Garcia}, \citenamefont
  {Hernandez-Cabezudo}, \citenamefont {Maltoni},\ and\ \citenamefont
  {Schwetz}}]{Esteban:2018azc}%
  \BibitemOpen
  \bibfield  {author} {\bibinfo {author} {\bibfnamefont {I.}~\bibnamefont
  {Esteban}}, \bibinfo {author} {\bibfnamefont {M.~C.}\ \bibnamefont
  {Gonzalez-Garcia}}, \bibinfo {author} {\bibfnamefont {A.}~\bibnamefont
  {Hernandez-Cabezudo}}, \bibinfo {author} {\bibfnamefont {M.}~\bibnamefont
  {Maltoni}}, \ and\ \bibinfo {author} {\bibfnamefont {T.}~\bibnamefont
  {Schwetz}},\ }\href {\doibase 10.1007/JHEP01(2019)106} {\bibfield  {journal}
  {\bibinfo  {journal} {JHEP}\ }\textbf {\bibinfo {volume} {01}},\ \bibinfo
  {pages} {106} (\bibinfo {year} {2019})},\ \Eprint
  {http://arxiv.org/abs/1811.05487} {arXiv:1811.05487 [hep-ph]} \BibitemShut
  {NoStop}%
\bibitem [{\citenamefont {Aghanim}\ \emph {et~al.}(2020)\citenamefont {Aghanim}
  \emph {et~al.}}]{Aghanim:2018eyx}%
  \BibitemOpen
  \bibfield  {author} {\bibinfo {author} {\bibfnamefont {N.}~\bibnamefont
  {Aghanim}} \emph {et~al.} (\bibinfo {collaboration} {Planck}),\ }\href
  {\doibase 10.1051/0004-6361/201833910} {\bibfield  {journal} {\bibinfo
  {journal} {Astron. Astrophys.}\ }\textbf {\bibinfo {volume} {641}},\ \bibinfo
  {pages} {A6} (\bibinfo {year} {2020})},\ \Eprint
  {http://arxiv.org/abs/1807.06209} {arXiv:1807.06209 [astro-ph.CO]}
  \BibitemShut {NoStop}%
\bibitem [{\citenamefont {Gando}\ \emph {et~al.}(2016)\citenamefont {Gando}
  \emph {et~al.}}]{KamLAND-Zen:2016pfg}%
  \BibitemOpen
  \bibfield  {author} {\bibinfo {author} {\bibfnamefont {A.}~\bibnamefont
  {Gando}} \emph {et~al.} (\bibinfo {collaboration} {KamLAND-Zen}),\ }\href
  {\doibase 10.1103/PhysRevLett.117.082503} {\bibfield  {journal} {\bibinfo
  {journal} {Phys. Rev. Lett.}\ }\textbf {\bibinfo {volume} {117}},\ \bibinfo
  {pages} {082503} (\bibinfo {year} {2016})},\ \bibinfo {note} {[Addendum:
  Phys.Rev.Lett. 117, 109903 (2016)]},\ \Eprint
  {http://arxiv.org/abs/1605.02889} {arXiv:1605.02889 [hep-ex]} \BibitemShut
  {NoStop}%
\bibitem [{\citenamefont {Brdar}\ \emph {et~al.}(2021)\citenamefont {Brdar},
  \citenamefont {Greljo}, \citenamefont {Kopp},\ and\ \citenamefont
  {Opferkuch}}]{Brdar:2020quo}%
  \BibitemOpen
  \bibfield  {author} {\bibinfo {author} {\bibfnamefont {V.}~\bibnamefont
  {Brdar}}, \bibinfo {author} {\bibfnamefont {A.}~\bibnamefont {Greljo}},
  \bibinfo {author} {\bibfnamefont {J.}~\bibnamefont {Kopp}}, \ and\ \bibinfo
  {author} {\bibfnamefont {T.}~\bibnamefont {Opferkuch}},\ }\href {\doibase
  10.1088/1475-7516/2021/01/039} {\bibfield  {journal} {\bibinfo  {journal}
  {JCAP}\ }\textbf {\bibinfo {volume} {01}},\ \bibinfo {pages} {039} (\bibinfo
  {year} {2021})},\ \Eprint {http://arxiv.org/abs/2007.15563} {arXiv:2007.15563
  [hep-ph]} \BibitemShut {NoStop}%
\bibitem [{\citenamefont {Arnold}\ \emph {et~al.}(2013)\citenamefont {Arnold},
  \citenamefont {Fornal},\ and\ \citenamefont {Wise}}]{Arnold:2013cva}%
  \BibitemOpen
  \bibfield  {author} {\bibinfo {author} {\bibfnamefont {J.~M.}\ \bibnamefont
  {Arnold}}, \bibinfo {author} {\bibfnamefont {B.}~\bibnamefont {Fornal}}, \
  and\ \bibinfo {author} {\bibfnamefont {M.~B.}\ \bibnamefont {Wise}},\ }\href
  {\doibase 10.1103/PhysRevD.88.035009} {\bibfield  {journal} {\bibinfo
  {journal} {Phys. Rev. D}\ }\textbf {\bibinfo {volume} {88}},\ \bibinfo
  {pages} {035009} (\bibinfo {year} {2013})},\ \Eprint
  {http://arxiv.org/abs/1304.6119} {arXiv:1304.6119 [hep-ph]} \BibitemShut
  {NoStop}%
\bibitem [{\citenamefont {Assad}\ \emph {et~al.}(2018)\citenamefont {Assad},
  \citenamefont {Fornal},\ and\ \citenamefont {Grinstein}}]{Assad:2017iib}%
  \BibitemOpen
  \bibfield  {author} {\bibinfo {author} {\bibfnamefont {N.}~\bibnamefont
  {Assad}}, \bibinfo {author} {\bibfnamefont {B.}~\bibnamefont {Fornal}}, \
  and\ \bibinfo {author} {\bibfnamefont {B.}~\bibnamefont {Grinstein}},\ }\href
  {\doibase 10.1016/j.physletb.2017.12.042} {\bibfield  {journal} {\bibinfo
  {journal} {Phys. Lett. B}\ }\textbf {\bibinfo {volume} {777}},\ \bibinfo
  {pages} {324} (\bibinfo {year} {2018})},\ \Eprint
  {http://arxiv.org/abs/1708.06350} {arXiv:1708.06350 [hep-ph]} \BibitemShut
  {NoStop}%
\bibitem [{\citenamefont {Banks}\ and\ \citenamefont
  {Seiberg}(2011)}]{Banks:2010zn}%
  \BibitemOpen
  \bibfield  {author} {\bibinfo {author} {\bibfnamefont {T.}~\bibnamefont
  {Banks}}\ and\ \bibinfo {author} {\bibfnamefont {N.}~\bibnamefont
  {Seiberg}},\ }\href {\doibase 10.1103/PhysRevD.83.084019} {\bibfield
  {journal} {\bibinfo  {journal} {Phys. Rev. D}\ }\textbf {\bibinfo {volume}
  {83}},\ \bibinfo {pages} {084019} (\bibinfo {year} {2011})},\ \Eprint
  {http://arxiv.org/abs/1011.5120} {arXiv:1011.5120 [hep-th]} \BibitemShut
  {NoStop}%
\bibitem [{\citenamefont {Di~Luzio}\ \emph {et~al.}(2017)\citenamefont
  {Di~Luzio}, \citenamefont {Greljo},\ and\ \citenamefont
  {Nardecchia}}]{DiLuzio:2017vat}%
  \BibitemOpen
  \bibfield  {author} {\bibinfo {author} {\bibfnamefont {L.}~\bibnamefont
  {Di~Luzio}}, \bibinfo {author} {\bibfnamefont {A.}~\bibnamefont {Greljo}}, \
  and\ \bibinfo {author} {\bibfnamefont {M.}~\bibnamefont {Nardecchia}},\
  }\href {\doibase 10.1103/PhysRevD.96.115011} {\bibfield  {journal} {\bibinfo
  {journal} {Phys. Rev. D}\ }\textbf {\bibinfo {volume} {96}},\ \bibinfo
  {pages} {115011} (\bibinfo {year} {2017})},\ \Eprint
  {http://arxiv.org/abs/1708.08450} {arXiv:1708.08450 [hep-ph]} \BibitemShut
  {NoStop}%
\bibitem [{\citenamefont {Greljo}\ and\ \citenamefont
  {Stefanek}(2018)}]{Greljo:2018tuh}%
  \BibitemOpen
  \bibfield  {author} {\bibinfo {author} {\bibfnamefont {A.}~\bibnamefont
  {Greljo}}\ and\ \bibinfo {author} {\bibfnamefont {B.~A.}\ \bibnamefont
  {Stefanek}},\ }\href {\doibase 10.1016/j.physletb.2018.05.033} {\bibfield
  {journal} {\bibinfo  {journal} {Phys. Lett. B}\ }\textbf {\bibinfo {volume}
  {782}},\ \bibinfo {pages} {131} (\bibinfo {year} {2018})},\ \Eprint
  {http://arxiv.org/abs/1802.04274} {arXiv:1802.04274 [hep-ph]} \BibitemShut
  {NoStop}%
\bibitem [{\citenamefont {Bordone}\ \emph
  {et~al.}(2018{\natexlab{a}})\citenamefont {Bordone}, \citenamefont
  {Cornella}, \citenamefont {Fuentes-Martin},\ and\ \citenamefont
  {Isidori}}]{Bordone:2017bld}%
  \BibitemOpen
  \bibfield  {author} {\bibinfo {author} {\bibfnamefont {M.}~\bibnamefont
  {Bordone}}, \bibinfo {author} {\bibfnamefont {C.}~\bibnamefont {Cornella}},
  \bibinfo {author} {\bibfnamefont {J.}~\bibnamefont {Fuentes-Martin}}, \ and\
  \bibinfo {author} {\bibfnamefont {G.}~\bibnamefont {Isidori}},\ }\href
  {\doibase 10.1016/j.physletb.2018.02.011} {\bibfield  {journal} {\bibinfo
  {journal} {Phys. Lett. B}\ }\textbf {\bibinfo {volume} {779}},\ \bibinfo
  {pages} {317} (\bibinfo {year} {2018}{\natexlab{a}})},\ \Eprint
  {http://arxiv.org/abs/1712.01368} {arXiv:1712.01368 [hep-ph]} \BibitemShut
  {NoStop}%
\bibitem [{\citenamefont {Bordone}\ \emph
  {et~al.}(2018{\natexlab{b}})\citenamefont {Bordone}, \citenamefont
  {Cornella}, \citenamefont {Fuentes-Mart\'\i{}n},\ and\ \citenamefont
  {Isidori}}]{Bordone:2018nbg}%
  \BibitemOpen
  \bibfield  {author} {\bibinfo {author} {\bibfnamefont {M.}~\bibnamefont
  {Bordone}}, \bibinfo {author} {\bibfnamefont {C.}~\bibnamefont {Cornella}},
  \bibinfo {author} {\bibfnamefont {J.}~\bibnamefont {Fuentes-Mart\'\i{}n}}, \
  and\ \bibinfo {author} {\bibfnamefont {G.}~\bibnamefont {Isidori}},\ }\href
  {\doibase 10.1007/JHEP10(2018)148} {\bibfield  {journal} {\bibinfo  {journal}
  {JHEP}\ }\textbf {\bibinfo {volume} {10}},\ \bibinfo {pages} {148} (\bibinfo
  {year} {2018}{\natexlab{b}})},\ \Eprint {http://arxiv.org/abs/1805.09328}
  {arXiv:1805.09328 [hep-ph]} \BibitemShut {NoStop}%
\bibitem [{\citenamefont {Cornella}\ \emph {et~al.}(2019)\citenamefont
  {Cornella}, \citenamefont {Fuentes-Martin},\ and\ \citenamefont
  {Isidori}}]{Cornella:2019hct}%
  \BibitemOpen
  \bibfield  {author} {\bibinfo {author} {\bibfnamefont {C.}~\bibnamefont
  {Cornella}}, \bibinfo {author} {\bibfnamefont {J.}~\bibnamefont
  {Fuentes-Martin}}, \ and\ \bibinfo {author} {\bibfnamefont {G.}~\bibnamefont
  {Isidori}},\ }\href {\doibase 10.1007/JHEP07(2019)168} {\bibfield  {journal}
  {\bibinfo  {journal} {JHEP}\ }\textbf {\bibinfo {volume} {07}},\ \bibinfo
  {pages} {168} (\bibinfo {year} {2019})},\ \Eprint
  {http://arxiv.org/abs/1903.11517} {arXiv:1903.11517 [hep-ph]} \BibitemShut
  {NoStop}%
\bibitem [{\citenamefont {Fornal}\ \emph {et~al.}(2019)\citenamefont {Fornal},
  \citenamefont {Gadam},\ and\ \citenamefont {Grinstein}}]{Fornal:2018dqn}%
  \BibitemOpen
  \bibfield  {author} {\bibinfo {author} {\bibfnamefont {B.}~\bibnamefont
  {Fornal}}, \bibinfo {author} {\bibfnamefont {S.~A.}\ \bibnamefont {Gadam}}, \
  and\ \bibinfo {author} {\bibfnamefont {B.}~\bibnamefont {Grinstein}},\ }\href
  {\doibase 10.1103/PhysRevD.99.055025} {\bibfield  {journal} {\bibinfo
  {journal} {Phys. Rev. D}\ }\textbf {\bibinfo {volume} {99}},\ \bibinfo
  {pages} {055025} (\bibinfo {year} {2019})},\ \Eprint
  {http://arxiv.org/abs/1812.01603} {arXiv:1812.01603 [hep-ph]} \BibitemShut
  {NoStop}%
\bibitem [{\citenamefont {Blanke}\ and\ \citenamefont
  {Crivellin}(2018)}]{Blanke:2018sro}%
  \BibitemOpen
  \bibfield  {author} {\bibinfo {author} {\bibfnamefont {M.}~\bibnamefont
  {Blanke}}\ and\ \bibinfo {author} {\bibfnamefont {A.}~\bibnamefont
  {Crivellin}},\ }\href {\doibase 10.1103/PhysRevLett.121.011801} {\bibfield
  {journal} {\bibinfo  {journal} {Phys. Rev. Lett.}\ }\textbf {\bibinfo
  {volume} {121}},\ \bibinfo {pages} {011801} (\bibinfo {year} {2018})},\
  \Eprint {http://arxiv.org/abs/1801.07256} {arXiv:1801.07256 [hep-ph]}
  \BibitemShut {NoStop}%
\bibitem [{\citenamefont {Fuentes-Mart\'\i{}n}\ \emph
  {et~al.}(2020{\natexlab{b}})\citenamefont {Fuentes-Mart\'\i{}n},
  \citenamefont {Isidori}, \citenamefont {K\"onig},\ and\ \citenamefont
  {Selimovi\'c}}]{Fuentes-Martin:2019ign}%
  \BibitemOpen
  \bibfield  {author} {\bibinfo {author} {\bibfnamefont {J.}~\bibnamefont
  {Fuentes-Mart\'\i{}n}}, \bibinfo {author} {\bibfnamefont {G.}~\bibnamefont
  {Isidori}}, \bibinfo {author} {\bibfnamefont {M.}~\bibnamefont {K\"onig}}, \
  and\ \bibinfo {author} {\bibfnamefont {N.}~\bibnamefont {Selimovi\'c}},\
  }\href {\doibase 10.1103/PhysRevD.101.035024} {\bibfield  {journal} {\bibinfo
   {journal} {Phys. Rev. D}\ }\textbf {\bibinfo {volume} {101}},\ \bibinfo
  {pages} {035024} (\bibinfo {year} {2020}{\natexlab{b}})},\ \Eprint
  {http://arxiv.org/abs/1910.13474} {arXiv:1910.13474 [hep-ph]} \BibitemShut
  {NoStop}%
\bibitem [{\citenamefont {Guadagnoli}\ \emph {et~al.}(2020)\citenamefont
  {Guadagnoli}, \citenamefont {Reboud},\ and\ \citenamefont
  {Stangl}}]{Guadagnoli:2020tlx}%
  \BibitemOpen
  \bibfield  {author} {\bibinfo {author} {\bibfnamefont {D.}~\bibnamefont
  {Guadagnoli}}, \bibinfo {author} {\bibfnamefont {M.}~\bibnamefont {Reboud}},
  \ and\ \bibinfo {author} {\bibfnamefont {P.}~\bibnamefont {Stangl}},\ }\href
  {\doibase 10.1007/JHEP10(2020)084} {\bibfield  {journal} {\bibinfo  {journal}
  {JHEP}\ }\textbf {\bibinfo {volume} {10}},\ \bibinfo {pages} {084} (\bibinfo
  {year} {2020})},\ \Eprint {http://arxiv.org/abs/2005.10117} {arXiv:2005.10117
  [hep-ph]} \BibitemShut {NoStop}%
\bibitem [{\citenamefont {Heeck}\ and\ \citenamefont
  {Teresi}(2018)}]{Heeck:2018ntp}%
  \BibitemOpen
  \bibfield  {author} {\bibinfo {author} {\bibfnamefont {J.}~\bibnamefont
  {Heeck}}\ and\ \bibinfo {author} {\bibfnamefont {D.}~\bibnamefont {Teresi}},\
  }\href {\doibase 10.1007/JHEP12(2018)103} {\bibfield  {journal} {\bibinfo
  {journal} {JHEP}\ }\textbf {\bibinfo {volume} {12}},\ \bibinfo {pages} {103}
  (\bibinfo {year} {2018})},\ \Eprint {http://arxiv.org/abs/1808.07492}
  {arXiv:1808.07492 [hep-ph]} \BibitemShut {NoStop}%
\bibitem [{\citenamefont {Fuentes-Mart\'\i{}n}\ and\ \citenamefont
  {Stangl}(2020)}]{Fuentes-Martin:2020bnh}%
  \BibitemOpen
  \bibfield  {author} {\bibinfo {author} {\bibfnamefont {J.}~\bibnamefont
  {Fuentes-Mart\'\i{}n}}\ and\ \bibinfo {author} {\bibfnamefont
  {P.}~\bibnamefont {Stangl}},\ }\href {\doibase
  10.1016/j.physletb.2020.135953} {\bibfield  {journal} {\bibinfo  {journal}
  {Phys. Lett. B}\ }\textbf {\bibinfo {volume} {811}},\ \bibinfo {pages}
  {135953} (\bibinfo {year} {2020})},\ \Eprint
  {http://arxiv.org/abs/2004.11376} {arXiv:2004.11376 [hep-ph]} \BibitemShut
  {NoStop}%
\bibitem [{\citenamefont {Fuentes-Mart\'\i{}n}\ \emph
  {et~al.}(2019)\citenamefont {Fuentes-Mart\'\i{}n}, \citenamefont {Reig},\
  and\ \citenamefont {Vicente}}]{Fuentes-Martin:2019bue}%
  \BibitemOpen
  \bibfield  {author} {\bibinfo {author} {\bibfnamefont {J.}~\bibnamefont
  {Fuentes-Mart\'\i{}n}}, \bibinfo {author} {\bibfnamefont {M.}~\bibnamefont
  {Reig}}, \ and\ \bibinfo {author} {\bibfnamefont {A.}~\bibnamefont
  {Vicente}},\ }\href {\doibase 10.1103/PhysRevD.100.115028} {\bibfield
  {journal} {\bibinfo  {journal} {Phys. Rev. D}\ }\textbf {\bibinfo {volume}
  {100}},\ \bibinfo {pages} {115028} (\bibinfo {year} {2019})},\ \Eprint
  {http://arxiv.org/abs/1907.02550} {arXiv:1907.02550 [hep-ph]} \BibitemShut
  {NoStop}%
\bibitem [{\citenamefont {Fuentes-Mart\'\i{}n}\ \emph
  {et~al.}(2020{\natexlab{c}})\citenamefont {Fuentes-Mart\'\i{}n},
  \citenamefont {Isidori}, \citenamefont {K\"onig},\ and\ \citenamefont
  {Selimovi\'c}}]{Fuentes-Martin:2020luw}%
  \BibitemOpen
  \bibfield  {author} {\bibinfo {author} {\bibfnamefont {J.}~\bibnamefont
  {Fuentes-Mart\'\i{}n}}, \bibinfo {author} {\bibfnamefont {G.}~\bibnamefont
  {Isidori}}, \bibinfo {author} {\bibfnamefont {M.}~\bibnamefont {K\"onig}}, \
  and\ \bibinfo {author} {\bibfnamefont {N.}~\bibnamefont {Selimovi\'c}},\
  }\href {\doibase 10.1103/PhysRevD.102.035021} {\bibfield  {journal} {\bibinfo
   {journal} {Phys. Rev. D}\ }\textbf {\bibinfo {volume} {102}},\ \bibinfo
  {pages} {035021} (\bibinfo {year} {2020}{\natexlab{c}})},\ \Eprint
  {http://arxiv.org/abs/2006.16250} {arXiv:2006.16250 [hep-ph]} \BibitemShut
  {NoStop}%
\bibitem [{\citenamefont {Fuentes-Mart\'\i{}n}\ \emph
  {et~al.}(2020{\natexlab{d}})\citenamefont {Fuentes-Mart\'\i{}n},
  \citenamefont {Isidori}, \citenamefont {K\"onig},\ and\ \citenamefont
  {Selimovi\'c}}]{Fuentes-Martin:2020hvc}%
  \BibitemOpen
  \bibfield  {author} {\bibinfo {author} {\bibfnamefont {J.}~\bibnamefont
  {Fuentes-Mart\'\i{}n}}, \bibinfo {author} {\bibfnamefont {G.}~\bibnamefont
  {Isidori}}, \bibinfo {author} {\bibfnamefont {M.}~\bibnamefont {K\"onig}}, \
  and\ \bibinfo {author} {\bibfnamefont {N.}~\bibnamefont {Selimovi\'c}},\
  }\href {\doibase 10.1103/PhysRevD.102.115015} {\bibfield  {journal} {\bibinfo
   {journal} {Phys. Rev. D}\ }\textbf {\bibinfo {volume} {102}},\ \bibinfo
  {pages} {115015} (\bibinfo {year} {2020}{\natexlab{d}})},\ \Eprint
  {http://arxiv.org/abs/2009.11296} {arXiv:2009.11296 [hep-ph]} \BibitemShut
  {NoStop}%
\bibitem [{\citenamefont {Heeck}\ and\ \citenamefont
  {Rodejohann}(2011)}]{Heeck:2011wj}%
  \BibitemOpen
  \bibfield  {author} {\bibinfo {author} {\bibfnamefont {J.}~\bibnamefont
  {Heeck}}\ and\ \bibinfo {author} {\bibfnamefont {W.}~\bibnamefont
  {Rodejohann}},\ }\href {\doibase 10.1103/PhysRevD.84.075007} {\bibfield
  {journal} {\bibinfo  {journal} {Phys. Rev. D}\ }\textbf {\bibinfo {volume}
  {84}},\ \bibinfo {pages} {075007} (\bibinfo {year} {2011})},\ \Eprint
  {http://arxiv.org/abs/1107.5238} {arXiv:1107.5238 [hep-ph]} \BibitemShut
  {NoStop}%
\bibitem [{\citenamefont {Crivellin}\ \emph
  {et~al.}(2015{\natexlab{b}})\citenamefont {Crivellin}, \citenamefont
  {D'Ambrosio},\ and\ \citenamefont {Heeck}}]{Crivellin:2015lwa}%
  \BibitemOpen
  \bibfield  {author} {\bibinfo {author} {\bibfnamefont {A.}~\bibnamefont
  {Crivellin}}, \bibinfo {author} {\bibfnamefont {G.}~\bibnamefont
  {D'Ambrosio}}, \ and\ \bibinfo {author} {\bibfnamefont {J.}~\bibnamefont
  {Heeck}},\ }\href {\doibase 10.1103/PhysRevD.91.075006} {\bibfield  {journal}
  {\bibinfo  {journal} {Phys. Rev. D}\ }\textbf {\bibinfo {volume} {91}},\
  \bibinfo {pages} {075006} (\bibinfo {year} {2015}{\natexlab{b}})},\ \Eprint
  {http://arxiv.org/abs/1503.03477} {arXiv:1503.03477 [hep-ph]} \BibitemShut
  {NoStop}%
\bibitem [{\citenamefont {Nomura}\ and\ \citenamefont
  {Okada}(2018)}]{Nomura:2018cle}%
  \BibitemOpen
  \bibfield  {author} {\bibinfo {author} {\bibfnamefont {T.}~\bibnamefont
  {Nomura}}\ and\ \bibinfo {author} {\bibfnamefont {H.}~\bibnamefont {Okada}},\
  }\href {\doibase 10.1016/j.physletb.2018.07.011} {\bibfield  {journal}
  {\bibinfo  {journal} {Phys. Lett. B}\ }\textbf {\bibinfo {volume} {783}},\
  \bibinfo {pages} {381} (\bibinfo {year} {2018})},\ \Eprint
  {http://arxiv.org/abs/1805.03942} {arXiv:1805.03942 [hep-ph]} \BibitemShut
  {NoStop}%
\bibitem [{\citenamefont {Araki}\ \emph {et~al.}(2019)\citenamefont {Araki},
  \citenamefont {Asai}, \citenamefont {Sato},\ and\ \citenamefont
  {Shimomura}}]{Araki:2019rmw}%
  \BibitemOpen
  \bibfield  {author} {\bibinfo {author} {\bibfnamefont {T.}~\bibnamefont
  {Araki}}, \bibinfo {author} {\bibfnamefont {K.}~\bibnamefont {Asai}},
  \bibinfo {author} {\bibfnamefont {J.}~\bibnamefont {Sato}}, \ and\ \bibinfo
  {author} {\bibfnamefont {T.}~\bibnamefont {Shimomura}},\ }\href {\doibase
  10.1103/PhysRevD.100.095012} {\bibfield  {journal} {\bibinfo  {journal}
  {Phys. Rev. D}\ }\textbf {\bibinfo {volume} {100}},\ \bibinfo {pages}
  {095012} (\bibinfo {year} {2019})},\ \Eprint
  {http://arxiv.org/abs/1909.08827} {arXiv:1909.08827 [hep-ph]} \BibitemShut
  {NoStop}%
\bibitem [{\citenamefont {Gninenko}\ and\ \citenamefont
  {Krasnikov}(2001)}]{Gninenko:2001hx}%
  \BibitemOpen
  \bibfield  {author} {\bibinfo {author} {\bibfnamefont {S.~N.}\ \bibnamefont
  {Gninenko}}\ and\ \bibinfo {author} {\bibfnamefont {N.~V.}\ \bibnamefont
  {Krasnikov}},\ }\href {\doibase 10.1016/S0370-2693(01)00693-1} {\bibfield
  {journal} {\bibinfo  {journal} {Phys. Lett. B}\ }\textbf {\bibinfo {volume}
  {513}},\ \bibinfo {pages} {119} (\bibinfo {year} {2001})},\ \Eprint
  {http://arxiv.org/abs/hep-ph/0102222} {arXiv:hep-ph/0102222} \BibitemShut
  {NoStop}%
\bibitem [{\citenamefont {Araki}\ \emph {et~al.}(2021)\citenamefont {Araki},
  \citenamefont {Asai}, \citenamefont {Honda}, \citenamefont {Kasuya},
  \citenamefont {Sato}, \citenamefont {Shimomura},\ and\ \citenamefont
  {Yang}}]{Araki:2021xdk}%
  \BibitemOpen
  \bibfield  {author} {\bibinfo {author} {\bibfnamefont {T.}~\bibnamefont
  {Araki}}, \bibinfo {author} {\bibfnamefont {K.}~\bibnamefont {Asai}},
  \bibinfo {author} {\bibfnamefont {K.}~\bibnamefont {Honda}}, \bibinfo
  {author} {\bibfnamefont {R.}~\bibnamefont {Kasuya}}, \bibinfo {author}
  {\bibfnamefont {J.}~\bibnamefont {Sato}}, \bibinfo {author} {\bibfnamefont
  {T.}~\bibnamefont {Shimomura}}, \ and\ \bibinfo {author} {\bibfnamefont
  {M.~J.~S.}\ \bibnamefont {Yang}},\ }\href@noop {} {\  (\bibinfo {year}
  {2021})},\ \Eprint {http://arxiv.org/abs/2103.07167} {arXiv:2103.07167
  [hep-ph]} \BibitemShut {NoStop}%
\bibitem [{\citenamefont {Agostini}\ \emph {et~al.}(2019)\citenamefont
  {Agostini} \emph {et~al.}}]{Agostini:2017ixy}%
  \BibitemOpen
  \bibfield  {author} {\bibinfo {author} {\bibfnamefont {M.}~\bibnamefont
  {Agostini}} \emph {et~al.} (\bibinfo {collaboration} {Borexino}),\ }\href
  {\doibase 10.1103/PhysRevD.100.082004} {\bibfield  {journal} {\bibinfo
  {journal} {Phys. Rev. D}\ }\textbf {\bibinfo {volume} {100}},\ \bibinfo
  {pages} {082004} (\bibinfo {year} {2019})},\ \Eprint
  {http://arxiv.org/abs/1707.09279} {arXiv:1707.09279 [hep-ex]} \BibitemShut
  {NoStop}%
\bibitem [{\citenamefont {Mishra}\ \emph {et~al.}(1991)\citenamefont {Mishra}
  \emph {et~al.}}]{Mishra:1991bv}%
  \BibitemOpen
  \bibfield  {author} {\bibinfo {author} {\bibfnamefont {S.~R.}\ \bibnamefont
  {Mishra}} \emph {et~al.} (\bibinfo {collaboration} {CCFR}),\ }\href {\doibase
  10.1103/PhysRevLett.66.3117} {\bibfield  {journal} {\bibinfo  {journal}
  {Phys. Rev. Lett.}\ }\textbf {\bibinfo {volume} {66}},\ \bibinfo {pages}
  {3117} (\bibinfo {year} {1991})}\BibitemShut {NoStop}%
\bibitem [{\citenamefont {Ilten}\ \emph {et~al.}(2018)\citenamefont {Ilten},
  \citenamefont {Soreq}, \citenamefont {Williams},\ and\ \citenamefont
  {Xue}}]{Ilten:2018crw}%
  \BibitemOpen
  \bibfield  {author} {\bibinfo {author} {\bibfnamefont {P.}~\bibnamefont
  {Ilten}}, \bibinfo {author} {\bibfnamefont {Y.}~\bibnamefont {Soreq}},
  \bibinfo {author} {\bibfnamefont {M.}~\bibnamefont {Williams}}, \ and\
  \bibinfo {author} {\bibfnamefont {W.}~\bibnamefont {Xue}},\ }\href {\doibase
  10.1007/JHEP06(2018)004} {\bibfield  {journal} {\bibinfo  {journal} {JHEP}\
  }\textbf {\bibinfo {volume} {06}},\ \bibinfo {pages} {004} (\bibinfo {year}
  {2018})},\ \Eprint {http://arxiv.org/abs/1801.04847} {arXiv:1801.04847
  [hep-ph]} \BibitemShut {NoStop}%
\bibitem [{\citenamefont {Lees}\ \emph {et~al.}(2012)\citenamefont {Lees} \emph
  {et~al.}}]{Lees:2012xj}%
  \BibitemOpen
  \bibfield  {author} {\bibinfo {author} {\bibfnamefont {J.~P.}\ \bibnamefont
  {Lees}} \emph {et~al.} (\bibinfo {collaboration} {BaBar}),\ }\href {\doibase
  10.1103/PhysRevLett.109.101802} {\bibfield  {journal} {\bibinfo  {journal}
  {Phys. Rev. Lett.}\ }\textbf {\bibinfo {volume} {109}},\ \bibinfo {pages}
  {101802} (\bibinfo {year} {2012})},\ \Eprint {http://arxiv.org/abs/1205.5442}
  {arXiv:1205.5442 [hep-ex]} \BibitemShut {NoStop}%
\bibitem [{\citenamefont {Lees}\ \emph {et~al.}(2013)\citenamefont {Lees} \emph
  {et~al.}}]{Lees:2013uzd}%
  \BibitemOpen
  \bibfield  {author} {\bibinfo {author} {\bibfnamefont {J.~P.}\ \bibnamefont
  {Lees}} \emph {et~al.} (\bibinfo {collaboration} {BaBar}),\ }\href {\doibase
  10.1103/PhysRevD.88.072012} {\bibfield  {journal} {\bibinfo  {journal} {Phys.
  Rev. D}\ }\textbf {\bibinfo {volume} {88}},\ \bibinfo {pages} {072012}
  (\bibinfo {year} {2013})},\ \Eprint {http://arxiv.org/abs/1303.0571}
  {arXiv:1303.0571 [hep-ex]} \BibitemShut {NoStop}%
\bibitem [{\citenamefont {Huschle}\ \emph {et~al.}(2015)\citenamefont {Huschle}
  \emph {et~al.}}]{Huschle:2015rga}%
  \BibitemOpen
  \bibfield  {author} {\bibinfo {author} {\bibfnamefont {M.}~\bibnamefont
  {Huschle}} \emph {et~al.} (\bibinfo {collaboration} {Belle}),\ }\href
  {\doibase 10.1103/PhysRevD.92.072014} {\bibfield  {journal} {\bibinfo
  {journal} {Phys. Rev. D}\ }\textbf {\bibinfo {volume} {92}},\ \bibinfo
  {pages} {072014} (\bibinfo {year} {2015})},\ \Eprint
  {http://arxiv.org/abs/1507.03233} {arXiv:1507.03233 [hep-ex]} \BibitemShut
  {NoStop}%
\bibitem [{\citenamefont {Sato}\ \emph {et~al.}(2016)\citenamefont {Sato} \emph
  {et~al.}}]{Sato:2016svk}%
  \BibitemOpen
  \bibfield  {author} {\bibinfo {author} {\bibfnamefont {Y.}~\bibnamefont
  {Sato}} \emph {et~al.} (\bibinfo {collaboration} {Belle}),\ }\href {\doibase
  10.1103/PhysRevD.94.072007} {\bibfield  {journal} {\bibinfo  {journal} {Phys.
  Rev. D}\ }\textbf {\bibinfo {volume} {94}},\ \bibinfo {pages} {072007}
  (\bibinfo {year} {2016})},\ \Eprint {http://arxiv.org/abs/1607.07923}
  {arXiv:1607.07923 [hep-ex]} \BibitemShut {NoStop}%
\bibitem [{\citenamefont {Aaij}\ \emph {et~al.}(2015)\citenamefont {Aaij} \emph
  {et~al.}}]{Aaij:2015yra}%
  \BibitemOpen
  \bibfield  {author} {\bibinfo {author} {\bibfnamefont {R.}~\bibnamefont
  {Aaij}} \emph {et~al.} (\bibinfo {collaboration} {LHCb}),\ }\href {\doibase
  10.1103/PhysRevLett.115.111803} {\bibfield  {journal} {\bibinfo  {journal}
  {Phys. Rev. Lett.}\ }\textbf {\bibinfo {volume} {115}},\ \bibinfo {pages}
  {111803} (\bibinfo {year} {2015})},\ \bibinfo {note} {[Erratum:
  Phys.Rev.Lett. 115, 159901 (2015)]},\ \Eprint
  {http://arxiv.org/abs/1506.08614} {arXiv:1506.08614 [hep-ex]} \BibitemShut
  {NoStop}%
\bibitem [{\citenamefont {Hirose}\ \emph {et~al.}(2017)\citenamefont {Hirose}
  \emph {et~al.}}]{Hirose:2016wfn}%
  \BibitemOpen
  \bibfield  {author} {\bibinfo {author} {\bibfnamefont {S.}~\bibnamefont
  {Hirose}} \emph {et~al.} (\bibinfo {collaboration} {Belle}),\ }\href
  {\doibase 10.1103/PhysRevLett.118.211801} {\bibfield  {journal} {\bibinfo
  {journal} {Phys. Rev. Lett.}\ }\textbf {\bibinfo {volume} {118}},\ \bibinfo
  {pages} {211801} (\bibinfo {year} {2017})},\ \Eprint
  {http://arxiv.org/abs/1612.00529} {arXiv:1612.00529 [hep-ex]} \BibitemShut
  {NoStop}%
\bibitem [{\citenamefont {Hirose}\ \emph {et~al.}(2018)\citenamefont {Hirose}
  \emph {et~al.}}]{Hirose:2017dxl}%
  \BibitemOpen
  \bibfield  {author} {\bibinfo {author} {\bibfnamefont {S.}~\bibnamefont
  {Hirose}} \emph {et~al.} (\bibinfo {collaboration} {Belle}),\ }\href
  {\doibase 10.1103/PhysRevD.97.012004} {\bibfield  {journal} {\bibinfo
  {journal} {Phys. Rev. D}\ }\textbf {\bibinfo {volume} {97}},\ \bibinfo
  {pages} {012004} (\bibinfo {year} {2018})},\ \Eprint
  {http://arxiv.org/abs/1709.00129} {arXiv:1709.00129 [hep-ex]} \BibitemShut
  {NoStop}%
\bibitem [{\citenamefont {Aaij}\ \emph
  {et~al.}(2018{\natexlab{a}})\citenamefont {Aaij} \emph
  {et~al.}}]{Aaij:2017uff}%
  \BibitemOpen
  \bibfield  {author} {\bibinfo {author} {\bibfnamefont {R.}~\bibnamefont
  {Aaij}} \emph {et~al.} (\bibinfo {collaboration} {LHCb}),\ }\href {\doibase
  10.1103/PhysRevLett.120.171802} {\bibfield  {journal} {\bibinfo  {journal}
  {Phys. Rev. Lett.}\ }\textbf {\bibinfo {volume} {120}},\ \bibinfo {pages}
  {171802} (\bibinfo {year} {2018}{\natexlab{a}})},\ \Eprint
  {http://arxiv.org/abs/1708.08856} {arXiv:1708.08856 [hep-ex]} \BibitemShut
  {NoStop}%
\bibitem [{\citenamefont {Aaij}\ \emph
  {et~al.}(2018{\natexlab{b}})\citenamefont {Aaij} \emph
  {et~al.}}]{Aaij:2017deq}%
  \BibitemOpen
  \bibfield  {author} {\bibinfo {author} {\bibfnamefont {R.}~\bibnamefont
  {Aaij}} \emph {et~al.} (\bibinfo {collaboration} {LHCb}),\ }\href {\doibase
  10.1103/PhysRevD.97.072013} {\bibfield  {journal} {\bibinfo  {journal} {Phys.
  Rev. D}\ }\textbf {\bibinfo {volume} {97}},\ \bibinfo {pages} {072013}
  (\bibinfo {year} {2018}{\natexlab{b}})},\ \Eprint
  {http://arxiv.org/abs/1711.02505} {arXiv:1711.02505 [hep-ex]} \BibitemShut
  {NoStop}%
\bibitem [{\citenamefont {Aaij}\ \emph
  {et~al.}(2018{\natexlab{c}})\citenamefont {Aaij} \emph
  {et~al.}}]{Aaij:2017tyk}%
  \BibitemOpen
  \bibfield  {author} {\bibinfo {author} {\bibfnamefont {R.}~\bibnamefont
  {Aaij}} \emph {et~al.} (\bibinfo {collaboration} {LHCb}),\ }\href {\doibase
  10.1103/PhysRevLett.120.121801} {\bibfield  {journal} {\bibinfo  {journal}
  {Phys. Rev. Lett.}\ }\textbf {\bibinfo {volume} {120}},\ \bibinfo {pages}
  {121801} (\bibinfo {year} {2018}{\natexlab{c}})},\ \Eprint
  {http://arxiv.org/abs/1711.05623} {arXiv:1711.05623 [hep-ex]} \BibitemShut
  {NoStop}%
\bibitem [{\citenamefont {Aoki}\ \emph {et~al.}(2017)\citenamefont {Aoki} \emph
  {et~al.}}]{Aoki:2016frl}%
  \BibitemOpen
  \bibfield  {author} {\bibinfo {author} {\bibfnamefont {S.}~\bibnamefont
  {Aoki}} \emph {et~al.},\ }\href {\doibase 10.1140/epjc/s10052-016-4509-7}
  {\bibfield  {journal} {\bibinfo  {journal} {Eur. Phys. J. C}\ }\textbf
  {\bibinfo {volume} {77}},\ \bibinfo {pages} {112} (\bibinfo {year} {2017})},\
  \Eprint {http://arxiv.org/abs/1607.00299} {arXiv:1607.00299 [hep-lat]}
  \BibitemShut {NoStop}%
\end{thebibliography}%

\end{document}